\documentclass[twoside]{IEEEtran}
\usepackage[cmex10]{amsmath}
\usepackage{amsfonts}
\usepackage{amssymb}
\usepackage{amsbsy}
\usepackage{amsthm}
\usepackage{mathrsfs}
\usepackage{cases}
\usepackage{graphicx}
\usepackage{times}
\usepackage{rotating}
\usepackage{bm}
\usepackage{bbm}
\usepackage[table]{xcolor}
\usepackage{cite}
\usepackage{enumitem} 
\usepackage{centernot}

\allowdisplaybreaks[4]

\newtheoremstyle{unnumbered}
  {3pt plus 1pt minus 2pt} 
  {3pt plus 1pt minus 2pt} 
  {} 
  {1em} 
  {\itshape} 
  {:} 
  {.3em} 
  {#1} 

\newtheoremstyle{numbered}
  {3pt plus 1pt minus 2pt} 
  {3pt plus 1pt minus 2pt} 
  {} 
  {1em} 
  {\itshape} 
  {:} 
  {.3em} 
  {\thmname{#1}\thmnumber{ #2}\thmnote{ (#3)}} 

\theoremstyle{numbered}
\newtheorem{theorem}{Theorem}
\newtheorem{corollary}[theorem]{Corollary}
\newtheorem{lemma}[theorem]{Lemma}
\newtheorem{definition}[theorem]{Definition}
\newtheorem{remark}[theorem]{Remark}
\newtheorem{example}[theorem]{Example}

\theoremstyle{unnumbered}
\newcommand{\be}{\begin{equation}}
\newcommand{\ee}{\end{equation}}

\def\ba#1\ea{\begin{align}#1\end{align}}
\def\bas#1\eas{\begin{align*}#1\end{align*}}

\DeclareSymbolFont{symbolsC}{U}{txsyc}{m}{n}
\SetSymbolFont{symbolsC}{bold}{U}{txsyc}{bx}{n}
\DeclareFontSubstitution{U}{txsyc}{m}{n}
\DeclareMathSymbol{\multimapboth}{\mathrel}{symbolsC}{"13}

\IEEEoverridecommandlockouts

\makeatletter
\renewcommand\section{\@startsection{section}{1}{\z@}{2ex plus 1.5ex minus 0.8ex}{1ex plus 1ex minus 0.5ex}{\normalfont\normalsize\centering\scshape}} 
\renewcommand\subsection{\@startsection{subsection}{2}{\z@}{2ex plus 1.5ex minus 0.8ex}{1ex plus 1ex minus 0.5ex}{\normalfont\normalsize\itshape}}

\makeatother

\def\N{\mathbb{N}}
\def\R{\mathbb{R}}

\def\sA{\mathcal{A}}
\def\sB{\mathcal{B}}

\def\cliq{C}

\def\sW{\mathcal{W}}
\def\sX{\mathcal{N}}
\def\sY{\mathcal{Y}}

\def\0v{\boldsymbol{0}}
\def\xv{\boldsymbol{x}}
\def\xvt{\widetilde{\xv}}
\def\yv{\boldsymbol{y}}

\def\wv{\boldsymbol{w}}
\def\yvt{\widetilde{\yv}}
\def\qv{\boldsymbol{q}}

\def\rX{\mathsf{X}}

\def\rY{\mathsf{Y}}

\def\rxv{\boldsymbol{\mathsf{x}}}
\def\ryv{\boldsymbol{\mathsf{y}}}
\def\rwv{\boldsymbol{\mathsf{w}}}

\def\rx{\mathsf{x}}

\def\rt{\mathsf{t}}

\def\expcard{\nu} 
\def\krX{k}
\def\krY{\ell}
\def\kX{k}
\def\kY{\ell}

\providecommand{\card}[1]{\lvert#1\rvert}

\providecommand{\norm}[1]{\lVert#1\rVert}
\providecommand{\bignorm}[1]{\big\lVert#1\big\rVert}

\def\E{\mathbb{E}}

\DeclareMathOperator*{\argmin}{arg\,min}

\def\dist{\rho}

\def\trans{{\operatorname{T}}}

\def\Leb{\mathscr{L}}

\def\ind{\mathbbm{1}}

\def\intd{\mathrm{d}}

\def\divKL{D_{\mathrm{KL}}}

\def\gammat{\widetilde{\gamma}}

\def\condi{\,|\,}
\def\bcondi{\,\big\vert\,}

\def\squr{Q}

\def\codew{M}

\begin{document}


\title{
Rate-Distortion Theory of Finite 
 Point Processes
}

\author{
G\"unther~Koliander, 
Dominic Schuhmacher, 
and~Franz~Hlawatsch,~\IEEEmembership{Fellow,~IEEE}
\thanks{G. Koliander is with the Acoustics Research Institute, Austrian Academy of Sciences, 1040 Vienna, Austria (e-mail: gkoliander@kfs.oeaw.ac.at).
}
\thanks{
D.~Schuhmacher is with the Institute of Mathematical Stochastics,
Georg-August-Universit\"at G\"ottingen, 37077 G\"ottingen,
Germany (e-mail: dominic.schuhmacher@mathematik.uni-goettingen.de).}%
\thanks{
F. Hlawatsch is with the Institute of Telecommunications, TU Wien, 1040 Vienna, Austria (e-mail: franz.hlawatsch@nt.tuwien.ac.at).
}%
\thanks{This work was supported by the  FWF under grant P27370-N30 and by the WWTF under grant  MA16-053.}
\thanks{Copyright   (c)  2018  IEEE.  Personal  use  of  this  material   is  permitted.
However,  permission  to  use  this  material  for  any  other  purposes  must  be
obtained  from the IEEE by sending a request to pubs-permissions@ieee.org.}}%

\maketitle
\begin{abstract}
We study the compression of data in the case where the useful  information is contained in a set rather than a vector, i.e., the ordering of the data points is  irrelevant and  the number of data points is  unknown.
Our analysis is based on rate-distortion theory and the theory of finite point processes.
We introduce fundamental information-theoretic concepts and quantities for point processes and
present general lower and upper bounds on the rate-distortion function.
To enable a comparison with the vector setting, we  concretize our bounds for point processes of fixed cardinality.
In particular, we analyze a fixed number of unordered Gaussian data points and show that we can significantly reduce the required rates compared to the best possible compression strategy for Gaussian vectors.
As an example of point processes with variable cardinality, we study 
the best possible compression of  
Poisson point processes. 
For the 
specific 
case of a Poisson point process with uniform intensity on the unit square,  our lower and
upper bounds are separated by only a small gap and thus provide a good characterization of the rate-distortion function.%
\end{abstract}

\section{Introduction} \label{sec:intro}   
The continuing growth of the amount of data   to be stored and analyzed in many applications calls for efficient methods for representing and compressing large data records \cite{chmali14}.
In the literature on data compression, one
aspect  was hardly considered: the fact that we are often interested in \textit{sets} and not in  ordered lists, i.e., vectors, of data points.
Our goal in this paper is to study the optimal compression of \emph{finite sets}, also called \emph{point patterns}, in an information-theoretic framework.
More specifically, we consider 
a sequence of independent and identically distributed (i.i.d.) point patterns, and we want to calculate the minimal rate---i.e., number of representation bits per point pattern---for a given upper bound on the expected distortion. 
For this 
analysis, we need  distributions on  point patterns. 
Fortunately, these and other relevant mathematical tools are provided  by the 
theory of (finite) point processes (PPs) \cite{dave03,dave08}.

The theory and applications of PPs have a long history, and  in most fields using this concept---such as, e.g., forestry \cite{stpe00}, epidemiology \cite{gaba96},
and astronomy \cite{masa01}---significant amounts of data in the form of point patterns are collected, stored, and processed.
Thus, we believe that lossy source coding may be of great interest in these fields.
Furthermore, the recently studied problem of super-resolution \cite{cafe14,heso17} or more generally atomic norm minimization \cite{bhtare13} results in a point pattern in a continuous alphabet and is often described by some statistical properties.
In this setting, one frequently deals with noisy signals, and thus
an additional distortion resulting from lossy compression may be acceptable.

As a more explicit example, consider a database of minutiae patterns in fingerprints \cite{howaja98,pegatr15}. 
Minutiae are endpoints and bifurcations of ridge lines on the finger. 
Typical data consists of $x$- and $y$-positions of points in some relative coordinate system and may well include further information such as the angle of the orientation field at the
minutiae or the minutia types. 
For simplicity, we consider here only the positions;
any
additional information  can be incorporated by a suitable adaptation of the
distortion measure. 
A fingerprint of good quality typically contains about 40--100
minutiae \cite{howaja98}. 
For many minutia-based algorithms for fingerprint matching, the order in which the minutiae
are stored is irrelevant \cite{pegatr15} and a fingerprint can thus be represented as a point pattern.
Furthermore, different pressures applied during the acquisition of a fingerprint  lead to varying
local deformations and thus varying minutiae for the same finger.
Hence, in most applications, a small additional distortion due to compression will be acceptable. 
Because  the exact locations as well as the number of minutiae acquired for the same finger  may vary,  the squared OSPA metric as defined further below in \eqref{eq:ospagen} appears well suited for measuring the distortion between minutiae patterns.

\subsection{Prior Work}
Information-theoretic work on PPs is scarce.
An extension of entropy to PPs is available \cite[Sec.~14.8]{dave08}, but 
apparently
the mutual information between PPs was never analyzed in detail (although it is defined by the general quantization-based definition of mutual information \cite[eq.~(8.54)]{Cover91} or its equivalent form in \eqref{eq:gyptheorem} below).
A similar quantity was recently considered for a special case in  \cite[Th.~VI.1]{bawo16}. 
However, this  quantity  deviates from the general definition of mutual information, because the joint distribution 
in \cite[eq.~(5)]{bawo16} implies a fixed association between the points in the two PPs involved.

Source coding results for PPs are available almost exclusively for (infinite) Poisson PPs on $\R$  \cite{ru74, ga76, ve96, saka87}. 
However, that setting  considers only a single PP rather than an i.i.d. sequence of PPs.
More specifically, the sequence considered for rate-distortion (RD) analysis in  \cite{ru74, ga76, ve96, saka87} is the growing vector of the smallest $n$ points of the PP. 
This approach was also adopted in \cite{vago07}, where the  motivation was similar to that of the present paper but the main objective was to study the asymptotic behavior of the RD function as the cardinality of the data set grows infinite. 
It was  shown in  \cite{vago07} that the expected distortion divided by the number of points in the data set converges to zero even for zero-rate coding.
Although we are interested in the nonasymptotic scenario, the motivation given in  \cite{vago07} and the fact that the per-element distortion increases significantly less fast than in the vector case are of relevance to our work.
 
In channel coding,  PPs were  used in optical communications \cite{masa76, wy88} and  for general timing channels \cite{anvu96}.
However, the PPs considered are again  on $\R$ and in most cases Poisson PPs.

A different 
source coding setting for point patterns was presented in%
 \cite{lamawa15,mawa12} . 
There, the goal was not to reconstruct the points, but to find a  covering  (consisting of intervals) of all points. 
There is a tradeoff between the description length of the covering set and its Lebesgue measure, both of which are desired to be as small as possible.

For discrete alphabets, an algorithm compressing multisets was presented in \cite{st15}.
However, from an information-theoretic viewpoint, the collection of all (multi-)sets in a discrete alphabet is just another discrete set and thus sufficiently addressed by the standard theory for discrete sources.

To the best of our knowledge,   the RD function for i.i.d.\ sequences of PPs has not been studied previously.
In full generality, such a study requires the definition of a distortion function between sets of possibly different cardinality. 
A pertinent and convenient definition of a distortion function between point patterns was proposed in \cite{scvovo08} in the context of target tracking  (see \eqref{eq:ospagen} below) .

\subsection{Contribution and Paper Organization}
In this paper, we are interested in lossy compression of i.i.d.\ sequences of PPs of possibly varying cardinality. 
We 
obtain  bounds on the RD function in a  general setting and 
analyze the benefits that a set-theoretic viewpoint provides over a vector setting.
Our results and methods are based on the measure-theoretic fundamentals of RD theory \cite{berger71}.

As the information-theoretic analysis of PPs is not well established, we present expressions of the mutual information between dependent PPs, which can be used in upper bounds on the RD function.
Our main contribution is the establishment of upper and lower bounds on the RD function of finite PPs on $\R^d$. 
The upper bounds are based either on the RD theorem for memoryless sources \cite{berger71} or on codebooks constructed by a variant of the Linde-Buzo-Gray algorithm \cite{libugr80}.
The lower bounds  are based on the characterization of mutual information as a supremum \cite{csiszar74}, which is closely related to the Shannon lower bound \cite[eq.~(4.8.8)]{gr90}.
To illustrate our results, we compare the setting of a PP of fixed cardinality with  that of a vector of the same dimension and find that the RD function in the PP setting is significantly lower.
Furthermore, we concretize our bounds for Poisson PPs and, in particular,  consider   a Poisson PP on the unit square in $\R^2$, for which our bounds convey an accurate characterization of the RD function.

The paper is organized as follows.
In Section~\ref{sec:pointprocessint}, we present some fundamentals of PP theory. 
In particular, we introduce pairs of dependent PPs, which are relevant to an information-theoretic analysis but not common in the statistical literature.
In Section \ref{sec:mutinfpp}, the mutual information between PPs is studied in detail, and some tools from measure-theoretic RD theory that can be used in the analysis of PPs are introduced.
In Sections~\ref{sec:lowerbounds} and \ref{sec:upperbounds}, we present lower and upper bounds on the RD function of PPs in a  general setting. 
These bounds are applied to PPs of fixed cardinality in Section~\ref{sec:fcpp} and to Poisson PPs in Section~\ref{sec:ppp}.
In Section~\ref{sec:conclusion}, we summarize our results and suggest future research directions.

\subsection{Notation}\label{sec:notation}
%
Boldface  lowercase letters denote vectors. 
A vector $\xv=(\xv_1^{\trans}\, \cdots\, \xv_k^{\trans})^{\trans}\in (\R^d)^k$  with $\xv_i \in \R^d$ will often be denoted as $(\xv_1, \dots, \xv_k)$ or, more compactly, as $\xv_{1:k}$.
Sets
 are denoted by capital letters, e.g.,~$A$. 
The set $A+\xv$ is 
defined as
 $\{\yv+\xv: \yv\in A\}$.
The complement of a set $A$ is denoted as $A^c$, and the cardinality as
$\card{A}$.
The indicator function $\ind_{A}$ is given by  $\ind_{A}(x)=1$ if $x\in A$ and $\ind_{A}(x)=0$ if $x\notin A$.
The Cartesian product $A_1 \times A_2 \times \cdots \times A_k$ of sets $A_i$, $i=1, \dots, k$ is denoted as $\prod_{i=1}^k A_i$.
Sets of sets are denoted by calligraphic letters (e.g., $\sA$).
Multisets, i.e., sets with not necessarily distinct elements, are 
distinguished from sets in that
we
use 
$A,B,C$ to denote sets and 
$X,Y,Z$ to denote multisets.
For a set $A$ and a multiset $X$, we denote by $X\cap A$ the multiset $\{x\in X: x\in A\}$, which conforms to the classical intersection if $X$ is a set but  contains $x\in A$ more than once if $X$ contains  $x$ more than once.
Similarly, the cardinality $\card{X}$ of a multiset $X$ gives the total number of the (not necessarily distinct) elements in $X$. 
The set of 
nonnegative integers $\{0\}\cup \N$ is denoted as $\N_0$,
the set of
 positive real numbers  as $\R^+$, and the set of 
nonnegative real numbers  as $\R_{\geq 0}$.
Sans serif letters denote random quantities,  e.g., 
$\rxv$ is a random vector and $\rX$ is a random multiset (or PP). 
We write $\E_{\rxv}[\cdot]$ for the expectation operator with respect to the random variable $\rxv$ and simply $\E[\cdot]$ for the expectation operator with respect to all random variables in the argument.
$\Pr[\rxv\in A]$ denotes the probability that $\rxv\in A$.
$\Leb^d$ denotes the $d$-dimensional Lebesgue measure and
$\sB_d$ the Borel $\sigma$-algebra on $\R^d$.
For measures $\mu, \nu$ on the same measurable space,  $\mu \ll \nu$ means that $\mu$ is absolutely continuous with respect to $\nu$, i.e., that $\nu(A)=0$ implies $\mu(A)=0$ for any measurable set $A$.
A random vector $\rxv$ on $\R^d$ is understood to be measurable with respect to $\sB_d$.
The differential entropy of a continuous random vector $\rxv$ with probability density function $g$ is denoted as $h(\rxv)$ or  $h(g)$, and the entropy of a discrete random variable $\rx$ is denoted as $H(\rx)$.
The logarithm to the base $e$ is denoted $\log$.
For a function $f\colon A \to B$ and a set $C\subseteq B$, $f^{-1}(C)$ denotes the inverse image $\{x\in A: f(x)\in C\}$. 
Finally, we indicate by,
e.g., $\stackrel{(\text{42})}=$ 
that the equality holds due to  
(42).

\section{Point Processes as Random Sets of Vectors} \label{sec:pointprocessint}   
%
%
In this section, we present basic definitions and results from PP theory.
In the classical literature on this subject, PPs are defined as random counting measures  \cite[Def. 9.1.VI]{dave08}. 
Although this approach is very general and mathematically elegant, we will use a more applied viewpoint and interpret PPs as random multisets, i.e., collections of a random number of random vectors that are not necessarily distinct.
These multisets are assumed to be finite in the sense that they  have a finite cardinality with probability one. 
\begin{definition}
A \textit{point pattern} $X$ on $\R^d$ is defined as a finite multiset
 $X\subseteq \R^d$, i.e.,
$
\card{X}< \infty\,.
$
The collection of all point patterns $X$ on $\R^d$ is denoted as $\sX$.
\end{definition}
Our goal is to compress point patterns under certain constraints limiting an expected distortion. 
To this end, we have to define random elements $\rX$ on $\sX$
 and, in turn, a $\sigma$-algebra.
\begin{definition} 
We denote by $\mathfrak{S}$ the $\sigma$-algebra on $\sX$ generated by the collections of multisets 
 $\sX_k(B)\triangleq\big\{X\in \sX:\card{X \cap B}=k \big\}$ for all $B\in \sB_d$ and all $k\in \N_0$.
\end{definition}

\subsection{Finite Point Processes}\label{sec:fpp}

The random variables
$\rX$ on $(\sX, \mathfrak{S})$ are called \textit{finite (spatial) PPs on $\R^d$}, hereafter simply referred to as PPs.
Following \cite[Sec.~5.3]{dave03}, a  PP $\rX$ can be constructed by three steps:
\begin{enumerate}
	\item Let $\card{\rX}$ be a discrete random variable on $\N_0$ with probability mass function $p_{\card{\rX}}$.
	\item For each $k\in \N$, let $\rxv_{\rX}^{(k)}$ be a random vector on $(\R^d)^{k}$ with probability measure $P_{\rX}^{(k)}$ and the following symmetry property:  $\rxv_{\rX}^{(k)}= (\rxv_1, \dots, \rxv_k)$ with $\rxv_i \in \R^d$ has the same distribution as $(\rxv_{\tau(1)}, \dots,  \rxv_{\tau(k)})$ for any permutation $\tau$ on $\{1, \dots, k\}$.
	\item The random variable $\rX$ is defined by first choosing a realization $k$ of the random cardinality $\card{\rX}$ according to $p_{\card{\rX}}$.
	Then, for $\card{\rX}=k \neq 0$,  a realization $\xv_{1:k}=(\xv_1, \dots, \xv_k)$ of $\rxv_{\rX}^{(k)}$ is chosen according to $P_{\rX}^{(k)}$, and this realization  is converted to a point pattern via the 
	mapping
	\be\label{eq:defiotasingle}
	\phi_k \colon (\R^{d})^k  \to \sX; 
	\qquad
	\xv_{1:k}  \mapsto\{\xv_1, \dots, \xv_k\}\,.
	\ee
	For $\card{\rX}=0$, we set $\rX=\emptyset$.
	More compactly, this procedure corresponds to constructing $\rX$ as
	\be\notag 
	\rX=\begin{cases}
	\emptyset & \text{ if } \card{\rX}=0 \\[-1mm]
	\phi_k\big(\rxv_{\rX}^{(k)}\big)& \text{ if } \card{\rX}=k\,.
	\end{cases}
	\ee
\end{enumerate}
 
\begin{remark}
In principle, it is not necessary to start with  symmetric random vectors $\rxv_{\rX}^{(k)}$. 
Indeed, the mapping $\phi_k$ erases any order information the vector $\rxv_{\rX}^{(k)}$ might have, and thus we would obtain a PP even for nonsymmetric $\rxv_{\rX}^{(k)}$.
However, for our information-theoretic analysis, it will turn out to be useful to have access to the symmetric random vectors $\rxv_{\rX}^{(k)}$ and the symmetric probability measures $P_{\rX}^{(k)}$.
Note that this does not imply a restriction on the PPs we consider, as any random vector can be symmetrized before using it in the PP construction.
\end{remark}

The probability measure on $(\sX,\mathfrak{S})$ induced by $\rX$ is denoted as $P_{\rX}$ and satisfies 
\ba
P_{\rX}(\sA) 
& = \Pr[\rX \in \sA] 
\notag \\ 
&= 
p_{\card{\rX}}(0) \ind_{\sA}(\emptyset)
+ \sum_{k\in \N} p_{\card{\rX}}(k) P_{\rX}^{(k)}\big(\phi_{k}^{-1}(\sA)\big)
\label{eq:pxa}
\ea
for any measurable set $\sA \subseteq \sX$ (i.e., $\sA \in \mathfrak{S}$).
This construction 
indeed results in a measurable $\rX$ \cite[Ch.~9]{dave08}. 
According to \eqref{eq:pxa}, an integral with respect to $P_{\rX}$ (or, equivalently, an expectation with respect to $\rX$) can be calculated as%
\footnote{
This expression can be shown by the standard measure-theoretic approach of defining an integral in turn for indicator functions, simple functions, nonnegative measurable functions, and finally all integrable functions 
\cite[Sec.~A1.4]{dave03}.
}
\ba
& \int_{\sX}g(X)\,\intd P_{\rX}(X) 
= \E[g(\rX)] \notag \\
& = p_{\card{\rX}}(0) g(\emptyset) + \sum_{k\in \N}p_{\card{\rX}}(k) \, \E\big[g\big(\phi_k(\rxv_{\rX}^{(k)})\big)\big] \notag \\
& = p_{\card{\rX}}(0) g(\emptyset) +
\sum_{k\in \N}p_{\card{\rX}}(k) \int_{(\R^{d})^k}g(\phi_k(\xv_{1:k})) \,\intd P_{\rX}^{(k)}(\xv_{1:k})\label{eq:pxaint}
\ea
for any integrable function $g\colon \sX \to \R$.
In particular, by \eqref{eq:pxaint} with $g(X)=\ind_{\sA}(X)\widetilde{g}(X)$, we 
obtain
\ba
\int_{\sA}\widetilde{g}(X)\,\intd P_{\rX}(X)
& = p_{\card{\rX}}(0) \ind_{\sA}(\emptyset) \widetilde{g}(\emptyset) 
 +
\sum_{k\in \N}p_{\card{\rX}}(k) 
\notag \\ & \quad \times
\int_{\phi_k^{-1}(\sA)}\widetilde{g}(\phi_k(\xv_{1:k})) \,\intd P_{\rX}^{(k)}(\xv_{1:k})
\label{eq:pxaintsa} \\[-8mm] \notag 
\ea
for any $\sA \in \mathfrak{S}$.

\subsection{Pairs of Point Processes}\label{sec:fppp}

For information-theoretic considerations, it is convenient to have a simple definition of the joint distribution of two PPs.
Thus, similar to the construction of $\rX$, we define a pair of (generally dependent) PPs $(\rX,\rY)$ as random elements on the product space $\sX \times \sX$ as follows.
\begin{enumerate}
	\item Let $(\card{\rX}, \card{\rY})$ be a discrete random variable on $\N_0\times \N_0=\N_0^2$ with probability mass function $p_{\card{\rX},\card{\rY}}$.	
	\item For each $(\krX ,\krY )\in \N_0^2\setminus \{(0,0)\}$, let $(\rxv,\ryv)_{\rX,\rY}^{(\krX ,\krY )}$ be a random vector on $(\R^d)^{\krX +\krY }$ with probability measure $P_{\rX,\rY}^{(\krX ,\krY )}$ and the following symmetry property: 
	$(\rxv,\ryv)_{\rX,\rY}^{(\krX ,\krY )}= (\rxv_1, \allowbreak\dots, \allowbreak\rxv_{\krX }, \allowbreak\ryv_1, \allowbreak\dots, \allowbreak\ryv_{\krY })$ with $\rxv_i, \ryv_j \in \R^d$ has the same distribution as 
	$(\rxv_{\tau_{\rX}(1)}, \allowbreak\dots, \allowbreak\rxv_{\tau_{\rX}(\krX )}, \allowbreak\ryv_{\tau_{\rY}(1)}, \allowbreak\dots, \allowbreak\ryv_{\tau_{\rY}(\krY )})$ 
	for any permutations $\tau_{\rX}$   on $\{1, \dots, \krX \}$ and $\tau_{\rY}$  on $\{1, \dots, \krY \}$.
	  Note that for the cases  $\krX =0$ and $\krY =0$, we have 
	$(\rxv,\ryv)_{\rX,\rY}^{(0,\krY )}= (\ryv_1, \allowbreak\dots, \allowbreak\ryv_{\krY })$ and
	$(\rxv,\ryv)_{\rX,\rY}^{(\krX ,0)}= (\rxv_1, \dots, \rxv_{\krX })$, respectively.
	 
	\item The random variable $(\rX,\rY)$ is defined by first choosing a realization $(\krX , \krY )$ of the random cardinalities $(\card{\rX}, \card{\rY})$ according to $p_{\card{\rX},\card{\rY}}$. 
	Then, for $(\card{\rX}, \card{\rY})=(\krX ,\krY )$ with $\krX \neq 0$ or $\krY \neq 0$,  a realization $(\xv_{1:\krX }, \yv_{1:\krY })=(\xv_1, \dots, \xv_{\krX }, \yv_1, \dots, \yv_{\krY })$ of $(\rxv,\ryv)_{\rX,\rY}^{(\krX ,\krY )}$ is chosen according to $P_{\rX,\rY}^{(\krX ,\krY )}$, and
	 this realization is converted to a pair of point patterns via the 
mapping
	\be\label{eq:iotadefxy1}
	\begin{split}
	\phi_{\krX ,\krY }\colon (\R^{d})^{\krX +\krY }
	& \to \sX^2;  \\
	(\xv_{1:\krX }, \yv_{1:\krY })  & \mapsto \big(\{\xv_1, \dots, \xv_{\krX }\}, \{\yv_1, \dots, \yv_{\krY }\}\big)
	\end{split}
	\ee
	if $(\krX ,\krY )\in \N^2$, 
or
	\be\label{eq:iotadefxy2}
	\phi_{0,\krY }\colon (\R^{d})^{\krY }   \to \sX^2; \quad
	\yv_{1:\krY } \mapsto \big(\emptyset, \{\yv_1, \dots, \yv_{\krY }\}\big)
	\ee
	if $\krX =0$ and $\krY \in \N$, 
or 
	\be\label{eq:iotadefxy3}
	\phi_{\krX ,0}\colon (\R^{d})^{\krX }  \to \sX^2;  \quad
	\xv_{1:\krX } \mapsto \big(\{\xv_1, \dots, \xv_{\krX }\}, \emptyset\big)
	\ee
	if $\krX \in \N$ and  $\krY =0$.
	For $(\krX , \krY )=(0,0)$, we set $(\rX,\rY)=(\emptyset, \emptyset)$.
	More compactly, the overall  procedure corresponds to constructing $(\rX, \rY)$ as
	\be\notag 
	(\rX, \rY)=\begin{cases}
	(\emptyset, \emptyset) & \text{if } (\card{\rX}, \card{\rY})=(0,0) \\ 
	\phi_{\krX ,\krY }\big((\rxv,\ryv)_{\rX,\rY}^{(\krX ,\krY )}\big)
	&  \text{if } (\card{\rX}, \card{\rY})=(\krX ,\krY ) \\
	& \text{\phantom{if} } \phantom{(\card{\rX}, \card{\rY})}\neq (0,0)\,.
	\end{cases}
	\ee
\end{enumerate}
As we will often  use inverse images of the mapping $\phi_{\krX , \krY }$ in our proofs, we state some properties of $\phi^{-1}_{\krX , \krY }(\sA)$ for $\sA\subseteq \sX^2$ in Appendix~\ref{app:proofiotaprops}.

The probability measure on $(\sX^2, \mathfrak{S}\otimes \mathfrak{S})$ induced by $(\rX,\rY)$ will be denoted as $P_{\rX, \rY}$ and satisfies 
\ba
P_{\rX, \rY}(\sA) 
& =
p_{\card{\rX},\card{\rY}}(0,0) \ind_{\sA}\big((\emptyset, \emptyset)\big)
\notag \\ & \quad 
+\hspace{-1mm} \sum_{\substack{(\krX , \krY )\in \N_0^2 \\ 
(\krX , \krY )\neq (0,0)} 
} \hspace{-1mm} p_{\card{\rX},\card{\rY}}(\krX ,\krY ) P_{\rX,\rY}^{(\krX ,\krY )}\big(\phi_{\krX ,\krY }^{-1}(\sA)\big) \notag \\[-6mm]
\label{eq:prodprob}
\ea
for any measurable $\sA \subseteq \sX^2$ (i.e., $\sA\in \mathfrak{S}\otimes \mathfrak{S}$).
An integral with respect to $P_{\rX, \rY}$ (or, equivalently, an expectation with respect to $(\rX, \rY)$) can be calculated as
\ba
& \int_{\sX^2}g(X,Y)\,\intd P_{\rX, \rY}(X,Y)
= \E[g(\rX, \rY)] \notag \\
&\; =  p_{\card{\rX},\card{\rY}}(0,0) g(\emptyset, \emptyset) 
\notag \\* & \;\quad
+
\hspace{-1mm}\sum_{\substack{(\krX , \krY )\in \N_0^2 \\ 
(\krX , \krY )\neq (0,0)}}\hspace{-1mm} p_{\card{\rX},\card{\rY}}(\krX ,\krY ) 
\,\E\big[g\big(\phi_{\krX ,\krY }\big((\rxv,\ryv)_{\rX,\rY}^{(\krX ,\krY )}\big)\big)\big] \notag \\
&\; = p_{\card{\rX},\card{\rY}}(0,0) g(\emptyset, \emptyset) 
+
\hspace{-1mm}\sum_{\substack{(\krX , \krY )\in \N_0^2 \\ 
(\krX , \krY )\neq (0,0)}}\hspace{-1mm} p_{\card{\rX},\card{\rY}}(\krX ,\krY ) 
\notag \\* & \quad \times
\int_{(\R^{d})^{\krX +\krY }}g\big(\phi_{\krX , \krY }(\xv_{1:\krX }, \yv_{1:\krY })\big) \,\intd P_{\rX, \rY}^{(\krX , \krY )}(\xv_{1:\krX }, \yv_{1:\krY })\label{eq:prodprobint}
\ea
for any integrable function $g\colon \sX^2\to \R$.
As in the single-PP case, $g(X,Y)=\ind_{\sA}((X,Y))\widetilde{g}(X,Y)$ results in an integral expression similar to \eqref{eq:pxaintsa} for any measurable set $\sA\subseteq \sX^2$.

The symmetry of the random vectors $(\rxv,\ryv)_{\rX,\rY}^{(\krX ,\krY )}$ implies that the corresponding probability measures $P_{\rX,\rY}^{(\krX ,\krY )}$ are symmetric in the following sense.
Let $\tau_{\rX}$ and $\tau_{\rY}$ be permutations on $\{1, \dots, \krX \}$ and $\{1, \dots, \krY \}$, respectively, and define 
\ba
\psi_{\tau_{\rX}, \tau_{\rY}}\colon (\R^d)^{\krX +\krY } & \to (\R^d)^{\krX +\krY };  \notag\\
(\xv_{1:\krX }, \yv_{1:\krY }) & \mapsto (\xv_{\tau_{\rX}(1)}, \makebox[1em][c]{.\hfil.\hfil.}, \xv_{\tau_{\rX}(\krX )},  \yv_{\tau_{\rY}(1)},  \makebox[1em][c]{.\hfil.\hfil.}, \yv_{\tau_{\rY}(\krY )})\,. 
\notag \\[-1mm]
\label{eq:defpsi}
\\[-8mm] \notag 
\ea
Then, for any measurable $A\subseteq (\R^d)^{\krX +\krY }$
\be\label{eq:symmeasure}
P_{\rX,\rY}^{(\krX ,\krY )}(A)= P_{\rX,\rY}^{(\krX ,\krY )}(\psi_{\tau_{\rX}, \tau_{\rY}}(A))\,.
\ee

We will also be interested in marginal probabilities. 
For a pair of PPs $(\rX,\rY)$, the marginal PP $\rX$ is defined by the probability measure $P_{\rX}(\sA)=P_{\rX,\rY}(\sA\times \sX)$ for all measurable sets $\sA \subseteq \sX$.
The corresponding probability measures $P^{(\krX )}_{\rX}$ for $\krX \in \N$ satisfy
\ba
p_{\card{\rX}} (\krX )P^{(\krX )}_{\rX}(B)
 & =p_{\card{\rX},\card{\rY}}(\krX , 0)P_{\rX,\rY}^{(\krX ,0)}(B) 
\notag \\* & \quad
+  \sum_{\krY \in \N}p_{\card{\rX},\card{\rY}}(\krX , \krY )P_{\rX,\rY}^{(\krX ,\krY )}\big(B\times (\R^d)^{\krY }\big)
\notag \\[-3mm]
\label{eq:margdist}
\\[-7mm]\notag 
\ea
for Borel sets $B\subseteq (\R^d)^{\krX }$, where for $\krX \in \N_0$
\be\label{eq:margcarddist}
p_{\card{\rX}}(\krX )
 = \sum_{\krY \in \N_0}p_{\card{\rX},\card{\rY}}(\krX , \krY )\,.
\ee
The definition of the marginal PP $\rY$ is analogous.
We caution that the probability measures $P^{(\krX )}_{\rX}$ and $P^{(\krY )}_{\rY}$ are in general not the marginals of $P_{\rX,\rY}^{(\krX ,\krY )}$. 
Indeed,  $P^{(\krX )}_{\rX}$  depends on  $P_{\rX,\rY}^{(\krX ,\krY )}$ for all $\krY \in \N_0$ with $p_{\card{\rX},\card{\rY}}(\krX , \krY )\neq 0$ 
and, similarly, $P^{(\krY )}_{\rY}$ depends on $P_{\rX,\rY}^{(\krX ,\krY )}$ for all $\krX \in \N_0$ with $p_{\card{\rX},\card{\rY}}(\krX , \krY )\neq 0$.
In particular, the probability measures of the marginals $\rxv_{\rX,\rY}^{(\krX ,\krY )}$ and $\ryv_{\rX,\rY}^{(\krX ,\krY )}$ of $(\rxv,\ryv)_{\rX,\rY}^{(\krX ,\krY )}$ generally are not equal to $P^{(\krX )}_{\rX}$ and $P^{(\krY )}_{\rY}$, respectively.

We will often consider the case of i.i.d.\ PPs.
Two PPs $\rX$ and $\rY$ are  \textit{independent} if $P_{\rX,\rY}=P_{\rX}\times P_{\rY}$, i.e., 
$\Pr[(\rX,\rY)\in \sA_{\rX}\times \sA_{\rY}]=\Pr[\rX\in \sA_{\rX}]\Pr[\rY \in  \sA_{\rY}]$ for all $\sA_{\rX},\sA_{\rY}\in \mathfrak{S}$.
Furthermore, $\rX$ and $\rY$  are  \textit{identically distributed} if their  measures $P_{\rX}$ and $P_{\rY}$ are equal.

All definitions and results in this subsection can be readily generalized to more than two PPs.
In particular, we will consider sequences of i.i.d.\ PPs in Section~\ref{sec:rdfunction}.

\subsection{Point Processes of Fixed Cardinality}\label{sec:ppfc}

%
%
There are two major differences between spatial PPs and random vectors: 
first, the number of elements in a point pattern is a random quantity whereas the dimension of a random vector is deterministic;
second, there is no inherent order of the elements of a point pattern. 
PPs of fixed cardinality differ from random vectors only by the second property.
More specifically, 
we say that a PP $\rX$ is \emph{of fixed cardinality} $\card{\rX}=k$ if $p_{\card{\rX}}(k)=\Pr[\card{\rX}=k]=1$ for some given $k\in \N$ (we do not consider the trivial case $k=0$).
The set of all possible realizations of $\rX$ is denoted as $\sX_{k}$, i.e., $\sX_k\triangleq \{X\in \sX: \card{X}=k\}$.
The probability measure $P_{\rX}$ for a PP $\rX$ of fixed cardinality $\card{\rX}=k$ simplifies to (cf.~\eqref{eq:pxa}) 
$P_{\rX}(\sA) = P_{\rX}^{(k)}\big(\phi_{k}^{-1}(\sA)\big)$,
i.e., it is simply the induced measure of $P_{\rX}^{(k)}$ under the mapping $\phi_{k}$.

Similarly, a pair of PPs $(\rX, \rY)$ is called \emph{of fixed cardinality} $(\card{\rX}, \card{\rY})=(\krX , \krY )$ if $p_{\card{\rX},\card{\rY}}(\krX , \krY )=1$ for some given  $\krX , \krY \in \N$, i.e., $\Pr[\card{\rX}=\krX ]=1$ and $\Pr[\card{\rY}=\krY ]=1$.
The corresponding probability measure $P_{\rX, \rY}$ satisfies (cf.~\eqref{eq:prodprob})
$P_{\rX, \rY}(\sA) 
=
 P_{\rX,\rY}^{(\krX ,\krY )}\big(\phi_{\krX ,\krY }^{-1}(\sA)\big)$.
Because $p_{\card{\rX},\card{\rY}}(\krX' , \krY' )=0$ for $(\krX' , \krY' )\neq (\krX , \krY )$, \eqref{eq:margcarddist} implies $p_{\card{\rX}}(\krX )=p_{\card{\rX},\card{\rY}}(\krX , \krY )=1$ and, similarly, $p_{\card{\rY}}(\krY )=p_{\card{\rX},\card{\rY}}(\krX , \krY )=1$.
Thus, \eqref{eq:margdist} simplifies to
$P^{(\krX )}_{\rX}(B)
 =P_{\rX,\rY}^{(\krX , \krY )}\big(B\times (\R^d)^{\krY }\big)$
for Borel sets $B\subseteq (\R^d)^{\krX}$.
Analogously, we obtain 
$P^{(\krY )}_{\rY}(B)
 =P_{\rX,\rY}^{(\krX , \krY )}\big((\R^d)^{\krX } \times B\big)
$
for Borel sets $B\subseteq (\R^d)^{\krY}$.
Hence, the probability measures of the marginals $\rxv_{\rX,\rY}^{(\krX , \krY )}$ and $\ryv_{\rX,\rY}^{(\krX , \krY )}$ of $(\rxv,\ryv)_{\rX,\rY}^{(\krX , \krY )}$ are given by  $P^{(\krX )}_{\rX}$ and $P^{(\krY )}_{\rY}$, respectively.

\subsection{Point Processes of Equal Cardinality}\label{sec:ppec}
%
%
%
A setting  of particular interest to our study are pairs of PPs that have equal but not necessarily fixed cardinality. 
More specifically, we say that a pair of PPs $(\rX, \rY)$ has \emph{equal cardinality} if $p_{\card{\rX},\card{\rY}}(\krX , \krY )=0$ for $\krX \neq \krY $, i.e., $\Pr[\card{\rX}=\card{\rY}]=1$.
The corresponding probability measure $P_{\rX, \rY}$ satisfies (cf.~\eqref{eq:prodprob})
\ba
P_{\rX, \rY}(\sA) 
& =
p_{\card{\rX},\card{\rY}}(0,0) \ind_{\sA}\big((\emptyset, \emptyset)\big)
\notag \\* & \quad
+ \sum_{k\in \N}p_{\card{\rX},\card{\rY}}(k,k) P_{\rX,\rY}^{(k,k)}\big(\phi_{k,k}^{-1}(\sA)\big)
\,.\notag 
\ea
A significant simplification can be observed for the marginal probabilities. 
Because $p_{\card{\rX},\card{\rY}}(\krX , \krY )=0$ for $\krX \neq \krY $, \eqref{eq:margcarddist} implies $p_{\card{\rX}}(k)=p_{\card{\rX},\card{\rY}}(k, k)$, and \eqref{eq:margdist} simplifies to
$
P^{(k)}_{\rX}(B)
 =P_{\rX,\rY}^{(k,k)}\big(B\times (\R^d)^{k}\big)
$
for $k\in \N$.
Thus, the probability measure of the marginal $\rxv_{\rX,\rY}^{(k,k)}$ of  $(\rxv,\ryv)_{\rX,\rY}^{(k, k)}$ is given by $P^{(\krX )}_{\rX}$ and we will write more compactly $\rxv_{\rX}^{(k)}\triangleq \rxv_{\rX,\rY}^{(k,k)}$.
Analogously, we define $\ryv_{\rY}^{(k)}\triangleq \ryv_{\rX,\rY}^{(k,k)}$, and thus can rewrite $(\rxv,\ryv)_{\rX,\rY}^{(k,k)}$ as
\be\label{eq:ecmargchar}
(\rxv,\ryv)_{\rX,\rY}^{(k,k)}= \big(\rxv_{\rX}^{(k)},\ryv_{\rY}^{(k)}\big)\,.
\ee
\section{Mutual Information and Rate-Distortion Function for Point Processes} \label{sec:mutinfpp}   
%
Mutual information is a general concept that can be applied to arbitrary probability spaces although it is most commonly used for continuous or discrete random vectors.
To obtain an intuition about the mutual information between PPs, we will analyze several special settings that will also be relevant later.
The basic definition of mutual information is for discrete random variables  \cite[eq.~(2.28)]{Cover91}
and readily extended to arbitrary random variables by quantization \cite[eq.~(8.54)]{Cover91}.
By the Gelfand-Yaglom-Perez theorem \cite[Lem.~5.2.3]{Gray1990Entropy},  mutual information can be expressed in terms of a Radon-Nikodym derivative:
for two random variables%
\footnote{We will use  \eqref{eq:gyptheorem} mainly for PPs and thus use the notation of PPs. However, it is also valid  for random vectors.}
 $\rX$ and $\rY$ on the same probability space,
\be\label{eq:gyptheorem}
I(\rX;\rY) =
\int \log \bigg(\frac{\mathrm{d}P_{\rX,\rY}}{\mathrm{d}(P_{\rX}\times P_{\rY})}(X, Y)\bigg) \, \mathrm{d}P_{\rX,\rY}(X,Y)
\ee
if $P_{\rX,\rY}\ll  P_{\rX}\times P_{\rY} $
and $I(\rX;\rY) = \infty$ else.

\subsection{General Expression of   Mutual Information}

Using   \eqref{eq:gyptheorem}, we can express the mutual information between PPs as a sum of Kullback-Leibler divergences (KLDs).
We recall that the KLD between two probability measures $\mu$ and $\nu$ on  the same measurable space $\Omega$ is given as \cite[Sec.~1.3]{ku78}
\be\label{eq:kldiv}
\divKL(\mu \| \nu) 
= 
\begin{cases}
\int_{\Omega}\log \big(\frac{\intd\mu}{\intd\nu}(\xv)\big)\,\intd\mu(\xv) & \text{ if }\mu \ll \nu \\
\infty & \text{ else}\,.
\end{cases}
\ee
As a preliminary result, we present a characterization of the Radon-Nikodym derivative $\frac{\mathrm{d}P_{\rX,\rY}}{\mathrm{d}(P_{\rX}\times P_{\rY})}$ for a pair of PPs $(\rX, \rY)$.
A proof is given in Appendix~\ref{app:rndrelations}.
\begin{lemma}\label{lem:rndrelations}
Let $(\rX, \rY)$ be a pair of PPs.
The following two properties are equivalent:
\begin{enumerate}
\renewcommand{\theenumi}{(\roman{enumi})}
\renewcommand{\labelenumi}{(\roman{enumi})}
	\item \label{en:rdexists}$P_{\rX,\rY}\ll P_{\rX}\times P_{\rY}$;
	\item \label{en:allrdexist}For all $\krX ,\krY \in \N$ such that $p_{\card{\rX},\card{\rY}}(\krX ,\krY )\neq 0$, we have $P_{\rX,\rY}^{(\krX ,\krY )} \ll P_{\rX}^{(\krX )}\times P_{\rY}^{(\krY )}$; 
	for all $\krX \in \N$ such that $p_{\card{\rX},\card{\rY}}(\krX ,0)\neq 0$, we have $P_{\rX,\rY}^{(\krX ,0)} \ll P_{\rX}^{(\krX )}$;
	and for all $\krY \in \N$ such that $p_{\card{\rX},\card{\rY}}(0,\krY )\neq 0$, we have $P_{\rX,\rY}^{(0,\krY )} \ll  P_{\rY}^{(\krY )}$.
\end{enumerate}
Furthermore, if the equivalent properties \ref{en:rdexists} and \ref{en:allrdexist} hold, then 
\be \notag
\frac{\mathrm{d}P_{\rX,\rY}}{\mathrm{d}(P_{\rX}\times P_{\rY})} = \theta_{\rX, \rY}
\ee
where $\theta_{\rX, \rY} \colon \sX^2 \to \R_{\geq 0}$ satisfies%
\footnote{
Note that the functions $\phi_{k}$ are not one-to-one and thus, e.g.,  for $\xv_{1:\krX }\neq \widetilde{\xv}_{1:\krX }$ with $\phi_{\krX }(\xv_{1:\krX })=\phi_{\krX }(\widetilde{\xv}_{1:\krX })=X$,  \eqref{eq:condixygen2} might seem to give contradictory values for $\theta_{\rX,\rY}(X,\emptyset)$.
However, due to our  symmetry assumptions on $P_{\rX,\rY}^{(\krX ,\krY )}$, $P_{\rX}^{(\krX )}$,  and $P_{\rY}^{(\krY )}$ (see Sections~\ref{sec:fpp} and \ref{sec:fppp}), all Radon-Nikodym derivatives on the right-hand side of \eqref{eq:condixygen} can be chosen symmetric and thus the values of $\theta_{\rX,\rY}$ given in \eqref{eq:condixygen} are consistent.
}
\begin{subequations}
\label{eq:condixygen}
\ba
\theta_{\rX,\rY}(\emptyset,\emptyset) & = 
\frac{p_{\card{\rX},\card{\rY}}(0,0)}{p_{\card{\rX}}(0)p_{\card{\rY}}(0)} \label{eq:condixygen1} \\
\theta_{\rX,\rY}(\phi_{\krX }(\xv_{1:\krX }),\emptyset) & = 
\frac{p_{\card{\rX},\card{\rY}}(\krX ,0)}{p_{\card{\rX}}(\krX )p_{\card{\rY}}(0)}
\frac{\intd P_{\rX,\rY}^{(\krX ,0)}}{\intd P_{\rX}^{(\krX )}}(\xv_{1:\krX })\label{eq:condixygen2} \\
\theta_{\rX,\rY}(\emptyset,\phi_{\krY }(\yv_{1:\krY })) & = 
\frac{p_{\card{\rX},\card{\rY}}(0,\krY )}{p_{\card{\rX}}(0)p_{\card{\rY}}(\krY )}
\frac{\intd P_{\rX,\rY}^{(0,\krY )}}{\intd P_{\rY}^{(\krY )}}(\yv_{1:\krY })\label{eq:condixygen3} \\
\theta_{\rX,\rY}(\phi_{\krX }(\xv_{1:\krX }), \phi_{\krY }(\yv_{1:\krY })) 
& = \frac{p_{\card{\rX},\card{\rY}}(\krX ,\krY )}{p_{\card{\rX}}(\krX )p_{\card{\rY}}(\krY )}
\notag \\* & \quad \times
\frac{\intd P_{\rX,\rY}^{(\krX ,\krY )}}{\intd \big(P_{\rX}^{(\krX )}\times P_{\rY}^{(\krY )}\big)}(\xv_{1:\krX }, \yv_{1:\krY })\,.
\label{eq:condixygen4}
\ea
\end{subequations}
Here, the right-hand sides of  \eqref{eq:condixygen} are understood to be zero if $p_{\card{\rX},\card{\rY}}(\krX ,\krY )=0$.
\end{lemma}

Using Lemma~\ref{lem:rndrelations}, we can decompose the mutual information between PPs into KLDs between measures   associated with random vectors.
The following theorem is proved in Appendix~\ref{app:proofmutinfgen}.

\begin{theorem}\label{th:mutinfgen}
The mutual information $I(\rX;\rY)$ for a pair of PPs $(\rX, \rY)$ is given by
\ba
I(\rX;\rY) 
& = I(\card{\rX}; \card{\rY}) + 
\sum_{\krX \in \N}p_{\card{\rX},\card{\rY}}(\krX ,0) \divKL\big(P_{\rX,\rY}^{(\krX ,0)}\big\| P_{\rX}^{(\krX )}\big) 
\notag \\*
& \quad
+ \sum_{\krY \in \N}p_{\card{\rX},\card{\rY}}(0,\krY ) \divKL\big(P_{\rX,\rY}^{(0,\krY )}\big\| P_{\rY}^{(\krY )}\big) 
\notag \\*
& \quad
+ \sum_{\krX \in \N} \sum_{\krY \in \N}p_{\card{\rX},\card{\rY}}(\krX ,\krY ) 
\divKL\big(P_{\rX,\rY}^{(\krX ,\krY )}\big\| P_{\rX}^{(\krX )}\times P_{\rY}^{(\krY )}\big)\,.\notag\\[-4mm]
\label{eq:sepmutinf}
\\[-7mm] \notag
\ea
\end{theorem}

Note that in general $\divKL\big(P_{\rX,\rY}^{(\krX ,\krY )}\big\| P_{\rX}^{(\krX )}\times P_{\rY}^{(\krY )}\big)$ cannot be represented as a mutual information because the probability measures $P_{\rX}^{(\krX )}$ and $P_{\rY}^{(\krY )}$ are not the marginals of  $P_{\rX,\rY}^{(\krX ,\krY )}$. 
However, for a pair of PPs of fixed cardinality or of equal cardinality, a representation as mutual information is possible.

\subsection{Mutual Information for Point Processes of Fixed \\ Cardinality}

For a pair of PPs $(\rX, \rY)$ of fixed cardinality, i.e., $p_{\card{\rX},\card{\rY}}(\krX , \krY )=1$ for some $\krX , \krY \in \N$ (see Section~\ref{sec:ppfc}), we can relate the mutual information to the mutual information between random vectors. 
Indeed, since $P_{\rX,\rY}^{(\krX ,\krY )}$, $P_{\rX}^{(\krX )}$, and $P_{\rY}^{(\krY )}$ are the probability measures of $(\rxv,\ryv)_{\rX,\rY}^{(\krX , \krY )}$ and its marginals, $\rxv_{\rX,\rY}^{(\krX , \krY )}$ and $\ryv_{\rX,\rY}^{(\krX , \krY )}$, respectively, 
we obtain
$\divKL\big(P_{\rX,\rY}^{(\krX ,\krY )}\big\| P_{\rX}^{(\krX )}\times P_{\rY}^{(\krY )}\big) = I\big( \rxv_{\rX,\rY}^{(\krX , \krY )}; \ryv_{\rX,\rY}^{(\krX , \krY )}\big)$.
Inserting this into \eqref{eq:sepmutinf} while recalling that $p_{\card{\rX},\card{\rY}}(\krX' ,\krY' )=0$ for $(\krX' ,\krY' )\neq (\krX ,\krY )$ and noting that $I(\card{\rX};\card{\rY})=0$, Theorem~\ref{th:mutinfgen} simplifies significantly.
\begin{corollary}\label{cor:mutinfeqcard}
Let $(\rX,\rY)$ be  a pair of PPs of fixed cardinality $(\card{\rX},\card{\rY})=(\krX , \krY )$ for some  $\krX , \krY  \in \N$.
Then
\be \notag 
I(\rX;\rY) = I\big( \rxv_{\rX,\rY}^{(\krX , \krY )}; \ryv_{\rX,\rY}^{(\krX , \krY )}\big)\,.
\ee
\end{corollary}

We can also start with an arbitrary random vector $\big(\rxv^{(\krX )}, \ryv^{(\krY )}\big)$ on $(\R^d)^{\krX +\krY }$ without assuming any symmetry properties.  
In that case, the mutual information between $\rxv^{(\krX )}$ and $\ryv^{(\krY )}$ cannot be completely described by the associated pair of PPs $\big(\phi_{\krX }(\rxv^{(\krX )}), \phi_{\krY }(\ryv^{(\krY )})\big)$ but we also have to consider \emph{random permutations}, i.e., discrete random variables $\rt_{\rxv}$ and $\rt_{\ryv}$ that specify the  order of the vectors in $\rxv^{(\krX )}$ and $\ryv^{(\krY )}$, respectively.
More specifically, for a point pattern $X= \{\xv_1, \dots, \xv_{\krX }\}$ where the indices are chosen according to a predefined total order (e.g., lexicographical) of the elements, a permutation $\tau$ specifies the vector%
\footnote{  Here and in what follows, we use the same symbol $\tau$ for both the permutation on $\{1, \dots, {\krX }\}$ and the associated mapping $\tau\colon \sX \to (\R^d)^{\krX }$,  and we refer to both as permutation. } 
$\tau(X)\triangleq (\xv_{\tau(1)},  \dots,  \xv_{\tau(\krX )})\in (\R^d)^{\krX }$. 
Using this convention, the random vector $\rxv^{(\krX )}$ can be equivalently represented by the associated PP $\phi_{\krX }(\rxv^{(\krX )})$ and a random permutation $\rt_{\rxv}$ specifying the order of the elements relative to the predefined total order, i.e., $\rxv^{(\krX )}= \rt_{\rxv}\big(\phi_{\krX }(\rxv^{(\krX )})\big)$.
Applying further the tie-break rule that $\rt_{\rxv}(i)<\rt_{\rxv}(j)$ if $\rxv^{(\krX )}_i=\rxv^{(\krX )}_j$ and $i<j$, there is a one-to-one relation between the random vector $\rxv^{(\krX )}$ and the pair $\big(\phi_{\krX }(\rxv^{(\krX )}), \rt_{\rxv}\big)$.
Similarly, we can represent $\ryv^{(\krY )}$ by the pair $\big(\phi_{\krY }(\ryv^{(\krY )}), \rt_{\ryv}\big)$.
This leads to the following expression of the mutual information between PPs of fixed cardinality.
\begin{lemma}\label{lem:splitmi}
Let $(\rX,\rY)$ be a pair of PPs of fixed cardinality $(\card{\rX},\card{\rY})=(\krX , \krY )$ for some  $\krX , \krY  \in \N$.
Furthermore, let $\big(\rxv^{(\krX )}, \ryv^{(\krY )}\big)$ be a random vector on $(\R^d)^{\krX +\krY }$ such that $(\rX,\rY)$ has the same distribution as $\big(\phi_{\krX }(\rxv^{(\krX )}), \phi_{\krY }(\ryv^{(\krY )})\big)$.
Then
\ba   %
I(\rX;\rY) 
& = I\big(\rxv^{(\krX )}; \ryv^{(\krY )}\big) 
- I\big(\rt_{\rxv}; \phi_{\krY }(\ryv^{(\krY )}) \bcondi \phi_{\krX }(\rxv^{(\krX )})\big)
\notag \\*
& \quad
- I\big(\rxv^{(\krX )}; \rt_{\ryv} \bcondi \phi_{\krY }(\ryv^{(\krY )})\big)  \label{eq:misort1} \\
& = I\big(\rxv^{(\krX )}; \ryv^{(\krY )}\big)
- I\big(\rt_{\rxv}; \phi_{\krY }(\ryv^{(\krY )}) \bcondi \phi_{\krX }(\rxv^{(\krX )})\big)
\notag \\*
& \quad
- I\big(\phi_{\krX }(\rxv^{(\krX )}); \rt_{\ryv} \bcondi \phi_{\krY }(\ryv^{(\krY )})\big) 
\notag \\*
& \quad 
- I\big(\rt_{\rxv}; \rt_{\ryv} \bcondi \phi_{\krX }(\rxv^{(\krX )}), \phi_{\krY }(\ryv^{(\krY )})\big) \label{eq:misort2}
\ea
where $\rt_{\rxv}$ and $\rt_{\ryv}$ are the random permutations associated with the vectors in $\rxv^{(\krX )}$ and $\ryv^{(\krY )}$, respectively.
\end{lemma}
\begin{IEEEproof}
Due to the one-to-one relation between $\rxv^{(\krX )}$ and $\big(\phi_{\krX }(\rxv^{(\krX )}), \rt_{\rxv}\big)$, and between $\ryv^{(\krY )}$ and $\big(\phi_{\krY }(\ryv^{(\krY )}), \rt_{\ryv}\big)$, we have $I\big(\rxv^{(\krX )}; \ryv^{(\krY )}\big)
 = I\big(\phi_{\krX }(\rxv^{(\krX )}), \rt_{\rxv}; \phi_{\krY }(\ryv^{(\krY )}), \rt_{\ryv}\big)$.
Using the chain rule for mutual information \cite[Cor.~5.5.3]{Gray1990Entropy} three times, we thus obtain
\ba
& I\big(\rxv^{(\krX )}; \ryv^{(\krY )}\big) \notag \\
& = I\big(\phi_{\krX }(\rxv^{(\krX )});\phi_{\krY }(\ryv^{(\krY )})\big) 
+ I\big(\rt_{\rxv}; \phi_{\krY }(\ryv^{(\krY )}) \bcondi \phi_{\krX }(\rxv^{(\krX )})\big)  
\notag \\*
& \quad
+ I\big(\rxv^{(\krX )}; \rt_{\ryv} \bcondi \phi_{\krY }(\ryv^{(\krY )})\big) \notag \\*[-4mm]
 \label{eq:misort1pr} \\
& =  I\big(\phi_{\krX }(\rxv^{(\krX )});\phi_{\krY }(\ryv^{(\krY )})\big)
+ I\big(\rt_{\rxv}; \phi_{\krY }(\ryv^{(\krY )}) \bcondi \phi_{\krX }(\rxv^{(\krX )})\big)
\notag \\*
& \quad
+ I\big(\phi_{\krX }(\rxv^{(\krX )}); \rt_{\ryv} \bcondi \phi_{\krY }(\ryv^{(\krY )})\big) 
+ I\big(\rt_{\rxv}; \rt_{\ryv} \bcondi \phi_{\krX }(\rxv^{(\krX )}), \phi_{\krY }(\ryv^{(\krY )})\big).\label{eq:misort2pr}
\ea
Because 
the distributions of $(\rX,\rY)$ and $\big(\phi_{\krX }(\rxv^{(\krX )}), \phi_{\krY }(\ryv^{(\krY )})\big)$ are equal, we have 
$I(\rX;\rY) = I\big(\phi_{\krX }(\rxv^{(\krX )}); \phi_{\krY }(\ryv^{(\krY )})\big)$.
Hence, \eqref{eq:misort1pr} implies \eqref{eq:misort1} and \eqref{eq:misort2pr} implies \eqref{eq:misort2}.
\end{IEEEproof}
 

\subsection{Mutual Information for Point Processes of Equal \\ Cardinality}

If $(\rX,\rY)$ is a pair of PPs of equal cardinality  (see Section~\ref{sec:ppec}), the mutual information $I(\rX;\rY)$ still simplifies significantly compared to the general case.
\begin{corollary}\label{cor:mutinfec}
Let $(\rX, \rY)$ be a pair of PPs of equal cardinality, i.e., $p_{\card{\rX},\card{\rY}}(\krX , \krY )=0$ for $\krX \neq \krY $.
Then
\be \label{eq:mutinfec}
I(\rX;\rY) = H(\card{\rX})
+ \sum_{k\in \N} p_{\card{\rX}}(k) \,
I\big(\rxv_{\rX}^{(k)};\ryv_{\rY}^{(k)}\big)\,.
\ee
\end{corollary}
\begin{IEEEproof}
We have $\card{\rX}=\card{\rY}$ and thus (see~\cite[eq.~(2.42)]{Cover91}) $I(\card{\rX}; \card{\rY}) = H(\card{\rX})$.
Furthermore, we have $p_{\card{\rX},\card{\rY}}(\krX , \krY )=0$ for $\krX \neq \krY $.
Thus, by Theorem~\ref{th:mutinfgen}, we obtain
\ba %
I(\rX;\rY) 
& =
H(\card{\rX}) 
\notag \\* & \quad 
+ \sum_{k\in \N} p_{\card{\rX},\card{\rY}}(k,k) 
\divKL\big(P_{\rX,\rY}^{(k,k)}\big\| P_{\rX}^{(k)}\times P_{\rY}^{(k)}\big)\,.\notag 
\ea
According to~\eqref{eq:margcarddist}, $p_{\card{\rX}}(k)=p_{\card{\rX},\card{\rY}}(k,k)$.
Because $P_{\rX,\rY}^{(k,k)}$, $P_{\rX}^{(k)}$, and $P_{\rY}^{(k)}$ are the probability measures of $(\rxv,\ryv)_{\rX,\rY}^{(k, k)}$ and its marginals, $\rxv_{\rX}^{(k)}$ and $\ryv_{\rY}^{(k)}$, respectively, 
\eqref{eq:ecmargchar} implies 
$
\divKL\big(P_{\rX,\rY}^{(k,k)}\big\| P_{\rX}^{(k)}\times P_{\rY}^{(k)}\big)= I\big(\rxv_{\rX}^{(k)};\ryv_{\rY}^{(k)}\big)
$,
which concludes the proof.
\end{IEEEproof}

As in the case of fixed cardinality, we can start with arbitrary vectors $\big(\rxv^{(k)}, \ryv^{(k)}\big)$ without assuming symmetry.
Combining Corollary~\ref{cor:mutinfec} with the expression  of mutual information provided by Lemma~\ref{lem:splitmi}, this approach yields the following result.
\begin{theorem}\label{th:splitmutinf}
Let $(\rX, \rY)$ be a pair of PPs of equal cardinality, i.e., $p_{\card{\rX},\card{\rY}}(\krX , \krY )=0$ for $\krX \neq \krY $.
For every $k\in \N$, let $\big(\rxv^{(k)}, \ryv^{(k)}\big)$ be random vectors such that 
$(\rX^{(k)}, \rY^{(k)})\triangleq\big(\phi_{k}(\rxv^{(k)}), \phi_{k}(\ryv^{(k)})\big)$
and 
$\big(\phi_{k}(\rxv_{\rX}^{(k)}), \phi_{k}(\ryv_{\rY}^{(k)})\big)$ 
have the same distribution.
Then 
\ba
& I(\rX;\rY) \notag \\
& = H(\card{\rX})\,
+ \sum_{k\in \N} p_{\card{\rX}}(k) \,
\Big( 
I\big(\rxv^{(k)}; \ryv^{(k)}\big)
- I\big(\rt^{(k)}_{\rxv}; \rY^{(k)} \bcondi \rX^{(k)}\big)
\notag \\* &\quad 
- I\big(\rxv^{(k)}; \rt^{(k)}_{\ryv} \bcondi \rY^{(k)}\big)
\Big) 
  \label{eq:mutinfec2}\\
& = H(\card{\rX})\,
+ \sum_{k\in \N} p_{\card{\rX}}(k) \,
\Big(
I\big(\rxv^{(k)}; \ryv^{(k)}\big)
- I\big(\rt^{(k)}_{\rxv}; \rY^{(k)} \bcondi \rX^{(k)}\big)
\notag \\* 
& \quad 
- I\big(\rX^{(k)}; \rt^{(k)}_{\ryv} \bcondi \rY^{(k)}\big) 
- I\big(\rt^{(k)}_{\rxv}; \rt^{(k)}_{\ryv} \bcondi \rX^{(k)}, \rY^{(k)}\big)
\Big)\label{eq:mutinfec2b}
\ea
where $\rt^{(k)}_{\rxv}$ and $\rt^{(k)}_{\ryv}$ are the random permutations associated with the vectors in $\rxv^{(k)}$ and $\ryv^{(k)}$, respectively.
\end{theorem}
\begin{IEEEproof}
Because  the distributions of $(\rX^{(k)}, \rY^{(k)})$ and $\big(\phi_{k}(\rxv_{\rX}^{(k)}), \phi_{k}(\ryv_{\rY}^{(k)})\big)$ are equal, we have
$I(\rX^{(k)}; \rY^{(k)}) = I\big(\phi_{k}(\rxv_{\rX}^{(k)}); \phi_{k}(\ryv_{\rY}^{(k)})\big)$.
Using this equality and applying Lemma~\ref{lem:splitmi} to the pair of PPs $(\rX^{(k)}, \rY^{(k)})$, we obtain
\ba \notag
& I\big(\phi_{k}(\rxv_{\rX}^{(k)}); \phi_{k}(\ryv_{\rY}^{(k)})\big) \notag \\
& = I\big(\rxv^{(k)}; \ryv^{(k)}\big)
- I\big(\rt^{(k)}_{\rxv}; \rY^{(k)} \bcondi \rX^{(k)}\big)
- I\big(\rxv^{(k)}; \rt^{(k)}_{\ryv} \bcondi \rY^{(k)}\big) \notag \\
& = I\big(\rxv^{(k)}; \ryv^{(k)}\big)
- I\big(\rt^{(k)}_{\rxv}; \rY^{(k)} \bcondi \rX^{(k)}\big)
- I\big(\rX^{(k)}; \rt^{(k)}_{\ryv} \bcondi \rY^{(k)}\big)
\notag \\* 
& \quad 
- I\big(\rt^{(k)}_{\rxv}; \rt^{(k)}_{\ryv} \bcondi \rX^{(k)}, \rY^{(k)}\big)\,. \notag 
\ea
On the other hand, applying  Corollary~\ref{cor:mutinfeqcard} to the pair of PPs of fixed cardinality $\big(\phi_{k}(\rxv_{\rX}^{(k)}), \phi_{k}(\ryv_{\rY}^{(k)})\big)$, we have
$I\big(\phi_{k}(\rxv_{\rX}^{(k)}); \phi_{k}(\ryv_{\rY}^{(k)})\big) = I\big( \rxv^{(k)}_{\rX}; \ryv^{(k)}_{\rY}\big)$.
Combining these equalities and inserting into \eqref{eq:mutinfec} concludes the proof.
\end{IEEEproof}

\subsection{Rate-Distortion Function for Point Processes} \label{sec:rdfunction}   
%
  
We summarize the main concepts of RD theory \cite[Sec.~10]{Cover91} in the PP setting. 
For two point patterns $X,Y\in \sX$, let
 $\dist\colon \sX\times \sX\to \R_{\geq 0}$ be a measurable distortion function, i.e., $\dist(X,Y)$ quantifies the distortion incurred by changing $X$ to $Y$. 
A source generates  i.i.d.\ copies $\rX[j]$, $j\in \N$ of a PP on $\R^d$.
Loosely speaking, the RD function $R_{(\rX[j])_{j\in \N},\,\dist}(D)$ gives the smallest possible encoding rate for maximum expected distortion $D$.
In mathematical terms, $R_{(\rX[j])_{j\in \N},\,\dist}(D)$ is the infimum of all 
$R>0$ such that for all $\varepsilon>0$ there exists 
an $n\in \N$ and a \emph{source code}, i.e., a measurable mapping $g_n\colon \sX^n\to \sX^n$,
satisfying $\log (\card{g_n(\sX^n)})\leq nR$ and $\E\big[\frac{1}{n}\sum_{j=1}^n\dist\big(\rX[j], \rY[j]\big)\big] \leq D+\varepsilon$, where $(\rY[1], \dots, \rY[n]) = g_n(\rX[1], \dots, \rX[n])\in \sX^n$.
 
%
Following common practice, we 
will write $R_{(\rX[j])_{j\in \N},\,\dist}(D)$ briefly
as $R(D)$.
Furthermore, we specify the source by only one PP $\rX$ and tacitly assume that $(\rX[j])_{j\in \N}$ consists of i.i.d.\ PPs with the same distribution as $\rX$.

\begin{remark}
In the vector case, $\dist(\xv,\yv)$ is usually defined based on  $\xv-\yv$, e.g., the  squared-error distortion $\dist(\xv,\yv)=\norm{\xv-\yv}^2$.
However, in the case of point patterns $X$ and $Y$, this convenient construction is not possible because there is no meaningful definition of $X-Y$ as a difference between point patterns.
This results in a significantly more involved analysis and construction of source codes.
\end{remark}

For simplicity, we will assume $\dist(X,X)=0$ for all $X \in \sX$.
Moreover, we will use some general theorems for the characterization of RD functions, which   can also be applied to the setting of PPs.
These theorems require   that 
%
	there exists a reference point pattern $A^* \in  \sX$ such that $\E[\dist(\rX,A^*)]<\infty$.
This condition is satisfied, e.g., if the distortion between $X\in \sX$ and  the empty set is  a linear function of the cardinality $\card{X}$ (cf.~\eqref{eq:ospagen}), i.e.,  $\dist(X,\emptyset)=c \card{X}$, and the PP $\rX$ has finite expected cardinality $\E[\card{\rX}]<\infty$.
 
%

The RD theorem for general i.i.d.\ sources \cite[Th. 7.2.4 and Th. 7.2.5]{berger71} states that for a given source PP $\rX$ and distortion function $\dist$, the RD function 
can be calculated as
\be\label{eq:rdasinf}
R(D)
=\inf_{ (\widetilde\rX,\rY) :\, \E[\dist( \widetilde\rX ,\rY)]\leq D} I( \widetilde\rX ;\rY)
\ee
where the infimum is taken over all pairs of PPs $(\widetilde\rX,\rY)$ such that $\widetilde\rX$ has the same distribution as $\rX$ and $\E[\dist(\widetilde\rX,\rY)]\leq D$.  
The expression \eqref{eq:rdasinf} is useful for the derivation of upper bounds on the RD function (see Section~\ref{sec:upperbounds}).
Another characterization of the RD function that is  more useful for the derivation of lower bounds (see Section~\ref{sec:lowerbounds}) is  \cite[Th.~2.3]{csiszar74}
\be\label{eq:rdassup}
R(D)=\max_{s\geq 0} \max_{\alpha_s(\cdot)>0} \bigg( \int_{\sX} \log \alpha_s(X) \, \mathrm{d}P_{\rX}(X) -sD \bigg)
\ee
where the inner maximization is over all   positive functions $\alpha_s\colon \sX \to \R^{+}$ satisfying
\be\label{eq:cond1}
\int_{\sX} \alpha_s(X) e^{-s \dist(X,Y)}\, \mathrm{d}P_{\rX}(X) \leq 1
\ee
for all $Y\in \sX$.
Let us assume that the measures $P_{\rX}^{(k)}$ are absolutely continuous
 with respect to $(\Leb^{d})^{k}$ with Radon-Nikodym derivatives
$\frac{\intd P_{\rX}^{(k)}}{\intd (\Leb^{d})^{k}}=f_{\rX}^{(k)}$,
i.e.,  the $\rxv_{\rX}^{(k)}$ are continuous random vectors.
Then, \eqref{eq:rdassup} and \eqref{eq:cond1} can equivalently be written as follows.
Using \eqref{eq:pxaint} with $g(X)=\log \alpha_s(X)$, \eqref{eq:rdassup} becomes
\ba
R(D) & =\max_{s\geq 0} \max_{\alpha_s(\cdot)>0} \bigg(
p_{\card{\rX}}(0)\log \alpha_s(\emptyset) 
+
\sum_{k\in \N}p_{\card{\rX}}(k) 
\notag \\* & \quad \times 
\int_{(\R^{d})^k}\log \alpha_s(\phi_k(\xv_{1:k}))\, f_{\rX}^{(k)}(\xv_{1:k})\,\intd\xv_{1:k} -sD
\bigg)  \label{eq:rdassupac}
\\[-7mm] \notag
\ea
where $\intd\xv_{1:k}$ is short for $\intd(\Leb^{d})^k(\xv_{1:k})$ 
and the inner maximization is over all  positive functions $\alpha_s\colon \sX \to \R^{+}$ satisfying
(using \eqref{eq:pxaint} with $g(X)=\alpha_s(X) e^{-s \dist(X,Y)}$ in \eqref{eq:cond1})
\ba
& p_{\card{\rX}}(0)\alpha_s(\emptyset) e^{-s \dist(\emptyset,Y)}
+
\sum_{k\in \N}p_{\card{\rX}}(k) 
\notag \\* &  \times 
\int_{(\R^{d})^k} \alpha_s(\phi_k(\xv_{1:k})) e^{-s \dist(\phi_k(\xv_{1:k}),Y)} f_{\rX}^{(k)}(\xv_{1:k})\,\intd\xv_{1:k}
\leq 1\label{eq:cond1ac}
\\[-8mm] \notag
\ea
for all $Y\in \sX$.

\section{Lower Bounds} \label{sec:lowerbounds}   
%

Lower bounds on the RD function are notoriously hard to obtain.
The only well-established lower bound is the Shannon lower bound, which is 
based on the
characterization of the RD function given in \eqref{eq:rdassup}, \eqref{eq:cond1}.
More specifically, by omitting in  \eqref{eq:rdassup} the maximization over $\alpha_s$ and using  any specific positive function $\alpha_s$ satisfying \eqref{eq:cond1}   yields the lower bound
\be\notag 
R(D)\geq\max_{s\geq 0} \bigg( \int_{\sX} \log \alpha_s(X) \, \mathrm{d}P_{\rX}(X) -sD \bigg)\,.
\ee 
The standard approach \cite[Sec.~4]{gr90} is to set
$\alpha_s(X)\triangleq 
\frac{1}{f_{\rX}(X)\gamma(s)}$,  
where $f_{\rX}=\frac{\intd P_{\rX}}{\intd Q}$ is the Radon-Nikodym derivative of $P_{\rX}$ with respect to some background measure $Q$ on the given measurable space  $(\sX, \mathfrak{S})$ that satisfies $P_{\rX}\ll Q$, and $\gamma(s)$, $s\geq 0$ is a suitably chosen function.
  
In this standard approach, $\gamma(s)$ is chosen independently of the cardinality of $X$, which is too restrictive for the construction of useful lower bounds for PPs.
Hence, we take a slightly different approach and define
$\alpha_s(X)\triangleq \frac{1}{f_{\rX}(X)\gamma_{\card{X}}(s)}$
 with appropriate functions $\gamma_{\card{X}}(s)$ that depend on $\card{\rX}$.
More specifically, we propose the following bound.
\begin{theorem}\label{th:shlbpp}
Let $\rX$ be a PP on $\R^d$ and 
assume that for all $k\in \N$, the measures $P_{\rX}^{(k)}$ are absolutely continuous with respect to $(\Leb^{d})^{k}$ with Radon-Nikodym derivatives $\frac{\intd P_{\rX}^{(k)}}{\intd (\Leb^{d})^{k}}=f_{\rX}^{(k)}$, i.e., $\rxv_{\rX}^{(k)}$ are continuous random vectors with probability  density functions $f_{\rX}^{(k)}$.
For any measurable sets $A_k\subseteq (\R^d)^k$ satisfying $P_{\rX}^{(k)}(A_k)=1$, i.e., $f_{\rX}^{(k)}(\xv_{1:k})=0$ for $(\Leb^{d})^k$-almost all $\xv_{1:k}\in (A_k)^c$, the RD function is lower-bounded according 
to
\ba
R(D)& \geq
 \sum_{k\in \N} p_{\card{\rX}}(k) h\big(f_{\rX}^{(k)}\big) 
\notag \\* & \quad 
+ \max_{s\geq 0} \bigg({-} \sum_{k\in \N_0} p_{\card{\rX}}(k)\log \gamma_k(s) -sD \bigg) \label{eq:rdlbac}
\ea
where $\gamma_k$ are any functions satisfying
\be\label{eq:choosegamma}
\gamma_k(s)\geq 
\begin{cases}
  e^{-s \dist(\emptyset,Y)}& \text{ if } k=0 \\ 
  \int_{A_k} e^{-s \dist(\phi_k(\xv_{1:k}),Y)}\, \intd \xv_{1:k} & \text{ if } k\in \N
\end{cases}
\ee
for all $Y\in \sX$ and $s\geq 0$.
\end{theorem}
\begin{IEEEproof}
The characterization of the RD function in  \eqref{eq:rdassupac} implies that for any $\alpha_s$ satisfying \eqref{eq:cond1ac}, 
\ba
R(D) & \geq \max_{s\geq 0}   \bigg(
p_{\card{\rX}}(0)\log \alpha_s(\emptyset) 
+
\sum_{k\in \N}p_{\card{\rX}}(k) 
\notag \\* & \quad \times 
\int_{(\R^{d})^k}\log \alpha_s(\phi_k(\xv_{1:k}))\, f_{\rX}^{(k)}(\xv_{1:k})\,\intd\xv_{1:k} -sD
\bigg) \notag \\
&  \stackrel{\hidewidth (a) \hidewidth}= \max_{s\geq 0}   \bigg(
p_{\card{\rX}}(0)\log \alpha_s(\emptyset) 
+
\sum_{k\in \N}p_{\card{\rX}}(k) 
\notag \\* & \quad \times 
\int_{A_k}\log \alpha_s(\phi_k(\xv_{1:k}))\, f_{\rX}^{(k)}(\xv_{1:k})\,\intd\xv_{1:k} -sD
\bigg) 
  \label{eq:rdassupaclb}
\ea
where $(a)$ holds because we assumed that $f_{\rX}^{(k)}(\xv_{1:k})=0$ for $(\Leb^{d})^k$-almost all
$\xv_{1:k}\in (A_k)^c$. 
Using functions $\gamma_k$ satisfying \eqref{eq:choosegamma}, we construct $\alpha_s$  as
\be\label{eq:newchoicealpha}
\alpha_s(\phi_k(\xv_{1:k}))\triangleq 
\begin{cases}
\frac{1}{f^{(k)}_{\rX}(\xv_{1:k})\gamma_{k}(s)} & \text{ if } f^{(k)}_{\rX}(\xv_{1:k})\neq 0 \\[2mm]
1 &  \text{ if } f^{(k)}_{\rX}(\xv_{1:k})= 0 
\end{cases}
\ee
 for $\xv_{1:k}\in (\R^d)^k$ and
\be\label{eq:newchoicealpha0}
\alpha_s(\emptyset)\triangleq  \frac{1}{\gamma_{0}(s)}\,.
\ee
Due to \eqref{eq:choosegamma},
the functions $\gamma_k$ satisfy
\ba 
& p_{\card{\rX}}(0)\frac{1}{\gamma_{0}(s)} e^{-s \dist(\emptyset,Y)}
\notag \\* & 
\quad + \sum_{k\in \N}p_{\card{\rX}}(k) \int_{A_k} \frac{1}{\gamma_{k}(s)} e^{-s \dist(\phi_k(\xv_{1:k}),Y)}\,\intd\xv_{1:k}
\notag \\ 
&   \leq 
p_{\card{\rX}}(0) + \sum_{k\in \N}p_{\card{\rX}}(k) \notag \\
&  = 1 \notag 
\ea
for all $Y\in \sX$, which is recognized as the condition \eqref{eq:cond1ac} evaluated  for the functions $\alpha_s$ given by  \eqref{eq:newchoicealpha} and \eqref{eq:newchoicealpha0}.
Inserting \eqref{eq:newchoicealpha} and \eqref{eq:newchoicealpha0} into   \eqref{eq:rdassupaclb} gives  
\ba
 R(D) 
& \geq \max_{s\geq 0} \bigg(
p_{\card{\rX}}(0)\log \frac{1}{\gamma_{0}(s)} 
+
\sum_{k\in \N}p_{\card{\rX}}(k) 
\notag \\* &  \quad \times \hspace{-1mm}
\int_{A_k}  \hspace{-1mm} f_{\rX}^{(k)}(\xv_{1:k}) \log \bigg(\frac{1}{f^{(k)}_{\rX}(\xv_{1:k})\gamma_{k}(s)}\bigg)  \intd\xv_{1:k} -sD  \hspace{-0.5mm}
\bigg)
\notag \\ 
& = \max_{s\geq 0} \bigg(
-p_{\card{\rX}}(0)\log \gamma_{0}(s)
\notag \\* & \quad 
+
\sum_{k\in \N}p_{\card{\rX}}(k) \Big( h\big(f_{\rX}^{(k)}\big) 
-   \log  \gamma_{k}(s)  \Big)
 -sD
\bigg) \notag 
\ea
which is equivalent to \eqref{eq:rdlbac}.
\end{IEEEproof}

\section{Upper Bounds} \label{sec:upperbounds}   
%
We will use two different approaches to calculate upper bounds on the RD function. 
The first is based on the RD theorem, i.e.,  expression \eqref{eq:rdasinf}, whereas the second uses concrete codes and the operational interpretation of the RD function.

\subsection{Upper Bounds Based on the Rate-Distortion Theorem}
Let $\rX$ be a PP defined by the cardinality distribution $p_{\card{\rX}}$ and the random vectors $\rxv_{\rX}^{(k)}$ (see~Section~\ref{sec:fpp}).
To calculate upper bounds, we can construct  an arbitrary   pair of PPs $(\widetilde{\rX}, \rY)$ (see Section~\ref{sec:fppp})   
such that   $\widetilde{\rX}$ has the same distribution as $\rX$ and 
$\E[\dist(  \widetilde{\rX},\rY)]\leq D$.
According to \eqref{eq:rdasinf}, we then have $R(D)\leq I(  \widetilde{\rX};\rY)$.
However, it is often easier to construct vectors $(\rxv^{(k)},\ryv^{(k)})$ that do not satisfy the symmetry properties we assumed 
in the construction of 
  pairs of PPs in Section~\ref{sec:fppp}.  
The following   corollary to Theorem~\ref{th:splitmutinf}  shows that in the case where $\rxv_{\rX}^{(k)}$ is a ``symmetrized'' version of $\rxv^{(k)}$, we can  construct upper bounds on $R(D)$ based on $(\rxv^{(k)},\ryv^{(k)})$. 
\begin{corollary}\label{cor:upperbounddpi}
Let $\rX$ be a PP on $\R^d$ defined by the cardinality distribution $p_{\card{\rX}}$ and the random vectors $\rxv_{\rX}^{(k)}$.
Furthermore, for each $k\in \N$, let $(\rxv^{(k)}, \ryv^{(k)})$ be a random vector on $(\R^{d})^{2k}$ such that $ \rX^{(k)}\triangleq\phi_k(\rxv^{(k)})$ 
has the same distribution as $\phi_k(\rxv_{\rX}^{(k)})$.
Finally, assume 
that 
\be\label{eq:assexbounded}
\sum_{k\in \N} p_{\card{\rX}}(k)\, \E\big[\dist\big( \rX^{(k)},\rY^{(k)}  \big)\big]\leq D 
\ee
  with $\rY^{(k)}\triangleq \phi_k(\ryv^{(k)})$. 
Then 
\ba
& R(D)
\notag \\* 
& \leq H(\card{\rX})\,
+ \sum_{k\in \N} p_{\card{\rX}}(k) \,
\Big( 
I\big(\rxv^{(k)}; \ryv^{(k)}\big)
- I\big(\rt^{(k)}_{\rxv}; \rY^{(k)} \bcondi \rX^{(k)}\big)
\notag \\* & \quad 
- I\big(\rxv^{(k)}; \rt^{(k)}_{\ryv} \bcondi \rY^{(k)}\big)
\Big)\label{eq:rdboundvectors} \\
&  = H(\card{\rX})\,
+ \sum_{k\in \N} p_{\card{\rX}}(k) \,
\Big(
I\big(\rxv^{(k)}; \ryv^{(k)}\big)
- I\big(\rt^{(k)}_{\rxv}; \rY^{(k)} \bcondi \rX^{(k)}\big)
\notag \\* & \quad 
- I\big(\rX^{(k)}; \rt^{(k)}_{\ryv} \bcondi \rY^{(k)}\big) 
- I\big(\rt^{(k)}_{\rxv}; \rt^{(k)}_{\ryv} \bcondi \rX^{(k)}, \rY^{(k)}\big)
\Big)\notag 
\ea 
where $\rt^{(k)}_{\rxv}$ and $\rt^{(k)}_{\ryv}$ are the random permutations associated with the vectors in $\rxv^{(k)}$ and $\ryv^{(k)}$, respectively.%
\end{corollary}
\begin{IEEEproof}
We construct a pair of PPs $(\widetilde{\rX}, \rY)$ of equal cardinality.
First, we define the cardinality distribution as
$p_{\card{\widetilde{\rX}},\card{\rY}}(\krX, \krY )=0$ for $\krX\neq  \krY $ and $p_{\card{\widetilde{\rX}},\card{\rY}}(k, k)=p_{\card{\rX}}(k)$ for $k\in \N_0$.
Next, we define the random vectors $\big(\rxv^{(k)}_{\widetilde{\rX}},\ryv^{(k)}_{\rY}\big)$ such that $(\rX^{(k)}, \rY^{(k)}) = \big(\phi_{k}(\rxv^{(k)}), \phi_{k}(\ryv^{(k)})\big)$
and 
$\big(\phi_{k}(\rxv_{\widetilde{\rX}}^{(k)}), \phi_{k}(\ryv_{\rY}^{(k)})\big)$ 
have the same distribution.
By \eqref{eq:mutinfec2} and \eqref{eq:mutinfec2b}, we then obtain for the pair of PPs  $(\widetilde{\rX}, \rY)$
\ba
& I(\widetilde{\rX};\rY) 
\notag \\*
& = H(\card{\rX})\,
+ \sum_{k\in \N} p_{\card{\rX}}(k) \,
\Big( 
I\big(\rxv^{(k)}; \ryv^{(k)}\big)
- I\big(\rt^{(k)}_{\rxv}; \rY^{(k)} \bcondi \rX^{(k)}\big)
\notag \\* & \quad 
- I\big(\rxv^{(k)}; \rt^{(k)}_{\ryv} \bcondi \rY^{(k)}\big)
\Big) \label{eq:rdboundvectorspr1} \\
& = H(\card{\rX})\,
+ \sum_{k\in \N} p_{\card{\rX}}(k) \,
\Big(
I\big(\rxv^{(k)}; \ryv^{(k)}\big)
- I\big(\rt^{(k)}_{\rxv}; \rY^{(k)} \bcondi \rX^{(k)}\big)
\notag \\* & \quad 
- I\big(\rX^{(k)}; \rt^{(k)}_{\ryv} \bcondi \rY^{(k)}\big) 
- I\big(\rt^{(k)}_{\rxv}; \rt^{(k)}_{\ryv} \bcondi \rX^{(k)}, \rY^{(k)}\big)
\Big)\,. \label{eq:rdboundvectorspr2}
\ea
Furthermore,
because $\rX^{(k)}$ has the same distribution as $\phi_k(\rxv_{\rX}^{(k)})$, the construction of $(\widetilde{\rX}, \rY)$ implies that $\phi_{k}(\rxv_{\widetilde{\rX}}^{(k)})$ has the same distribution as $\phi_k(\rxv_{\rX}^{(k)})$ too, and, in turn, the PPs $\widetilde{\rX}$ and $\rX$ have the same distribution.
Since \eqref{eq:assexbounded} implies $\E[\dist(\widetilde{\rX},\rY)]= \sum_{k\in \N} p_{\card{\rX}}(k)\, \E\big[\dist\big(\rX^{(k)},\rY^{(k)}\big)\big]\leq D$, we obtain by \eqref{eq:rdasinf} that $R(D)\leq I(\widetilde{\rX};\rY)$,
which in combination with \eqref{eq:rdboundvectorspr1} and \eqref{eq:rdboundvectorspr2} concludes the proof.
\end{IEEEproof}

\subsection{Codebook-Based Upper Bounds}\label{sec:cbub}
 
It is well known \cite[Sec.~10.2]{Cover91} that 
   the RD function for a given PP can be easily upper-bounded based on its operational interpretation
	if we are able to construct good source codes. 
%
Let $\rX$ be a PP and assume that there exists a source code $g\colon \sX\to \sX$ 
such that
$\card{g(\sX)}=\codew$.
If $\E[\dist(\rX,g(\rX))]\leq \widetilde{D}$, then the RD function at $\widetilde{D}$ 
satisfies  
\be   \label{eq:scupperbound}
R(\widetilde{D})\leq \log\codew\,.
\ee

The construction of good source codes, even in the vector case, is a difficult optimization task.
In the case of PPs, this task is further complicated by the absence of a meaningful vector space structure of sets; even the definition of a ``mean'' of point patterns is not straightforward.
Our construction is motivated by the 
Lloyd algorithm \cite{ll82}, which, for a given number $\codew$ of codewords  (i.e., elements in~$g(\sX)$), 
alternately finds $\codew$ ``centers'' $X_j \in \sX$ and constructs an associated partition $\{\sA_j\}_{j=1, \dots, \codew}$ of $\sX$.
The resulting centers $X_1, \dots, X_{\codew}$ can be used as codewords 
 and the associated source code $g$ is defined as
\be\notag
g\colon \sX  \to \sX; \qquad 
X  \mapsto \argmin_{X_j\in \{X_1,  \dots, X_{\codew}\}} \dist(X,X_j)\,.
\ee
That is, a point pattern $X\in \sX$ is encoded into the center point pattern $X_j$ that is closest to $X$ in the sense of minimizing $\dist(X,X_j)$.
In our setting, the Lloyd algorithm can be formalized as follows:

\begin{itemize}
	\item Input: PP $\rX$; distortion function $\dist\colon \sX\times \sX \to \R_{\geq 0}$; number $\codew\in \N$ of codewords.
	\item Initialization: Draw $\codew$ different initial codewords $X_j\in \sX$  according to the distribution of $\rX$.
	\item Step 1: Find a partition of $\sX$ into $\codew$ disjoint subsets $\sA_j$ such that the distortion incurred by changing $X\in\sA_j$ to $X_j$ is less than or equal to the distortion incurred by changing $X$ to any other $X_{j'}$, $j'\neq j$, i.e.,  $\dist(X,X_j)\leq \dist(X,X_{j'})$ for all $j'\neq j$ and all $X\in \sA_j$.
	\item Step 2: For each $j\in \{1,\dots, \codew\}$, find a new codeword associated with $\sA_j$ that has the smallest expected distortion from all point patterns in $\sA_j$, i.e., a ``center point pattern'' $X_j$ (replacing the previous $X_j$) satisfying
	$
	 X_j =\argmin_{\widetilde{X}\in \sX} \E_{\rX|\rX\in\sA_j}[\dist(\rX,\widetilde{X})]
	$.
	\item Repeat Step 1 and Step 2 until some convergence criterion is satisfied.
	\item Output: codebook $\{X_1, \dots, X_{\codew}\}$.
\end{itemize}

Unfortunately, 
closed-form solutions   for Steps 1 and 2 do not exist in general.
A workaround 
is an approach known in vector quantization as Linde-Buzo-Gray (LBG) algorithm \cite{libugr80}. 
Here, a codebook is constructed based on a given set $\sA$ of source realizations.
We can generate the set $\sA$ by drawing i.i.d.\ samples of $\rX$.
Adapted to our setting, the algorithm can be stated as follows:

\begin{itemize}
	\item Input: a set $\sA\subseteq \sX$ containing $\card{\sA}< \infty$ point patterns; distortion function $\dist\colon \sX\times \sX \to \R_{\geq 0}$; number $\codew$ of codewords.
	\item Initialization: Randomly choose $\codew$ different initial codewords $X_j\in \sA$.
	\item Step 1: Find a partition of $\sA$ into $\codew$ disjoint subsets $\sA_j$ such that the distortion incurred by changing $X\in\sA_j$ to $X_j$ is less than or equal to the distortion incurred by changing $X$ to any other $X_{j'}$, $j'\neq j$, i.e.,  
	$\dist(X,X_j)\leq \dist(X,X_{j'})$ for all $j'\neq j$ and all $X\in \sA_j$.
	\item Step 2: For each $j\in\{1,\dots, \codew\}$, find a new codeword associated with $\sA_j$ that has the smallest average distortion from all point patterns in $\sA_j$, i.e., a ``center point pattern'' $X_j\in \sX$ (replacing the previous $X_j$) 
satisfying 
	\be\label{eq:centerpp}
	 X_j =\argmin_{\widetilde{X}\in \sX} \frac{1}{\card{\sA_j}}\sum_{X\in\sA_j}\dist(X,\widetilde{X})\,.
	\ee
	\item Repeat Step 1 and Step 2 until some convergence criterion is satisfied.
	\item Output: codebook $\{X_1, \dots, X_{\codew}\}$.
\end{itemize}

Step 1 can  be performed by calculating $\card{\sA} \codew$ times a distortion $\dist(X,X_j)$.
However, Step 2 is typically computationally unfeasible: in many cases,  finding a center point pattern of a finite collection $\sA_j$ of point patterns according to \eqref{eq:centerpp} is equivalent to solving a multi-dimensional assignment problem, which is known to be NP-hard.
Hence, we will have to resort to approximate or heuristic solutions.
Note that we do not have to solve the optimization problem exactly   to obtain upper bounds on the RD function.
We merely have  to construct a source code that can be analyzed, no matter what heuristics or approximations were used in its construction. 
A convergence analysis of the proposed algorithm appears to be difficult, as
even  the convergence behavior of the Lloyd algorithm in $\R^d$ is not
completely understood \cite{emjura08,luzh16}.

\section{Point Processes  of Fixed Cardinality}\label{sec:fcpp}

In this section, we present lower and upper bounds on the RD function for
PPs   of fixed cardinality as discussed in Section~\ref{sec:ppfc}.
We thus restrict our analysis to source codes and distortion functions on the subset $\sX_k=\{X\in \sX:\card{X}=k\}\subseteq \sX$.
The assumption of fixed cardinality leads to more concrete bounds and enables a comparison with the vector viewpoint.

\subsection{Lower Bound}
%
%
For a PP $\rX$ of fixed cardinality, the RD lower bound in Theorem~\ref{th:shlbpp} simplifies as follows.

\begin{corollary}\label{cor:rdlbfc}
Let $\rX$ be a PP on $\R^d$ of fixed cardinality $\card{\rX}=k$, i.e., $p_{\card{\rX}}(k)=1$ for some $k\in \N$.
Assume that the measure $P_{\rX}^{(k)}$ is absolutely continuous with respect to $(\Leb^{d})^{k}$ with Radon-Nikodym derivative $\frac{\intd P_{\rX}^{(k)}}{\intd (\Leb^{d})^{k}}=f_{\rX}^{(k)}$.
Then the RD function is lower-bounded according to
\be\notag 
R(D)\geq
h\big(f_{\rX}^{(k)}\big) + \max_{s\geq 0}  ( - \log \gamma_k(s) -sD )
\ee
where $\gamma_k$ is any function satisfying
\be
\gamma_k(s)\geq
  \int_{(\R^{d})^k} e^{-s \dist(\phi_k(\xv_{1:k}),Y)}\, \mathrm{d}
	\xv_{1:k}
	\notag
\ee
for all $Y\in \sX_k$. 
\end{corollary}

 
We can obtain a simpler bound by considering a specific distortion function.
For point patterns $X=\{\xv_1, \dots, \xv_k\}$ and $Y=\{\yv_1, \dots, \yv_k\}$ of equal cardinality $k$,
we  define the distortion function as
\be\label{eq:ospafc}
\dist_2(X,Y)\triangleq\min_{\tau}\sum_{i=1}^k\norm{\xv_i-\yv_{\tau(i)}}^2
\ee
 where  the minimum is taken over all permutations $\tau$ on $\{1, \dots, k\}$.
This is 
a natural counterpart of the classical squared-error distortion function of vectors.
We note that a generalization to sets $X,Y$ of different cardinalities (and the inclusion of a  normalization factor $1/k$) leads to the squared optimal subpattern assignment (OSPA) metric
%
defined in \cite{scvovo08}.
The idea of the following lower bound is that a source code  for PPs can be extended to a source code  for vectors, by additionally specifying an ordering. 
\begin{theorem}\label{th:lbspecific}
Let $\rX$ be a PP on $\R^d$ of fixed cardinality $\card{\rX}=k$
and let $\rxv^{(k)}$ be a random vector on $(\R^d)^k$ such that $\phi_k(\rxv^{(k)})$ has the same distribution as $\rX$.
Then the RD function  for $\rX$ and distortion function $\dist_2$ is lower-bounded in terms of the RD function $R_{\text{vec}}$ for $\rxv^{(k)}$ and squared-error distortion according to
\be\label{eq:veclowerbound}
R(D) 
\geq R_{\text{vec}}(D) - \log k!\,.
\ee
\end{theorem}
\begin{IEEEproof}
Let $D>0$ be fixed. 
For any $R>R(D)$ and $\varepsilon>0$, the operational definition of the RD function (see\ Section~\ref{sec:rdfunction}) implies that there exists an $n\in \N$ and a source code $g_n\colon \sX^n\to \sX^n$ such that
$\log (\card{g_n(\sX^n)})\leq nR$ 
and 
\be
\E\bigg[\frac{1}{n}\sum_{j=1}^n\dist_2\big(\rX[j], \rY[j]\big)\bigg] \leq D+\varepsilon
\label{eq:expcdisteps}
\ee
where $(\rY[1], \dots, \rY[n]) = g_n(\rX[1], \dots, \rX[n])\in \sX^n$ and  the $\rX[j]$ are i.i.d.\ copies of $\rX$.
We define a vector source code 
$g_{\text{vec}, n}\colon ((\R^d)^{k})^n \to ((\R^d)^{k})^n$ for sequences of length $n$ in $(\R^d)^{k}$ by the following procedure.
For a sequence $\big(\xv_{1:k}{[1]}, \allowbreak \dots,  \allowbreak \xv_{1:k}{[n]}\big)$ with $\xv_{1:k}{[j]}\in (\R^d)^{k}$,
we first map each vector $\xv_{1:k}{[j]}$ to the corresponding point pattern  $X[j]= \phi_k\big(\xv_{1:k}{[j]}\big) = \big\{\xv_1{[j]},  \allowbreak \dots,  \allowbreak \xv_k{[j]}\big\}$.
Then we use the source code $g_n$ to obtain an encoded sequence of point patterns 
$\big(Y[1],  \allowbreak \dots,  \allowbreak Y[n]\big)=g_n\big(X[1],  \allowbreak \dots,  \allowbreak X[n]\big)$.
Finally, we map each point pattern $Y[j]= \big\{\yv_1{[j]},  \allowbreak \dots,  \allowbreak \yv_k{[j]}\big\}$ to a vector $\big(\yv_{\tau{[j]}(1)}{[j]}, \allowbreak \dots, \yv_{\tau{[j]}(k)}{[j]}\big)$ by a permutation $\tau[j]$ such that the squared error $\sum_{i=1}^k\bignorm{\xv_i{[j]}-\yv_{\tau{[j]}(i)}{[j]}}^2$ is minimized, i.e., 
$\tau{[j]} = \argmin_{\tilde\tau{[j]}}\sum_{i=1}^k\bignorm{\xv_i{[j]}-\yv_{\tilde\tau{[j]}(i)}{[j]}}^2$.
Based on this construction, the elements 
$\big(\yv_{\tau{[1]}(1)}{[1]}, \allowbreak\dots,  \allowbreak\yv_{\tau{[1]}(k)}{[1]}, \allowbreak\dots, \allowbreak\yv_{\tau{[n]}(1)}{[n]}, \allowbreak\dots, \allowbreak\yv_{\tau{[n]}(k)}{[n]}\big)$ 
in the range of  $g_{\text{vec}, n}$ are sequences in $g_n(\sX^n)$ with the elements of  each component $Y[j]$ ordered according to some permutation $\tau{[j]}$. 
As there are $k!$ possible orderings for each component, we have $\big\lvert g_{\text{vec}, n}\big((\R^d)^{kn}\big)\big\rvert\leq (k!)^n \card{g_n(\sX^n)} \leq (k!)^n e^{nR} \leq e^{n(R+\log k!)}$.
Furthermore, for $\big(\yv_{1:k}{[1]}, \allowbreak\dots, \allowbreak\yv_{1:k}{[n]}\big)= g_{\text{vec}, n}\big(\xv_{1:k}{[1]}, \allowbreak\dots, \allowbreak\xv_{1:k}{[n]}\big)$, we have
\ba
\sum_{j=1}^n \bignorm{\xv_{1:k}{[j]}- \yv_{1:k}{[j]}}^2
& = \sum_{j=1}^n \min_{\tilde\tau{[j]}}\sum_{i=1}^k\bignorm{\xv_i{[j]}-\yv_{\tilde\tau{[j]}(i)}{[j]}}^2\notag \\ 
& \stackrel{\eqref{eq:ospafc}}= \sum_{j=1}^n \dist_2(X[j], Y[j]) \notag
\ea
and thus for $\big(\ryv^{(k)}[1], \allowbreak\dots, \allowbreak \ryv^{(k)}[n]\big) = g_{\text{vec}, n}\big(\rxv^{(k)}[1], \allowbreak\dots, \allowbreak\rxv^{(k)}[n]\big)$, where the $\rxv^{(k)}[j]$ are i.i.d.\ copies of $\rxv^{(k)}$,
\ba
\E\bigg[\frac{1}{n}\sum_{j=1}^n\bignorm{\rxv^{(k)}[j]- \ryv^{(k)}[j]}^2\bigg]
& = \E\bigg[\frac{1}{n}\sum_{j=1}^n\dist_2\big(\rX[j], \rY[j]\big)\bigg] \notag \\
& \stackrel{\hidewidth \eqref{eq:expcdisteps} \hidewidth}\leq D+\varepsilon\,.\notag
\ea
Hence, for an arbitrary $\tilde R\triangleq R+\log k!>R(D)+\log k!$ and $\varepsilon >0$, we constructed a source code $g_{\text{vec}, n}$ such that $\log \big(\big\lvert g_{\text{vec}, n}\big((\R^d)^{kn}\big)\big\rvert\big)\leq n\tilde R$ and the expected average distortion between $\big(\rxv^{(k)}[1], \dots, \rxv^{(k)}[n]\big)$ and $g_{\text{vec}, n}\big(\rxv^{(k)}[1], \dots, \rxv^{(k)}[n]\big)$ is less than or equal to $D+\varepsilon$.
According to the operational definition of the RD function in Section~\ref{sec:rdfunction}, we hence obtain $R_{\text{vec}}(D)\leq R(D)+\log k!$.
\end{IEEEproof}
The offset $\log k!$ 
corresponds to the maximal   information that a vector contains in addition to the information present in the set, i.e., the maximal information provided by the \emph{ordering} of the $k$ elements. 
Indeed, 
the ``information content'' of the   ordering is maximal if all of the $k!$ possible orderings are   equally likely, in which case it is given by $\log k!$.
For $D\to 0$, the bound   \eqref{eq:veclowerbound}  shows that the asymptotic behavior of the  RD function $R(D)$ for small distortions is similar to the vector case, i.e., $R_{\text{vec}}(D)$. 
In particular, we expect that an analysis of the RD dimension \cite{KaDe94} of PPs can be based on 
  \eqref{eq:veclowerbound}  and the asymptotic tightness of the Shannon lower bound in the vector case \cite{ko16}.
On the other hand,   \eqref{eq:veclowerbound}  does not allow us to analyze the RD function for $k\to \infty$, as the resulting bounds quickly fall below zero.

Let us   combine  the bound   \eqref{eq:veclowerbound}  with the classical Shannon lower bound for a random vector $\rxv^{(k)}$ with probability density function $f_{\rxv^{(k)}}$ and squared-error distortion, which is given by \cite[eq.~(4.8.8)]{gr90}
\be \label{eq:shlbvec}
R_{\text{vec}}(D) 
\geq h\big(f_{\rxv^{(k)}}\big) -\frac{kd}{2} \bigg(1+ \log \bigg(\frac{2\pi D}{kd}\bigg)\bigg)\,.
\ee
In particular, 
for a  PP $\rX$ on $\R^d$ of fixed cardinality $\card{\rX}=k$ whose measure $P_{\rX}^{(k)}$ is absolutely continuous with respect to $(\Leb^{d})^{k}$ with Radon-Nikodym derivative $\frac{\intd P_{\rX}^{(k)}}{\intd (\Leb^{d})^{k}}=f_{\rX}^{(k)}$,
setting $\rxv^{(k)}= \rxv_{\rX}^{(k)}$, and combining \eqref{eq:shlbvec} with Theorem~\ref{th:lbspecific} gives 
\be 
R(D) 
\geq h\big(f_{\rX}^{(k)}\big) -\frac{kd}{2}\bigg(1 + \log \bigg(\frac{2\pi D}{kd}\bigg) \bigg)- \log k!\,.
\ee
The same result can also be  obtained by concretizing Corollary \ref{cor:rdlbfc} for the distortion function $\dist_2$.

\subsection{Upper Bound Based on the Rate-Distortion Theorem} \label{sec:fcupperboundsrd}
%
%

We can also concretize the upper bounds from Section~\ref{sec:upperbounds} for PPs of fixed cardinality.
  Corollary~\ref{cor:upperbounddpi}  becomes particularly simple.
\begin{corollary}\label{cor:dpupperbound}
Let $\rX$ be a PP on $\R^d$ of fixed cardinality $\card{\rX}=k$.
Denote by $\rxv_{\rX}^{(k)}$ the associated symmetric random vector on $(\R^{d})^{k}$.
Furthermore, let $(\rxv^{(k)}, \ryv^{(k)})$ be any random vector on $(\R^{d})^{2k}$ such that $\phi_k(\rxv^{(k)})$ has the same distribution as $\phi_k(\rxv_{\rX}^{(k)})$.
Finally, assume that
\be\label{eq:boundfcexp}
\E\big[\dist_2\big(\phi_k(\rxv^{(k)}),\phi_k(\ryv^{(k)})\big)\big]\leq D\,.
\ee
Then the RD function for distortion function $\dist_2$, at distortion $D$, is upper-bounded according to
\ba
R(D)
& \leq 
I\big(\rxv^{(k)}; \ryv^{(k)}\big)   
- I\big(\rt^{(k)}_{\rxv}; \phi_k(\ryv^{(k)}) \bcondi \phi_k(\rxv^{(k)})\big)
\notag \\* & \quad
- I\big(\rxv^{(k)}; \rt^{(k)}_{\ryv} \bcondi \phi_k(\ryv^{(k)})\big)\label{eq:dpupperbound}
\ea
  where $\rt^{(k)}_{\rxv}$ and $\rt^{(k)}_{\ryv}$ are the random permutations associated with the vectors in $\rxv^{(k)}$ and $\ryv^{(k)}$, respectively.
\end{corollary}
We can simplify  \eqref{eq:boundfcexp}  by using the following relation of $\dist_2$ to the squared-error distortion of vectors:
\ba
\dist_2\big(\phi_k(\xv_{1:k}),\phi_k(\yv_{1:k})\big)
& = \min_{\tau}\sum_{i=1}^k\norm{\xv_i-\yv_{\tau(i)}}^2
\notag \\[-1mm]
& \leq \sum_{i=1}^k\norm{\xv_i-\yv_{i}}^2 \notag \\
& = \norm{\xv_{1:k}-\yv_{1:k}}^2\,.\label{eq:boundospase}
\ea
Thus, 
$
\E\big[\norm{\rxv^{(k)}-\ryv^{(k)}}^2\big]\leq D
$
implies \eqref{eq:boundfcexp}.
This shows that the upper bound $I\big(\rxv^{(k)}; \ryv^{(k)}\big)$ on the RD function of a random vector $\rxv^{(k)}$ based on an arbitrary random vector $\ryv^{(k)}$ satisfying $\E\big[\norm{\rxv^{(k)}-\ryv^{(k)}}^2\big]\leq D$, is also an upper bound on the RD function of the corresponding fixed-cardinality PP $\rX=\phi_k(\rxv^{(k)})$.

\subsection{Codebook-Based Upper Bounds} \label{sec:fcupperboundscb}
%

We can also obtain 
upper bounds by constructing explicit source codes using the variation of the 
LBG algorithm proposed in Section~\ref{sec:cbub}.
We will specify the two iteration steps of that algorithm for point patterns of fixed cardinality $k$. 
In Step 1, for a given set $\sA\subseteq \sX_k$ of point patterns and $\codew$ center point patterns $X^*_j$, we have to associate each point pattern  $X\in \sA$ with the 
 center point pattern $X_j^*$ 
with minimal distortion $\dist_2(X,X^*_j)$.
This requires an evaluation of $\dist_2(X,X^*_j)$ for each $X\in \sA$ and each $j\in \{1, \dots, \codew\}$.
If several distortions $\dist_2(X,X^*_j)$ are minimal for a given $X$,  we 
choose the one with the smallest index $j$.
All point patterns $X\in \sA$ that are associated with the center point  pattern $X^*_j$ are collected in the set $\sA_j$.
In Step 2, for each subset $\sA_j\subseteq \sA$, we have to find an updated center point pattern $X^*_j$ of minimal average distortion from all point patterns in $\sA_j$, i.e.,
\be\label{eq:sumospafc}
	X^*_j=\argmin_{\widetilde{X}\in \sX_k} \frac{1}{ \card{\sA_j}}\sum_{X\in\sA_j}\dist_2(X,\widetilde{X})\,.
\ee
We can reformulate 
\eqref{eq:sumospafc} as the task of finding an ``optimal'' permutation
 (corresponding to an ordering) $\tau^*_{X}$  of each point pattern $X\in \sA_j$  according to the following lemma.
A proof is given in Appendix~\ref{app:eqospabcap}.
\begin{lemma}\label{lem:eqospabcap}
Let $\sA_j\subseteq \sX_k$ be a finite collection of point patterns in $\R^d$ of fixed cardinality $k\in \N$, i.e., for all $X\in \sA_j$, we have  $X=\big\{\xv^{(X)}_1, \dots, \xv^{(X)}_k\big\}$ with $\xv^{(X)}_i\in \R^d$.
Then a center point pattern%
\footnote{The center point pattern is not necessarily unique.}
 $X^*_j=\argmin_{\widetilde{X}\in \sX_k} \frac{1}{ \card{\sA_j}}\sum_{X\in\sA_j}\dist_2(X,\widetilde{X})$ is given as
\be\label{eq:ospacenterpp}
X^*_j=\{\xv^*_1, \dots, \xv^*_k\} \quad \text{ with }\xv^*_i=\frac{1}{\card{\sA_j}}\sum_{X\in \sA_j}\xv^{(X)}_{\tau^*_{X}(i)}
\ee
where  the collection of permutations $\{\tau^*_{X}\}^{}_{X\in \sA_j}$ is given by
\be\label{eq:ospacenterasap}
\{\tau^*_{X}\}^{}_{X\in \sA_j}=\argmin_{\{\tau_{X}\}^{}_{X\in \sA_j}} \sum_{i=1}^k\sum_{X\in\sA_j}\sum_{X'\in\sA_j}\bignorm{\xv^{(X)}_{\tau_{X}(i)}-\xv^{(X')}_{\tau_{X'}(i)}}^2 \,.
\ee
\end{lemma}

\begin{figure}[t]
\centering
\includegraphics{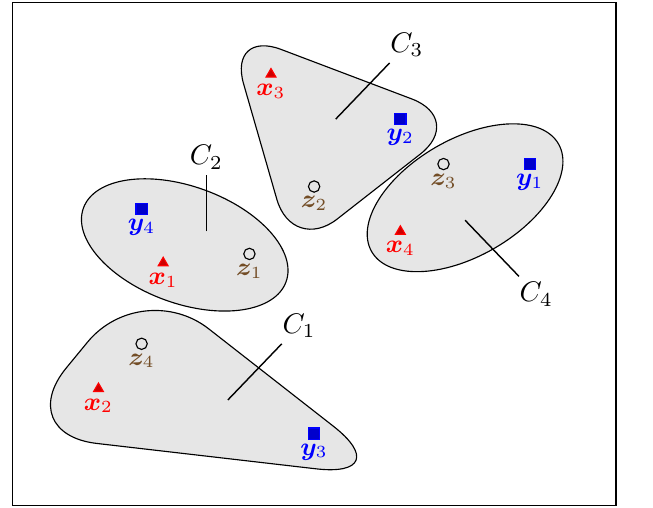}
\caption{Cliques $\cliq_i$ for $\sA_j=\{X,Y,Z\}$, $k=4$, and permutations $\{\tau_{X'}\}_{X'\in \sA_j}=\{\tau_{X}, \tau_{Y}, \tau_{Z}\}$ with 
$\tau_{X}(1)=2$,
$\tau_{X}(2)=1$,
$\tau_{X}(3)=3$,
$\tau_{X}(4)=4$;
$\tau_{Y}(1)=3$,
$\tau_{Y}(2)=4$,
$\tau_{Y}(3)=2$,
$\tau_{Y}(4)=1$; and 
$\tau_{Z}(1)=4$,
$\tau_{Z}(2)=1$,
$\tau_{Z}(3)=2$,
$\tau_{Z}(4)=3$.
Note that, e.g., $\tau_{Y}(4)=1$ expresses the fact that $\yv_1\in C_4$.
\vspace{-3mm}}
\label{fig:cliq}
\end{figure}

By Lemma~\ref{lem:eqospabcap}, the minimization problem in \eqref{eq:sumospafc} is equivalent to a \emph{multi-dimensional assignment problem} (MDAP).
Indeed, a collection of  permutations $\{\tau_{X}\}_{X\in \sA_j}$ corresponds to a choice of $k$ \emph{cliques}%
\footnote{
An 
MDAP
can also be formulated as a graph-theoretic problem where a \emph{clique} corresponds to a  complete subgraph
\cite{bacrsp94}.
}
\be \label{eq:defclique1}
\cliq_i\triangleq \big\{\xv^{(X)}_{\tau_{X}(i)}: X\in \sA_j\big\}, \qquad i=1, \dots, k\,.
\ee
Thus, for each point pattern $X\in \sA_j$, $\tau_{X}$ assigns each of the $k$ vectors in $X$ to one of $k$ different cliques, such that no two vectors in $X$ are assigned to the same clique.
Each resulting clique hence contains $\card{\cliq_i}=\card{\sA_j}$ vectors---one from each $X \in \sA_j$---and each vector $\xv \in X$ belongs to exactly one clique for all  $X \in \sA_j$.
Note that the union of all 
$X\in \sA_j$ is the same as the union of all 
cliques $\cliq_i$, i.e., $\bigcup_{X\in \sA_j}X=\bigcup_{i=1}^k \cliq_i$.
The relation between the cliques $\cliq_i$ and the point patterns $X\in \sA_j$ is illustrated in Figure~\ref{fig:cliq}.
We define the \emph{cost} of a clique $\cliq_i$ as  the sum 
\be\label{eq:costclique}
\sum_{\xv\in\cliq_i}\sum_{\xv'\in\cliq_i} \norm{\xv-\xv'}^2 
 = \sum_{X\in\sA_j}\sum_{X'\in\sA_j} \bignorm{\xv^{(X)}_{\tau_{X}(i)}-\xv^{(X')}_{\tau_{X'}(i)}}^2 
\,.
\ee
Finding the collection of cliques $\{\cliq^*_i\}_{i=1, \dots, k}$ with minimal sum cost, i.e.,
\be\label{eq:optcliques}
\{\cliq^*_i\}_{i=1, \dots, k} =
\argmin_{\{\cliq_i\}_{i=1, \dots, k}}\sum_{i=1}^k\sum_{\xv\in\cliq_i}\sum_{\xv'\in\cliq_i}\norm{\xv-\xv'}^2 
\ee
 is then equivalent to finding the optimal collection of permutations in \eqref{eq:ospacenterasap}.
Moreover, according to its definition in \eqref{eq:costclique}, the cost of a clique 
can be decomposed into a sum of squared distances, each between two of its members.
Thus, the minimization in \eqref{eq:optcliques} is an MDAP 
 with decomposable costs.

Although finding an exact solution to such an MDAP  is 
unfeasible for large $k$ and large clique sizes $\card{\cliq_i}=\card{\sA_j}$, there exist various heuristic algorithms producing approximate solutions \cite{bacrsp94,bamasp04, kuma09}. 
Because we are mainly interested in the case of a large clique size, we will merely use a variation of the basic single-hub heuristic and the multi-hub heuristic presented in \cite{bacrsp94}.
Since these heuristic algorithms are used in Step 2 of the proposed 
LBG-type algorithm, we will label the corresponding steps as 2.1--2.3.
The classical single-hub heuristic is based on assigning the vectors $\xv^{(X)}_i$ in each point pattern $X\in \sA_j$ to the $k$ different cliques (using a permutation $\tau_X$) by minimizing the sum of squared distances between the vectors $\xv^{(X)}_{\tau(i)}$ and the vectors $\xv^{(X_1)}_{i}$ of one ``hub'' point pattern $X_1$.
More specifically, the algorithm (corresponding to Step 2 of the 
LBG algorithm) is given as follows.
\begin{itemize}
	\item Input: a collection $\sA_j$ of point patterns; each point pattern $X\in \sA_j$ contains  $k$ points in $\R^d$.
	\item Initialization: Choose a point pattern $X_1\in \sA_j$ (called the \emph{hub}) and define $\tau_{X_1}(i)=i$ for $i=1, \dots, k$.
	\item Step 2.1: For each $X=\big\{\xv^{(X)}_1, \dots, \xv^{(X)}_k\big\}\in \sA_j\setminus \{X_1\}$, find the best 
	assignment between the vectors in $X$ and $X_1$, i.e., a 
permutation\vspace{-2mm}
	\be \label{eq:optassign}
	\tau_X= \argmin_{\tau}\sum_{i=1}^k\bignorm{\xv^{(X)}_{\tau(i)}-\xv^{(X_1)}_{i}}^2\,.
	\ee
	\item Step 2.2: Define the clique $\cliq_i$ according to \eqref{eq:defclique1}, i.e., 
	\be \label{eq:defclique}
	\cliq_i\triangleq \big\{\xv^{(X)}_{\tau_{X}(i)}:X\in \sA_j\big\}, \qquad i=1, \dots, k\,.
	\ee
	\item Step 2.3: Define the (approximate) center point pattern $\hat X^*_j$ as the union of the centers (arithmetic means) of all cliques, 
i.e., 
\be \label{eq:defcenter}
\hat X^*_j\triangleq \bigcup_{i=1}^k \big\{\bar{\xv}_{\cliq_i}\big\} \qquad \text{with } \bar{\xv}_{\cliq_i}\triangleq \frac{1}{\card{\cliq_i}}\sum_{\xv\in \cliq_i}\xv\,.
\ee
	\item Output: approximate center point pattern $\hat X^*_j$.
\end{itemize}

The above heuristic requires only $\card{\sA_j}-1$ optimal assignments between point patterns. 
However, the 	accuracy of the resulting approximate center point pattern $\hat X^*_j$ strongly depends on the choice of the hub $X_1$ and can be very poor for certain choices of $X_1$. 
The more robust multi-hub heuristic \cite{bacrsp94} performs the single-hub algorithm with all the $X\in \sA_j$ as alternative hubs $X_1$, which can be shown to  increase the complexity to $\frac{(\card{\sA_j}-1)\card{\sA_j}}{2}$  optimal assignments, i.e.,  by a factor of $\card{\sA_j}/2$. 
We here propose a different heuristic that has almost the same complexity as the single-hub heuristic but is more robust.
The idea of our approach is to replace the best assignment to the single hub $X_1$ by an optimal assignment to the approximate center point pattern of the subsets defined in the preceding steps.
More specifically, 
we start with a hub $X_1\in \sA_j$ and, as in the single-hub heuristic, search for the best assignment $\tau_{X_2}$ (see \eqref{eq:optassign} with $X=X_2$) between the vectors in  a randomly chosen $X_2\in \sA_j\setminus \{X_1\}$ and $X_1$.
Then, we calculate the center point pattern $\hat{X}_2$ of $X_1$ and $X_2$ as in \eqref{eq:defclique} and \eqref{eq:defcenter} but with $\sA_j$ replaced by $\{X_1, X_2\}$.
In the next step, we choose a random $X_3\in  \sA_j\setminus \{X_1, X_2\}$ and find the optimal assignment $\tau_{X_3}$ between the vectors in  $X_3$ and $\hat{X}_2$.
An approximate center point pattern $\hat{X}_3$  of  $X_1$, $X_2$, and $X_3$ is then   calculated  as in \eqref{eq:defclique} and \eqref{eq:defcenter} but with $\sA_j$ replaced by $\{X_1, X_2, X_3\}$.
Equivalently, we can calculate $\hat{X}_3$ as a ``weighted'' center point pattern of $X_3$ and $\hat{X}_2$.
We proceed in this way  with all the point patterns in $\sA_j$, always calculating the optimal assignment $\tau_{X_r}$ ($r=4, 5, \dots$) between the vectors in  $X_{r}$ and the approximate center point pattern $\hat{X}_{r-1}$ of the previous $r-1$ point patterns.
A formal statement of the algorithm is  as follows.
\begin{itemize}
	\item Input: a collection $\sA_j$ of point patterns; each point pattern $X\in \sA_j$ contains  $k$ points in $\R^d$.
	\item Initialization: (Randomly) order the point patterns $X\in \sA_j$, i.e., choose a sequence $(X_1, \dots, X_{\card{\sA_j}})$ where the $X_r$ are all the elements of $\sA_j$. Set the initial subset center point pattern $\hat{X}_1$ to $X_1$.
	\item For  $r=2, \dots, \card{\sA_j} $:
	\begin{itemize}
		\item Step 2.1: Find the best assignment between the vectors in $X_r$ and $\hat{X}_{r-1}$, i.e., a permutation 
$\tau_{X_r}= \argmin_{\tau}\sum_{i=1}^k\bignorm{\xv^{(X_r)}_{\tau(i)}-\xv^{(\hat{X}_{r-1})}_{i}}^2$.
		\item Step 2.2: Generate an updated (approximate) subset center point pattern $\hat{X}_{r}= \big\{\xv^{(\hat{X}_{r})}_{1}, \dots, \xv^{(\hat{X}_{r})}_{k}\big\}$ according 
		\vspace{-2.5mm}to 
			\be
				\xv^{(\hat{X}_{r})}_{i} 
				 = 
				\frac{1}{r} \sum_{s=1}^r \xv^{(X_s)}_{\tau_{X_s}(i)}
				 =
				\frac{(r-1)\xv^{(\hat{X}_{r-1})}_{i}+\xv^{(X_r)}_{\tau_{X_r}(i)}}{r}
				\label{eq:centerupdate}
			\ee
			for $ i=1, \dots, k$.
	\end{itemize}
	\item Output: approximate center point pattern $\hat X^*_j=\hat{X}_{\card{\sA_j}}$.
\end{itemize}
As in the case of the single-hub heuristic, we only have to perform $\card{\sA_j}-1$ optimal assignments. 
The complexity of the additional center update \eqref{eq:centerupdate} is negligible.
On the other hand, the multi-hub algorithm can be easily parallelized whereas our algorithm works only sequentially.

\subsection{Example: Gaussian Distribution}\label{sec:gaussdist}

As an example, we consider the case of a PP $\rX$ on $\R^{  d}$ of fixed cardinality $\card{\rX}=k$ whose points  are independently distributed according to a standard Gaussian distribution on~$\R^{d}$, i.e., $\rxv_{\rX}^{(k)}\in (\R^{  d})^k$ has i.i.d.\ zero-mean Gaussian entries with variance $1$.
We want to compare the 
lower bound \eqref{eq:veclowerbound} to the 
upper bounds presented in Sections~\ref{sec:fcupperboundsrd} and \ref{sec:fcupperboundscb}   and  to the RD function for the vector setting, i.e., to the RD function of a standard Gaussian vector $\rxv^{(k)}$ in $(\R^{  d})^k$.
In the PP setting, we use the distortion $\dist_2$ (see \eqref{eq:ospafc}), while in the vector setting, we use the conventional squared-error distortion.
The RD function for $\rxv^{(k)}$  in the vector setting can be calculated in closed form;
assuming $D\leq k{  d}$, it is equal to 
\be
R_{\text{vec}}(D) 
= \frac{k{  d}}{{  2}}\log \bigg(\frac{k{  d}}{D}\bigg)\,. \label{eq:rdvecgauss}
\ee
This result was shown (see~\cite[Sec.~10.3.2]{Cover91}) by using the RD theorem for the vector case, i.e., \eqref{eq:rdasinf} with obvious modifications, and choosing  $\rxv^{(k)}=\ryv^{(k)}+\rwv^{(k)}$, where $\rwv^{(k)}$ has i.i.d.\ zero-mean Gaussian entries with variance $D/(k{  d})$, $\ryv^{(k)}$   has i.i.d.\ zero-mean Gaussian entries with variance $1-D/(k{  d})$,   and $\ryv^{(k)}$ and $\rwv^{(k)}$ are independent.
This choice can be shown to achieve the infimum in the RD theorem and hence the mutual information $I(\rxv^{(k)}; \ryv^{(k)})$ is equal to the RD function.

In the PP setting, inserting \eqref{eq:rdvecgauss} into \eqref{eq:veclowerbound} results in the lower bound
\be
R(D) 
\geq  \frac{k d}{ 2}\log \bigg(\frac{kd}{D}\bigg) - \log k!\,. \label{eq:rdlowerboundgauss}
\ee

For the calculation of the upper bound \eqref{eq:dpupperbound}, we use a similar approach as in the vector case. 
Let $\rX$ be a PP of fixed cardinality $\card{\rX}=k$, where $\rxv_{\rX}^{(k)}$ has i.i.d.\ zero-mean Gaussian entries with variance $1$.
We choose $\rxv^{(k)}=\ryv^{(k)}+\rwv^{(k)}$, where $\rwv^{(k)}$ has i.i.d.\ zero-mean Gaussian entries with variance $\sigma^2 <1$, $\ryv^{(k)}$ has i.i.d.\ zero-mean Gaussian entries with variance $1-\sigma^2$,   and $\ryv^{(k)}$ and $\rwv^{(k)}$ are independent.
The random vectors $\rxv_{\rX}^{(k)}$ and $\rxv^{(k)}$ have the same distribution, and the
  first term on the 
 right-hand side in \eqref{eq:dpupperbound} is here obtained 
 as
\be\label{eq:ubscgaussfc1a}
I\big(\rxv^{(k)}; \ryv^{(k)}\big) = \frac{ k{  d}}{{  2}} \log \bigg(\frac{1}{\sigma^2}\bigg)\,.
\ee
The second term on the right-hand side in \eqref{eq:dpupperbound} can  be dropped, which in general results in a looser upper bound. 
However, in our example, this term can be shown to be zero and thus dropping it does not  loosen the bound.
The third term can be rewritten as 
\ba
- & I\big(\rxv^{(k)}; \rt^{(k)}_{\ryv} \bcondi \phi_k(\ryv^{(k)})\big) \notag \\
& = - H\big( \rt^{(k)}_{\ryv} \bcondi \phi_k(\ryv^{(k)})\big) 
+ H\big( \rt^{(k)}_{\ryv} \bcondi \phi_k(\ryv^{(k)}), \rxv^{(k)}\big)\,.\label{eq:ubscgaussfc1b}
\ea
Because all the elements $\ryv^{(k)}_i$ of $\ryv^{(k)}$ are i.i.d.\ and thus symmetric, the associated random permutation $\rt^{(k)}_{\ryv}$ is uniformly distributed.
Furthermore, this symmetry implies that $\rt^{(k)}_{\ryv}$ is independent of the  values of the elements of $\phi_k(\ryv^{(k)})$.
Thus,
\be\label{eq:htcondphiy}
H\big( \rt^{(k)}_{\ryv} \bcondi \phi_k(\ryv^{(k)})\big)= H\big( \rt^{(k)}_{\ryv}  \big)= \log k!\,.
\ee
Furthermore, 
 the entropy $H\big( \rt^{(k)}_{\ryv} \bcondi \phi_k(\ryv^{(k)}), \rxv^{(k)}\big)$ is shown in Appendix~\ref{app:boundent} to be bounded for any $\varepsilon >0$ according to 
\ba
H& \big( \rt^{(k)}_{\ryv} \bcondi \phi_k(\ryv^{(k)}), \rxv^{(k)}\big) 
\notag \\*
& \leq \bigg(\frac{k(k-1)}{2}\, F_{\chi^2}\bigg(\frac{9\varepsilon^2}{2(1-\sigma^2)}; d\bigg)
+ 1-F_{\chi^2}\bigg( \frac{\varepsilon^2}{ \sigma^2 }; kd\bigg)\bigg) 
\notag \\*
& \quad  \times
\log k!
+H_2(p_0(\varepsilon) )
+ (1-p_0(\varepsilon))\log (k!-1)
\label{eq:boundhtcondphiyx}  
\ea
where 
$F_{\chi^2}(\, \cdot\, ; d)$ denotes the  cumulative distribution function of a $\chi^2$ distribution with $d$ degrees of freedom, 
$H_2(\cdot)$ is the binary entropy function,
and 
$p_0(\varepsilon)= 1/\big( 1 + (k!-1)  \exp\big({-}\frac{3\varepsilon^2}{2\sigma^{2}}\big)\big)$.
Inserting \eqref{eq:htcondphiy} and  \eqref{eq:boundhtcondphiyx} into \eqref{eq:ubscgaussfc1b} and, in turn, inserting \eqref{eq:ubscgaussfc1a} and  \eqref{eq:ubscgaussfc1b} into \eqref{eq:dpupperbound}, we obtain
\ba
R(D) 
& \leq 
  \frac{kd}{2} \log \bigg(\frac{1}{\sigma^2}\bigg)
-  \log k! + \bigg(\frac{k(k-1)}{2}\notag \\* & \quad
\times F_{\chi^2}\bigg(\frac{9\varepsilon^2}{2(1-\sigma^2)}; d\bigg)
+ 1-F_{\chi^2}\bigg( \frac{\varepsilon^2}{ \sigma^2 }; kd\bigg)\bigg) 
  \log k!
	\notag \\*
& \quad
+H_2
(p_0(\varepsilon))
+ 
(1- p_0(\varepsilon))\log (k!-1)
\label{eq:ubgaussiancase}
\ea
provided that \eqref{eq:boundfcexp} is satisfied.
Due to \eqref{eq:boundospase}, this is the case if  
$\E\big[\norm{\rxv^{(k)}-\ryv^{(k)}}^2\big]\leq D$.
Because $\E\big[\norm{\rxv^{(k)}-\ryv^{(k)}}^2\big]= \E\big[\norm{\rwv^{(k)}}^2\big]=dk\sigma^2$, we thus choose
 $\sigma^2=D/(kd)$.

By choosing $\varepsilon$ appropriately, we can show that the upper bound \eqref{eq:ubgaussiancase} converges to the lower bound \eqref{eq:rdlowerboundgauss} as $D \to 0$, i.e., that the lower bound \eqref{eq:rdlowerboundgauss} is asymptotically tight.
Indeed, choosing $\varepsilon>0$ such that $\varepsilon\to 0$ and $\varepsilon/\sigma \to \infty$ as $D \to 0$, we obtain 
$F_{\chi^2}\big(\frac{9\varepsilon^2}{2(1-\sigma^2)}; d\big)\to 0$, 
$F_{\chi^2}\big( \frac{\varepsilon^2}{ \sigma^2 }; kd\big)\to 1$, 
and 
$p_0(\varepsilon)\to 1$.
Thus, \eqref{eq:ubgaussiancase} gives 
\be
R(D) 
\leq 
 \frac{kd}{2} \log \bigg(\frac{kd}{D}\bigg)
-  \log k! + o(1)
\ee
where $o(1)$ is a function that converges to zero as $D\to 0$.
\begin{figure*}[t]
\centering
\includegraphics{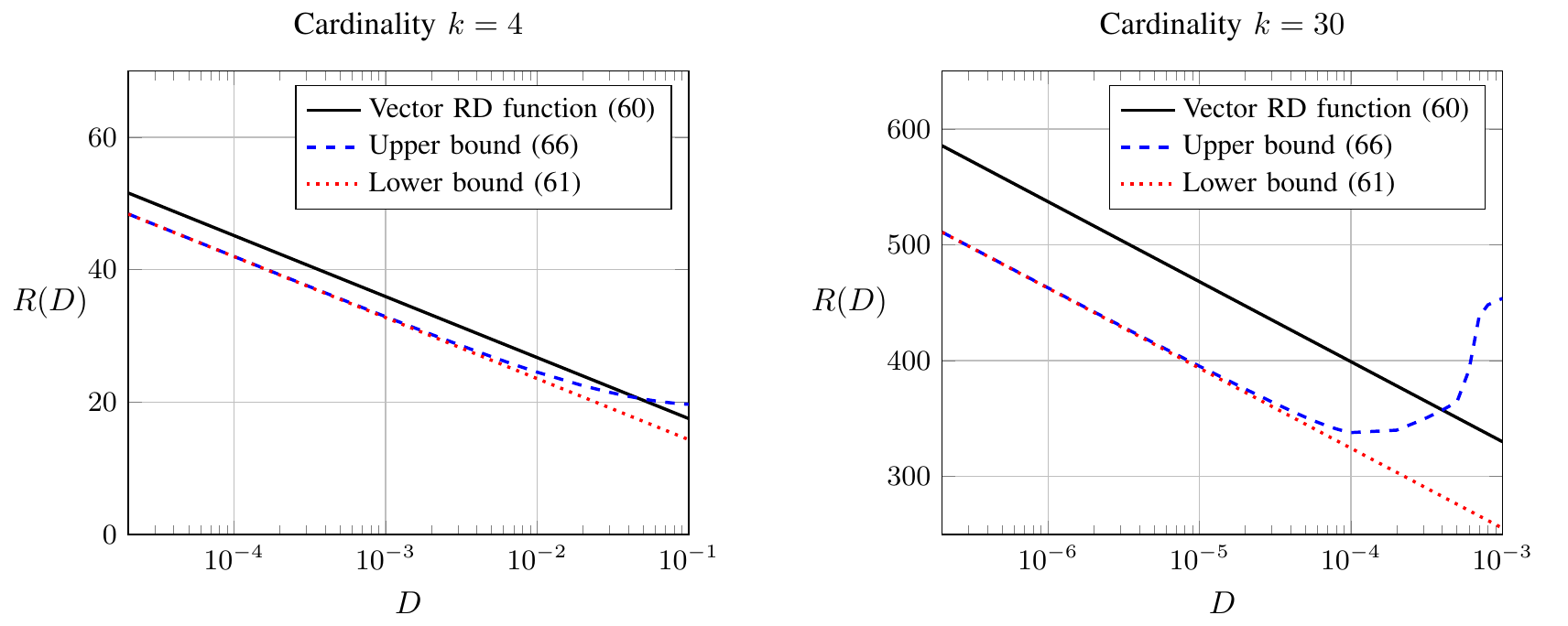}
\vspace{-3mm}
\caption{Lower bound on the RD function according to \eqref{eq:rdlowerboundgauss}  and upper bound  on the RD function according to \eqref{eq:ubgaussiancase} with $\varepsilon = \sigma^{3/4} =(D/(kd))^{3/8}$ for a PP of fixed cardinality $k$ (left: $k=4$; right: $k=30$) and with $\rxv_{\rX}^{(k)}$ following a multivariate standard normal distribution.
For comparison, also the corresponding vector RD function  \eqref{eq:rdvecgauss} is shown. 
}
\label{fig:gaussdpubtight}
\end{figure*} 
In Figure~\ref{fig:gaussdpubtight}, we show  the   upper bound  \eqref{eq:ubgaussiancase} for $d=2$, cardinalities $k=4$ and $k=30$, and $\varepsilon = \sigma^{3/4} =(D/(kd))^{3/8}$  in comparison to the lower bound \eqref{eq:rdlowerboundgauss} and the vector RD function \eqref{eq:rdvecgauss}.
We see that as $D \to 0$, our upper and lower bounds are tight.
However, as the upper bound was designed for small values of $D$, it is not useful for larger values of $D$.

\begin{figure*}[t]
\centering
\includegraphics{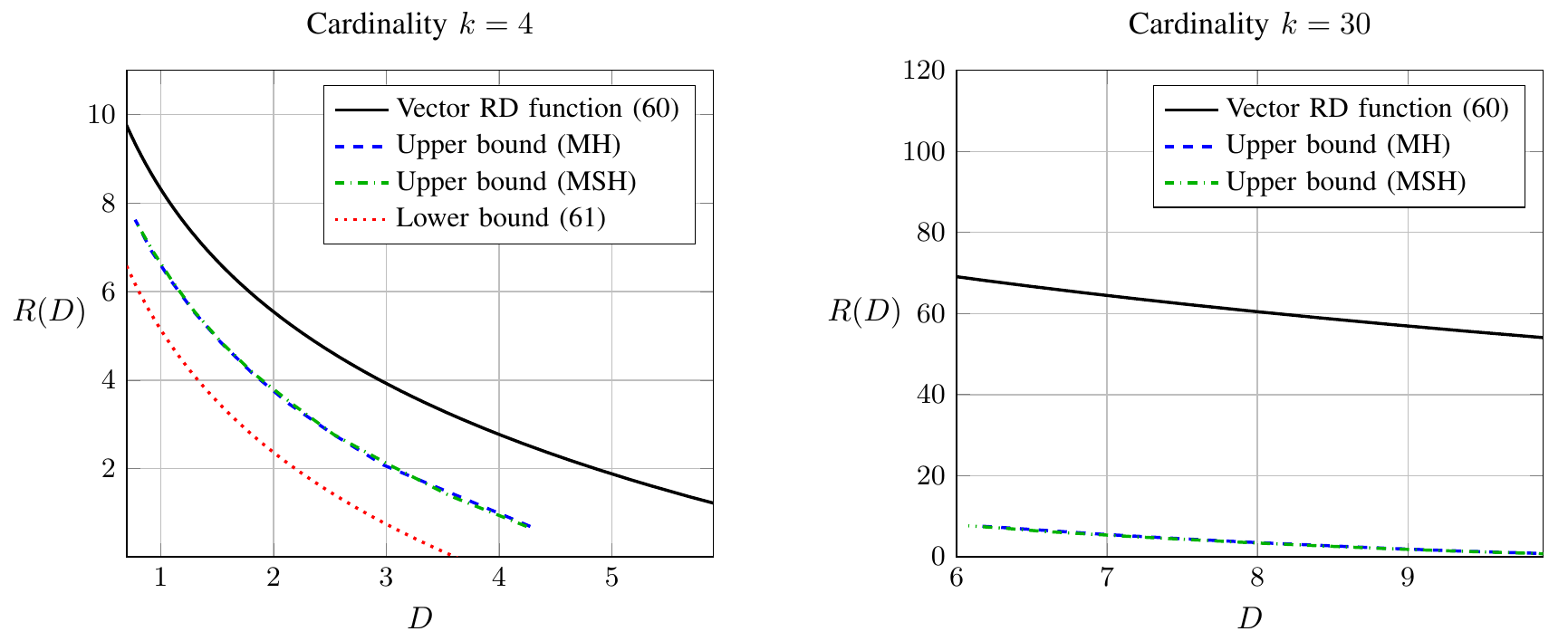}
\vspace{-2mm}
\caption{  Numerical codebook-based upper bounds on the RD function of a PP of fixed cardinality $k$ (left: $k=4$; right: $k=30$) and with $\rxv_{\rX}^{(k)}$ following a multivariate standard normal distribution.
The codebook construction underlying the upper bounds used the multi-hub heuristic (MH) or the proposed modified single-hub heuristic (MSH). 
For comparison, also the corresponding vector RD function  \eqref{eq:rdvecgauss} and 
the lower bound \eqref{eq:rdlowerboundgauss} are shown. (The latter is shown only for $k=4$, because  for $k=30$ it is below $0$ in the regime where code construction is feasible.)}
\label{fig:gaussdpubjoint}
\end{figure*}

We also considered codebook-based upper bounds following \eqref{eq:scupperbound}.
Using the LBG-type algorithm presented in Sections~\ref{sec:cbub} and \ref{sec:fcupperboundscb}, we constructed codebooks for fixed-cardinality PPs with $k=4$ and $k=30$ i.i.d.\ Gaussian points. 
As input to the  LBG algorithm, we used $\card{\sA}=100\cdot \codew$ random realizations of the source PP.
In Step 2 of the  algorithm, we employed the multi-hub heuristic as well as the modified single-hub heuristic proposed in Section~\ref{sec:fcupperboundscb}.
The expected distortion $\widetilde{D} = \E[\dist_2(\rX,g(\rX))]$ for each constructed source code $g$ was calculated using Monte Carlo integration \cite[Ch.~3]{roca04}.
In Figure~\ref{fig:gaussdpubjoint}, we show the resulting upper bounds %
on the RD function based on codebooks of up to $\codew=2048$ codewords  in comparison to  the lower bound \eqref{eq:rdlowerboundgauss} and the vector RD function \eqref{eq:rdvecgauss}.
Unfortunately, for larger values of $\codew$, Step 1 in the LBG algorithm becomes computationally unfeasible. 
It can be seen that the PP setting can significantly reduce the required rates compared to the vector setting   also for large values of $D$.
Furthermore, using the significantly less computationally demanding modified single-hub heuristic does not result in increased upper bounds compared to the multi-hub heuristic.

\section{Poisson Point Processes}\label{sec:ppp}

Poisson PPs, the most prominent and widely used class of PPs, are characterized by a complete randomness property.
A PP $\rX$ on $\R^d$ is a Poisson PP if the number of points in each Borel set $A\subseteq \R^d$ is Poisson distributed with parameter $\lambda(A)$---where the measure $\lambda$ is referred to as the \emph{intensity measure} of $\rX$---and for any disjoint Borel sets $A, B\subseteq \R^d$, the number of points in $A$ is independent of the number of points in $B$.
For a formal definition see  \cite[Sec.~2.4]{dave03}. 
We will assume that $\rX$ is a  finite Poisson PP, which satisfies $\lambda(\R^d)<\infty$.
In this setting, we  easily obtain the cardinality distribution $p_{\card{\rX}}$ and the measures $P_{\rX}^{(k)}$.
Let us express the intensity measure as $\lambda=\expcard \lambda_0$, where $\expcard\triangleq\lambda(\R^d)\in \R_{\geq 0}$ 
and $\lambda_0\triangleq\lambda/\expcard$ is a probability measure.
We then obtain
\be\label{eq:pppcard}
p_{\card{\rX}}(k)= \frac{e^{-\expcard}\expcard^k}{k!} \qquad \text{ for } k\in \N_0
\ee
and moreover it can be shown that (see~\cite[Sec.~5.3]{dave03})
\be\label{eq:pppprobmeasure}
P_{\rX}^{(k)}=\lambda_0^k \qquad \text{ for } k\in \N\,.
\ee
Note that \eqref{eq:pppprobmeasure} implies that for a given cardinality $\card{\rX}=k$, the vectors $\rxv_i\in \rX$ are i.i.d.\ with probability measure $\lambda_0$.

In the following, we will often consider Poisson PPs with intensity measure $\lambda=\expcard \lambda_0$, where $\lambda_0$ is absolutely continuous with respect to $\Leb^d$ with Radon-Nikodym derivative $g_{\rX}=\frac{\intd \lambda_0}{\intd \Leb^d}$. 
According to \eqref{eq:pppprobmeasure}, this implies that the probability measures $P_{\rX}^{(k)}$ are absolutely continuous with respect to $(\Leb^d)^k$ with Radon-Nikodym derivative
\be\label{eq:denspppk}
\frac{\intd P_{\rX}^{(k)}}{\intd (\Leb^d)^k}(\xv_{1:k})=f^{(k)}_{\rX}(\xv_{1:k})=\prod_{i=1}^k g_{\rX}(\xv_i)
\ee
i.e., the $\rxv_{\rX}^{(k)}$ are continuous random vectors with probability density function $\prod_{i=1}^k g_{\rX}(\xv_i)$.

\subsection{Distortion Function}

For a Poisson PP $\rX$, there is a nonzero probability that  $\card{\rX}=k$ for each $k\in \N_0$.
Thus, for an RD analysis, we have to define a distortion function between point patterns of different cardinalities.
We choose  the squared OSPA distance \cite{scvovo08} (up to a normalization factor%
\footnote{
In \cite{scvovo08}, the OSPA is normalized by the maximal number of points in either pattern. 
This normalization is unfavorable for our RD analysis as it would cause the distortion, and in turn the RD function, to  converge to zero for large patterns. 
}).
For $X=\{\xv_1, \dots, \xv_{\kX }\}$ and $Y=\{\yv_1, \dots, \yv_{\kY }\}$ with $\kX \leq \kY $, we define the unnormalized squared OSPA (USOSPA) distortion
\ba
\dist_2^{  (c) }(X,Y)\triangleq (\kY -\kX )\,c^2 + \min_{\tau}\sum_{i=1}^{\kX }\min\big\{\norm{\xv_i-\yv_{\tau(i)}}^2, c^2\big\}
\notag \\*[-3mm]
\label{eq:ospagen}
\\*[-7mm] \notag 
\ea
 where $c>0$ is a parameter (the \emph{cut-off value}) and the outer  minimum is taken over all permutations $\tau$ on $\{1, \dots, \kY \}$.
For $\kX  > \kY $, we define $\dist_2^{  (c) }(X,Y)\triangleq\dist_2^{  (c) }(Y,X)$.
According to \eqref{eq:ospagen}, the USOSPA distortion is constructed by first penalizing the difference in cardinalities via the term $\card{\kY -\kX }\,c^2$.
Then   an optimal assignment between the points of $X$ and $Y$ is established  based on the Euclidean distance, and   the minima of the squared distances and $c^2$ are summed.
To bound the RD function, we will use the following bounds on the USOSPA distortion, which are proved in Appendix~\ref{app:proofusospabounds}.
\begin{lemma}\label{lem:usospabounds}
Let $X=\{\xv_1, \dots, \xv_{\kX }\}\in \sX$ and  $Y=\{\yv_1, \dots, \yv_{\kY }\}\in \sX$. 
Then for $ \kX  \geq \kY $
\be
\dist_2^{  (c) }(X,Y)\geq 
\sum_{i=1}^{\kX }\min_{j=1}^{\kY }\min\{\norm{\xv_{i}-\yv_j}^2, c^2\} \label{eq:bounddist2a}
\ee
and for $ \kX  \leq \kY $
\be
\dist_2^{  (c) }(X,Y)\geq 
 (\kY -\kX ) c^2 + \sum_{i=1}^{\kX }\min_{j=1}^{\kY }\min\{\norm{\xv_{i}-\yv_j}^2, c^2\}\,. \label{eq:bounddist2b}
\ee
\end{lemma}

\subsection{Lower Bounds  for Poisson Point Processes}

Based on Theorem~\ref{th:shlbpp} and Lemma~\ref{lem:usospabounds}, we can formulate lower bounds on the RD function of Poisson PPs.
A proof of the following result is given in Appendix~\ref{app:proofppplowerbound}.
\begin{theorem}\label{th:ppplowerbound}
Let $\rX$ be a Poisson PP on $\R^d$ with intensity measure $\lambda=\expcard \lambda_0$, where $\lambda_0$ is absolutely continuous with respect to $\Leb^d$ with probability density function $g_{\rX}=\frac{\intd \lambda_0}{\intd \Leb^d}$.
Furthermore, let $A$ be a Borel set satisfying $\int_{A}g_{\rX}(\xv)\,\intd\xv=1$, i.e., $g_{\rX}(\xv)=0$ for $\Leb^{d}$-almost all $\xv\in A^c$.
Then the RD function of $\rX$ using distortion $\dist_2^{  (c) }$ is lower-bounded as
\ba
R(D)
& \geq
\expcard \,h(g_{\rX}) + \max_{s\geq 1/c^2} \bigg({-} \sum_{k\in \N} \frac{e^{-\expcard}\expcard^k}{k!} 
\notag \\*
& \quad \times
\log \big(\min\big\{(\Leb^{d}(A))^k,\gammat_k(s)\big\}\big)-sD \bigg)\label{eq:rdlbppp}
\ea
where
\ba
\gammat_k(s)
& \triangleq  \Bigg(e^{-s c^2} \Leb^{d}(A) 
\notag \\*[-1mm] & \quad
+k\bigg({-}e^{-s c^2}\Leb^d(U_c) + \int_{U_c} e^{-s  \norm{\xv}^2}  \mathrm{d}\xv \bigg) \Bigg)^k \label{eq:mygammappp}
\ea
with $U_c\triangleq \{\xv\in \R^d: \norm{\xv}\leq c\}$.
\end{theorem}
Note that the PP $\rX$ enters the bound \eqref{eq:rdlbppp} only via $\expcard$ and the differential entropy $h(g_{\rX})$. 
In particular, the functions $\gammat_k$ in  \eqref{eq:mygammappp} do not depend on  $\rX$. However, they do depend on the set $A$.

\begin{figure}[t]
\centering
\includegraphics{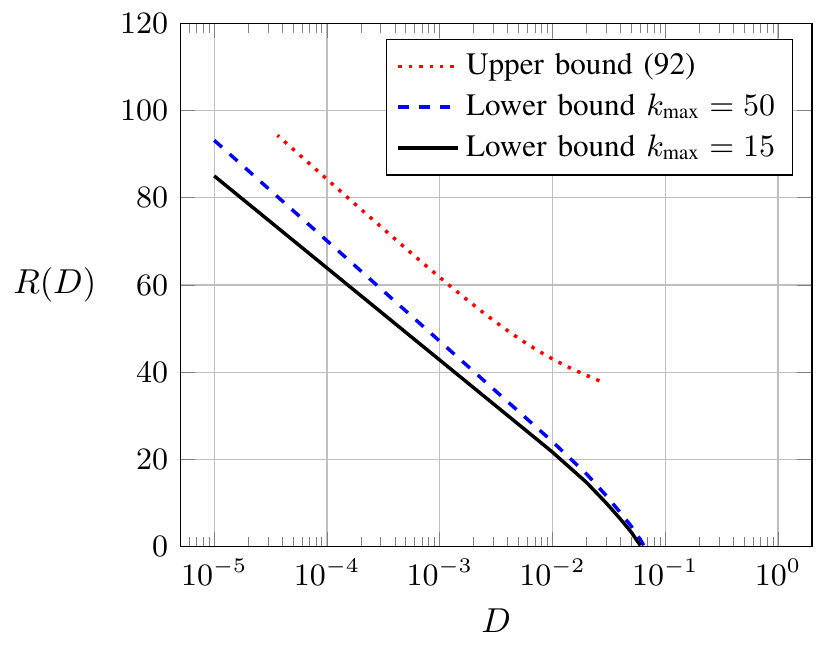}
\caption{Lower and upper bounds on the RD function for a Poisson PP with uniform intensity measure $\lambda=10 \cdot\Leb^2|_{[0,1)^2}$ on $[0,1)^2$, using the USOSPA distortion with cut-off value $c=0.1$.}
\label{fig:shlb10}
\end{figure}

\begin{example} \label{ex:pppsquare}
Let $\rX$ be a Poisson PP on $\R^2$ with intensity measure  $\lambda=\expcard\Leb^2|_{[0,1)^2}$, i.e., the points are independently and uniformly distributed on the unit square.
In this setting, we have $g_{\rX}=\ind_{[0,1)^2}$ and we can choose $A=[0,1)^2$ in Theorem~\ref{th:ppplowerbound}.
The differential entropy $h(g_{\rX})$ is zero, because the density $g_{\rX}=\ind_{[0,1)^2}$ takes on the values zero or one.
Furthermore, we have $d=2$, and thus, 
using $\Leb^2(A)=1$, $\Leb^2(U_c)= \pi c^2$, and $\int_{U_c} e^{-s  \norm{\xv}^2}  \mathrm{d}\xv=\frac{\pi}{s}\big( 1-e^{-s c^2} \big)$,  \eqref{eq:mygammappp} reduces to
\be
\gammat_k(s)
=  \Bigg(e^{-s c^2}\bigg(1-\pi c^2k-\frac{\pi k}{s}\bigg)+\frac{\pi k}{s}\Bigg)^k . \notag 
\ee
We further obtain 
\ba 
& \log\big(\min\big\{(\Leb^2(A))^k, \gammat_k(s)\big\}\big)
\notag \\
& \quad= \min\big\{k \log \Leb^2(A), \log \gammat_k(s)\big\} 
\notag \\
& \quad= \min\{0, \log \gammat_k(s)\} \notag 
\ea
i.e., we can upper-bound $\log\big(\min\big\{(\Leb^2(A))^k, \gammat_k(s)\big\}\big)$ either by zero (which corresponds to omitting the $k$th summand in \eqref{eq:rdlbppp}) or by $ \log \gammat_k(s)$.
In particular, we can omit all but the first $k_{\text{max}}\in \N_0$ summands in the lower bound \eqref{eq:rdlbppp} and, in  the remaining  summands, bound the factors $\log\big(\min\big\{(\Leb^2(A))^k, \gammat_k(s)\big\}\big)$ by  $ \log \gammat_k(s)$. 
We then obtain 
\ba
R(D)
& \geq
\max_{s\geq 1/c^2} \bigg({-} \sum_{k=1}^{k_{\text{max}}} \frac{e^{-\expcard}\expcard^k}{k!}\log \gammat_k(s) -sD \bigg) \notag \\
& = \max_{s\geq  1/c^2} \bigg({-} \sum_{k=1}^{k_{\text{max}}} \frac{e^{-\expcard}\expcard^k}{(k-1)!}
\notag \\* & \quad \times 
\log \bigg(e^{-s c^2}\bigg(1-\pi c^2k-\frac{\pi k}{s}\bigg)+\frac{\pi k}{s}\bigg) -sD \bigg) 
\label{eq:shlbunif}
\\[-9mm] \notag 
\ea
where  we used $h(g_{\rX})=0$.

Let us next investigate the convexity properties of the right-hand side in \eqref{eq:shlbunif}.
The second derivative of $\log \gammat_k(s)$ is obtained as
\ba
&  (\log \gammat_k(s))'' 
\notag \\
& \quad=
\!\Big(
\pi k^2  \big((1-\pi c^2k)
({c}^{4}{s}^{3}
- {c}^{2} {s}^{2}
-2  s{ e^{-s {c}^{2}}}
+2  s)
\notag \\* & \quad\quad 
- {c}^{2}{s}^{2}  
+e^{-s{c}^{2}}\pi k(1-e^{sc^2})^2\big)
\Big)
\notag \\* & \quad\quad \times 
{s}^{-2} e^{s{c}^{2}}
 \big( \pi  {c}^{2}ks+ \pi k- \pi k e^{s{c}^{2}}-s \big)^{-2}
\,.
\ea
It can  be shown that $(\log \gammat_k(s))''>0$ if $s\geq 3/c^2$ and $k\leq 1/(2\pi c^2)$.
Hence, $\log \gammat_k$ is convex in that case.
In particular, 
choosing $k_{\text{max}}\leq 1/(2\pi c^2)$, we have  that $\log \gammat_k$ is a convex function for $k\leq k_{\text{max}}$ and  $s\geq 3/c^2$, and thus the sum on the right-hand side in \eqref{eq:shlbunif} is---as a sum of concave functions---concave.
Hence,  if we restrict the maximization in \eqref{eq:shlbunif} to $s\geq 3/c^2$, we obtain a lower bound  for given values of $c$, $k_{\text{max}}\leq 1/(2\pi c^2)$, $\expcard$, and $D$ that we can compute using standard numerical algorithms.
In Figure~\ref{fig:shlb10}, we show this lower bound for $c=0.1$, $k_{\text{max}}=\lfloor 1/(2 \pi c^2)\rfloor=15$, $\expcard=10$, and various values of $D$.

For  $k_{\text{max}}>1/(2 \pi c^2)$, the right-hand side in \eqref{eq:shlbunif} is not guaranteed to be concave.
However, we can still use numerical solvers to try to find local maxima of \eqref{eq:shlbunif} that give even better results. 
In particular, we show in Figure~\ref{fig:shlb10} also an optimized lower bound for $k_{\text{max}}=50$.
\end{example}

%

\subsection{Upper Bound  for Poisson Point Processes}

Next, we establish an upper bound on the RD function of a Poisson PP $\rX$.
In the following theorem, which is proven in Appendix~\ref{app:proofubppp}, we apply   Corollary~\ref{cor:upperbounddpi}  to vectors $(\rxv^{(k)}, \ryv^{(k)})$ where $\rxv^{(k)}$ has the same distribution as  $\rxv_{\rX}^{(k)}$.
\begin{theorem}\label{th:ubppp}
Let $\rX$ be a Poisson PP on $\R^d$ with intensity measure $\lambda = \expcard \lambda_0$, where $\lambda_0$ is absolutely continuous with respect to $\Leb^d$ with probability density function $g_{\rX}=\frac{\intd \lambda_0}{\intd \Leb^d}$.
Furthermore, let $\lambda_{\rY}$ be a probability measure on $\R^d$ and let $\ryv^{(k)}$ be  random vectors on $(\R^d)^k$ with probability measure $(\lambda_{\rY})^k$ for each $k\in \N$.
Define the joint distribution of $(\rxv^{(k)}, \ryv^{(k)})$ by a given conditional probability density function $g_{\rxv^{(k)}\mid\ryv^{(k)}}(\xv_{1:k}\condi \yv_{1:k})$ on $(\R^d)^k$ for  each $\yv_{1:k}\in (\R^d)^k$.
Assume that the resulting random vector $\rxv^{(k)}$ has the same distribution as $\rxv_{\rX}^{(k)}$ (see~\eqref{eq:denspppk}), i.e., 
\be\label{eq:asscorrmarg}
\prod_{i=1}^k g_{\rX}(\xv_i)=\int_{(\R^d)^k}g_{\rxv^{(k)}\mid\ryv^{(k)}}(\xv_{1:k}\condi \yv_{1:k})\,\intd \lambda_{\rY}^k(\yv_{1:k})
\ee
and that
\be\label{eq:explconddistbound}
\sum_{k\in \N} \frac{e^{-\expcard}\expcard^k}{k!} \, \E\big[\dist_2^{  (c) }\big(\phi_k(\rxv^{(k)}),\phi_k(\ryv^{(k)})\big)\big]\leq D 
\ee
where
\ba
& \E\big[\dist_2^{  (c) }\big(\phi_k(\rxv^{(k)}),\phi_k(\ryv^{(k)})\big)\big]
\notag \\ & \rule{10mm}{0mm}
=  \int_{(\R^d)^k} \int_{(\R^d)^k} \dist_2^{  (c) }(\phi_k(\xv_{1:k}),\phi_k(\yv_{1:k}))\, 
\notag \\* & \rule{10mm}{0mm} \quad \times 
g_{\rxv^{(k)}\mid\ryv^{(k)}}(\xv_{1:k}\condi \yv_{1:k}) \,\intd \lambda_{\rY}^k(\yv_{1:k})\, \intd \xv_{1:k} \,. 
\notag 
\ea
Then 
\ba
 R(D)
& \leq
\expcard - \expcard \log \expcard + \expcard \,h(g_{\rX})
\notag \\* & \quad 
+ \sum_{k\in \N} \frac{e^{-\expcard} \expcard^k}{k!} \big(
\log k!
- h\big(\rxv^{(k)}\bcondi\ryv^{(k)}\big)
\big) \label{eq:rdubppp}
\ea
where 
\be\label{eq:condent}
h\big(\rxv^{(k)}\bcondi\ryv^{(k)}\big)
= \int_{(\R^d)^k} h\big(\rxv^{(k)}\bcondi\ryv^{(k)}=\yv_{1:k}\big)\,\intd \lambda_{\rY}^k(\yv_{1:k}) 
\ee
with
\ba
 h\big(\rxv^{(k)}\bcondi\ryv^{(k)}=\yv_{1:k}\big)
& =  - \int_{(\R^d)^k}  g_{\rxv^{(k)}\mid\ryv^{(k)}}(\xv_{1:k}\condi \yv_{1:k}) 
 \notag \\* & \quad \times 
\log g_{\rxv^{(k)}\mid\ryv^{(k)}}(\xv_{1:k}\condi \yv_{1:k}) \, \intd \xv_{1:k}\,. \label{eq:condentgiven}
\ea
\end{theorem}

In the proof of  Theorem~\ref{th:ubppp} in Appendix~\ref{app:proofubppp},  we do not make use of the conditional mutual informations in \eqref{eq:rdboundvectors}.
Although this loosens the bound in general, it does not if we use a conditional probability density function $g_{\rxv^{(k)}\mid\ryv^{(k)}}(\xv_{1:k}\condi \yv_{1:k})$ that does not depend on the ordering of the elements $\yv_{1}, \dots, \yv_{k}$.
Indeed, this additional assumption can be shown to imply that the conditional mutual informations in \eqref{eq:rdboundvectors} are zero and thus can be dropped without loosening the bound.
This is in stark contrast to the setting we encountered in Section \ref{sec:gaussdist}, where  the conditional mutual informations are required to obtain a useful upper bound.
Indeed, these two settings illustrate different proof strategies: 
Either the joint distribution of $\rxv^{(k)}$ and $\ryv^{(k)}$ is carefully constructed to gain conditional independence of the orderings, 
or we have to analyze the conditional mutual informations in  \eqref{eq:rdboundvectors} in detail. 
Next, we use Theorem \ref{th:ubppp} with such a carefully constructed conditional probability density function  to obtain upper bounds on the RD function of the Poisson PP  discussed in Example~\ref{ex:pppsquare}.

\begin{example}\label{ex:unifppp}
Let $\rX$ be a Poisson PP  on $\R^2$ with intensity measure $\lambda=\expcard\Leb^2|_{[0,1)^2}$, i.e., $g_{\rX} =\ind_{[0,1)^2} $.
To use Theorem~\ref{th:ubppp}, we have to define a measure $\lambda_{\rY}$ and conditional probability density functions $g_{\rxv^{(k)}\mid\ryv^{(k)}}(\xv_{1:k}\condi \yv_{1:k})$ such that \eqref{eq:asscorrmarg} is satified, i.e., in our case,
\be\label{eq:concmargdist}
\int_{(\R^2)^k}g_{\rxv^{(k)}\mid\ryv^{(k)}}(\xv_{1:k}\condi \yv_{1:k})\,\intd \lambda_{\rY}^k(\yv_{1:k})
= 
\prod_{i=1}^k \ind_{[0,1)^2}(\xv_i)\,.
\ee
To this end, for $N\in \N$ satisfying $N\geq 1/(\sqrt{2}c)$ (this condition will be used later), we define $\lambda_{\rY}$ as 
\be\label{eq:defyn}
\lambda_{\rY}(C) 
=   \frac{1}{N^2} \sum_{j_1=1}^N \sum_{j_2=1}^N \delta_{\qv_{j_1, j_2}}(C)
\ee
 for $C\in \sB_2$, where $\delta_{\xv}$ denotes the point measure at $\xv$ and $\qv_{j_1, j_2}=\big(\frac{2j_1-1}{2N}, \frac{2j_2-1}{2N}\big)$.
Hence, $\lambda_{\rY}$ corresponds to a discrete uniform distribution  with the $N^2$ possible realizations $\qv_{j_1, j_2}$, $j_1, j_2\in \{1, \dots, N\}$. 
Furthermore, consider a set of $k$ index pairs $\big\{\big(j_1^{(i)}, j_2^{(i)}\big)\big\}_{i=1, \dots, k}\subseteq \{1, \dots, N\}^2$.
The function $g_{\rxv^{(k)}\mid\ryv^{(k)}}(\xv_{1:k}\condi \yv_{1:k})$ is then defined for $\yv_{i}=\qv_{j_1^{(i)}, j_2^{(i)}}$ by
\ba
g_{\rxv^{(k)}\mid\ryv^{(k)}}\big(\xv_{1:k}& \bcondi \big(\qv_{j_1^{(1)}, j_2^{(1)}}, \dots, \qv_{j_1^{(k)}, j_2^{(k)}}\big)\big) \notag \\
& = \frac{1}{k!}\sum_{\tau} \prod_{i=1}^k  N^2 \ind_{\squr_{j_1^{(\tau(i))}, j_2^{(\tau(i))}}} (\xv_i) \label{eq:deftransprop}
 \\
& = \frac{N^{2k}}{k!}\sum_{\tau}   \ind_{\prod_{i=1}^k \squr_{j_1^{(\tau(i))}, j_2^{(\tau(i))}}} (\xv_{1:k}) \label{eq:deftranspropalt}
\ea
where $\squr_{j_1,j_2}\triangleq \big[{-}\frac{1}{2N},\frac{1}{2N}\big)^2 +\qv_{j_1, j_2}$ and the sum is over all permutations $\tau$ on $\{1, \dots, k\}$.
Note that $\{\squr_{j_1,j_2}\}_{j_1,j_2=1,\dots,N}$ constitutes a partition of $[0,1)^2$ into $N^2$ squares.
Furthermore, note that $N^2 \ind_{\squr_{j_1^{(\tau(i))}, j_2^{(\tau(i))}}}$ is the probability density function of a uniform random vector on the square $\squr_{j_1^{(\tau(i))}, j_2^{(\tau(i))}}$.
Thus, the distribution specified by \eqref{eq:deftransprop} can be interpreted as first randomly  choosing an assignment (represented by $\tau$) between $\{\rxv_i\}_{i=1,\dots, k}$ and $\big\{\qv_{j_1^{(i)}, j_2^{(i)}}\big\}_{i=1,\dots, k}$, 
and then distributing $\rxv_i$ uniformly and independently on the square $\squr_{j_1^{(\tau(i))}, j_2^{(\tau(i))}}$ with center $\qv_{j_1^{(\tau(i))}, j_2^{(\tau(i))}}$.
In particular, $g_{\rxv^{(k)}\mid\ryv^{(k)}}\big(\xv_{1:k}\bcondi \big(\qv_{j_1^{(1)}, j_2^{(1)}}, \dots, \allowbreak \qv_{j_1^{(k)}, j_2^{(k)}}\big)\big)$ does not depend on the ordering of the points $\qv_{j_1^{(1)}, j_2^{(1)}}, \allowbreak\dots, \allowbreak\qv_{j_1^{(k)}, j_2^{(k)}}$.

By Lemma~\ref{lem:glemcor} in Appendix~\ref{app:proofexunifppp},  
$\lambda_{\rY}$ defined by \eqref{eq:defyn}  
and $g_{\rxv^{(k)}\mid\ryv^{(k)}}$ defined by \eqref{eq:deftransprop}  
satisfy  \eqref{eq:concmargdist}.
Furthermore, by Lem\-ma~\ref{lem:expecdistunifppp}  in Appendix~\ref{app:proofexunifppp}, the left-hand side of \eqref{eq:explconddistbound} is given as
\be\label{eq:expecdistunifppp}
\sum_{k\in \N} \frac{e^{-\expcard}\expcard^k}{k!} \,\E\big[\dist_2^{  (c) }\big(\phi_k(\rxv^{(k)}), \phi_k(\ryv^{(k)})\big)\big]
 = \frac{\expcard}{6 N^{2}}\,.
\ee
Thus, condition \eqref{eq:explconddistbound} is satisfied for 
$D\geq \expcard/(6 N^{2})$ and, in particular, for
$
D=\expcard/(6 N^{2})$.

Finally, we will  simplify the  bound \eqref{eq:rdubppp} for our setting.
We first recall that the differential entropy $h(g_{\rX})$ is zero (see Example~\ref{ex:pppsquare}).
Furthermore, according to Lemma~\ref{lem:condentbound} in Appendix \ref{app:proofexunifppp}, the conditional differential entropy $h\big(\rxv^{(k)}\bcondi\ryv^{(k)}\big)$   can 
be lower-bounded by
\ba
h\big(\rxv^{(k)}\bcondi\ryv^{(k)}\big) 
& \geq  \frac{\binom{N^2}{k}k!}{N^{2k}} \log k!
- k \log N^{2} \,. \label{eq:mutinfboundxyn} 
\ea
Inserting $D=\expcard/(6 N^{2})$, $h(g_{\rX})=0$, and \eqref{eq:mutinfboundxyn} into \eqref{eq:rdubppp}, we obtain
\ba 
R\bigg(\frac{\expcard}{6 N^{2}}\bigg) 
& \leq
\expcard - \expcard \log \expcard 
+ \sum_{k\in \N} \frac{e^{-\expcard} \expcard^k}{k!} \big(
\log k!
+ k \log N^{2} 
\big) 
\notag \\ & \quad 
-  \sum_{k=1}^{N^2}  \frac{e^{-\expcard} \expcard^k \binom{N^2}{k} }{N^{2k}} \log k! \label{eq:rexpcardsixnbound1}
\ea
where we used that $\binom{N^2}{k}=0$ for $k> N^2$.
By Lemma~\ref{lem:boundrdnusixn}  in Appendix \ref{app:proofexunifppp} with $\widetilde{N} =N$, \eqref{eq:rexpcardsixnbound1} implies
\ba\label{eq:rateubunif}
R\bigg(\frac{\expcard}{6 N^{2}}\bigg) 
& \leq
  \expcard + \expcard \log \frac{N^2}{\expcard}  + \sum_{k=1}^{N^2} e^{-\expcard} \expcard^k \log k! \bigg(\frac{1}{k!}- \frac{\binom{N^2}{k}}{N^{2k}}\bigg)
	\notag \\* & \quad 
 + \bigg(1-\sum_{k=0}^{N^2-2} \frac{e^{-\expcard} \expcard^k}{k!} \bigg) \expcard^2\,.
\ea
The bound \eqref{eq:rateubunif} can   be calculated explicitly for various $N$.
However, for large $N$, this is   computationally intensive. 
The computational complexity can be reduced by 
omitting the summands with $k>N_{\max}^2$, where $N_{\max}\leq N$, in the last sum in \eqref{eq:rexpcardsixnbound1}, which results in
\ba \notag 
R\bigg(\frac{\expcard}{6 N^{2}}\bigg) 
&  \leq
\expcard - \expcard \log \expcard 
+ \sum_{k\in \N} \frac{e^{-\expcard} \expcard^k}{k!} \big(
\log k!
+ k \log N^{2} 
\big)  
\notag \\* & \quad 
-  \sum_{k=1}^{N_{\max}^2}  \frac{e^{-\expcard} \expcard^k \binom{N^2}{k} }{N^{2k}} \log k!\,.
\ea
Again using Lemma~\ref{lem:boundrdnusixn}  in Appendix \ref{app:proofexunifppp}, this time  with $\widetilde{N} =N_{\max}$, we finally obtain
\ba 
R\bigg(\frac{\expcard}{6 N^{2}}\bigg) 
& \leq \expcard + \expcard \log \frac{N^2}{\expcard}  + \sum_{k=1}^{\hidewidth N_{\max}^2 \hidewidth} e^{-\expcard} \expcard^k \log k! \bigg(\frac{1}{k!}- \frac{\binom{N^2}{k}}{N^{2k}}\bigg)
\notag \\* & \quad 
  + \bigg(1-\sum_{k=0}^{N_{\max}^2-2} \frac{e^{-\expcard} \expcard^k}{k!} \bigg) \expcard^2\,.
\label{eq:pppunifub}
\ea
In Figure~\ref{fig:shlb10}, this upper bound is depicted for the case $\expcard=10$, $c=0.1$, $N_{\max}=\min\{N, 10\}$, and $N$ ranging from $8$ to $207$ (corresponding to $D=\expcard/(6 N^{2})$ ranging from $3.9 \cdot 10^{-5}$ to $2.6 \cdot 10^{-2}$).
\end{example}

\section{Conclusion} \label{sec:conclusion}


We established  lower and upper  bounds on the RD function of finite PPs. 
Our bounds provide insights into the behavior of the RD function and demonstrate that the RD function based on the PP viewpoint can be significantly lower than the RD function based on the vector viewpoint.
Furthermore, the PP viewpoint allows sets of different sizes to be considered in a single source coding scenario.
Our lower bounds are based on the general  RD characterization in \cite{csiszar74}.
Our upper bounds are based either on the RD theorem   and an expression of the mutual information between PPs  or on a concrete source code.

To enable a comparison with the vector viewpoint, we considered PPs of fixed cardinality with a specific distortion function.
For consistency with the classical squared-error distortion,
we used a squared-error distortion between optimally assigned point patterns.
To obtain upper bounds, we 
established a relation between the mutual informations for random vectors and for PPs.
We further
proposed a Lloyd-type algorithm for  the construction of source codes. 
We applied our upper bounds to a PP  of fixed cardinality where all points are Gaussian and i.i.d.
The result implies  that the RD function in the PP setting is significantly smaller than that of a Gaussian vector of the same dimension. 
Furthermore, we showed that our upper bound converges to the lower bound as the distortion goes to zero.

The complexity of our proposed  Lloyd-type algorithm does not scale well in the codebook size and the cardinality of the point patterns. 
An efficient heuristic scheme for computing the ``center point pattern'' for a large collection of point patterns would significantly reduce the complexity but does not seem to be available.
We note that our algorithm can be easily generalized to PPs of variable cardinality by sorting the collection of point patterns  representing the source according to their cardinality and then performing the algorithm for each cardinality separately.
However, an  algorithm that is able to find center point patterns directly for point patterns of different cardinality may result in  better source codes.
A first approach in this direction was presented in \cite{babawi15}.
Another possible extension of our codebook construction is to encode several successive point patterns jointly, resulting in a source code of length greater than one.
This is expected to yield  tighter upper bounds, but also to result in a  higher computational  complexity.

As an example of PPs with variable cardinality, we studied  Poisson PPs
along with an unnormalized squared OSPA distortion function. 
For a  Poisson PP of uniform intensity on the unit square in $\R^2$, our lower and upper bounds are separated by only a small gap and thus provide a good characterization of the RD function.
For the construction of the upper bound, we used a uniform quantization. 
This quantization can also be employed to construct source codes and is a first, simple  constructive approach to the generation of source codes for PPs.
We expect that---similar to the vector case---finding a good systematic source code construction for general PPs is  challenging.

The specific PPs we considered in this paper were only the most basic ones.
A large variety of other PPs have been defined in the literature \cite[Ch.~3]{ilpest08}. 
In particular, statistical dependencies between the points should result in even lower RD functions but will also require a significantly more complicated analysis.
Furthermore, in certain applications,  distortion functions that are not based on optimal assignments (e.g., the Hausdorff distance \cite{scvovo08}) may be more appropriate.
%
%
Finally, we restricted our analysis to memoryless sources, i.e., i.i.d.\ sequences of PPs.
Modeling sources with memory would require mathematical results on random sequences of point patterns (e.g., Markov chains \cite[Sec.~7]{mowa04}).
An information-theoretic analysis of these sequences 
appears to be 
an interesting direction for future research.

\appendices
\renewcommand*\thesubsectiondis{\thesection.\arabic{subsection}}
\renewcommand*\thesubsection{\thesection.\arabic{subsection}}

\section{Properties of $\phi_{\krX , \krY }$} \label{app:proofiotaprops}
%
\begin{lemma}\label{lem:iotaprops}
For $k\in \N$, let $\phi_{k}$ be defined as in \eqref{eq:defiotasingle}, and for $(\krX , \krY )\in \N_0^2\setminus \{(0,0)\}$, let $\phi_{\krX , \krY }$ be defined as in \eqref{eq:iotadefxy1}--\eqref{eq:iotadefxy3}.
Then for $\sA_{\rX}, \sA_{\rY}\in \mathfrak{S}$, we have  
\begin{subnumcases}{\phi_{\krX , \krY }^{-1}(\sA_{\rX} \times \sA_{\rY})
=\label{eq:iotachar}} 
\phi_{\krX }^{-1}(\sA_{\rX})
\times \phi_{\krY }^{-1}(\sA_{\rY}) \hspace{-23mm}&  \notag \\
&  if   $\krX , \krY \neq 0$ \label{eq:iotachar1}\\ 
\phi_{\krX }^{-1}(\sA_{\rX}) &    if    $\krX \neq 0,  \krY = 0, \emptyset \in  \sA_{\rY}$ \notag \\[-1mm] \label{eq:iotachar2}\\ 
\emptyset &    if   $\krX \neq 0,  \krY = 0, \emptyset \notin  \sA_{\rY}$ \notag \\[-1mm] \label{eq:iotachar3}\\ 
\phi_{\krY }^{-1}(\sA_{\rY}) &    if   $\krX = 0,  \krY \neq 0, \emptyset \in  \sA_{\rX}$ \notag \\[-1mm] \label{eq:iotachar4}\\ 
\emptyset &   if     $\krX = 0,  \krY \neq 0, \emptyset \notin  \sA_{\rX}$\,. \notag \\[-1mm] \label{eq:iotachar5}
\end{subnumcases}

\end{lemma}
\begin{IEEEproof}
\textit{Case $\krX , \krY \neq 0$}:
According to \eqref{eq:iotadefxy1}, a vector $(\xv_{1:\krX }, \yv_{1:\krY })$ belongs to $\phi_{\krX , \krY }^{-1}(\sA_{\rX} \times \sA_{\rY})$ if and only if 
$\{\xv_1, \dots, \xv_{\krX }\}\in \sA_{\rX}$ and $\{\yv_1, \dots, \yv_{\krY }\} \in  \sA_{\rY}$.
By~\eqref{eq:defiotasingle}, this in turn is equivalent to 
$\xv_{1:\krX }\in \phi_{\krX }^{-1}(\sA_{\rX})$ and $\yv_{1:\krY }\in \phi_{\krY }^{-1}(\sA_{\rY})$. This proves \eqref{eq:iotachar1}.

\textit{Case $\krX \neq 0, \krY = 0, \emptyset \in  \sA_{\rY}$}:
According to \eqref{eq:iotadefxy3}, a vector $\xv_{1:\krX }$ belongs to $\phi_{\krX , 0}^{-1}(\sA_{\rX} \times \sA_{\rY})$ if and only if 
$\{\xv_1, \dots, \xv_{\krX }\}\in \sA_{\rX}$ and $\emptyset \in  \sA_{\rY}$.
As we assumed $\emptyset \in  \sA_{\rY}$, this is, by~\eqref{eq:defiotasingle}, equivalent to $\xv_{1:\krX }\in \phi_{\krX }^{-1}(\sA_{\rX})$. This proves \eqref{eq:iotachar2}.

\textit{Case $\krX \neq 0, \krY = 0, \emptyset \notin  \sA_{\rY}$}:
According to \eqref{eq:iotadefxy3}, a vector $\xv_{1:\krX }$ belongs to $\phi_{\krX , 0}^{-1}(\sA_{\rX} \times \sA_{\rY})$ if and only if 
$\{\xv_1, \dots, \xv_{\krX }\}\in \sA_{\rX}$ and $\emptyset \in  \sA_{\rY}$.
Because we assumed $\emptyset \notin  \sA_{\rY}$, there is no $\xv_{1:\krX }$ that belongs to $\phi_{\krX , 0}^{-1}(\sA_{\rX} \times \sA_{\rY})$. This proves \eqref{eq:iotachar3}.

The remaining cases, \eqref{eq:iotachar4} and  \eqref{eq:iotachar5},  follow by symmetry.
\end{IEEEproof}

\begin{lemma}
Let $A\subseteq (\R^d)^{\krX +\krY }$.
Then 
\be\label{eq:phiinvphieqpsi}
\phi_{\krX , \krY }^{-1}(\phi_{\krX , \krY }(A))
= \bigcup_{\tau_{\rX}, \tau_{\rY}} \psi_{\tau_{\rX}, \tau_{\rY}}(A)
\ee
where $\psi_{\tau_{\rX}, \tau_{\rY}}$ is given by \eqref{eq:defpsi}, 
and the union is over all permutations $\tau_{\rX}$ and $\tau_{\rY}$ on $\{1, \dots, \krX \}$ and $\{1, \dots, \krY \}$, respectively.
\end{lemma}
\begin{IEEEproof}
We first show $\phi_{\krX , \krY }^{-1}(\phi_{\krX , \krY }(A))
\subseteq \bigcup_{\tau_{\rX}, \tau_{\rY}} \psi_{\tau_{\rX}, \tau_{\rY}}(A)$.
To this end, let 
\be\label{eq:xyinphiphiinv}
(\xv_{1:\krX }, \yv_{1:\krY })\in \phi_{\krX , \krY }^{-1}(\phi_{\krX , \krY }(A))\,.
\ee
We have to show that $(\xv_{1:\krX }, \yv_{1:\krY })\in \psi_{\tau_{\rX}, \tau_{\rY}}(A)$ for some permutations $\tau_{\rX}, \tau_{\rY}$.
By the definition of the inverse image, \eqref{eq:xyinphiphiinv} implies that  
$\phi_{\krX , \krY }(\xv_{1:\krX }, \yv_{1:\krY })$
belongs to $\phi_{\krX , \krY }(A)$.
This does not necessarily imply $(\xv_{1:\krX }, \yv_{1:\krY })\in A$, but  there must exist a vector $(\xvt_{1:\krX }, \yvt_{1:\krY })\in A$ such that
$\phi_{\krX , \krY }(\xvt_{1:\krX },  \yvt_{1:\krY })= \phi_{\krX , \krY }(\xv_{1:\krX }, \yv_{1:\krY })$, i.e., 
$(\{\xvt_1, \allowbreak\dots, \allowbreak\xvt_{\krX }\}, \{\yvt_1, \allowbreak\dots, \allowbreak\yvt_{\krY }\})=(\{\xv_1, \allowbreak\dots, \allowbreak\xv_{\krX }\}, \{\yv_1, \allowbreak\dots, \allowbreak\yv_{\krY }\})$.
This equality implies that there exist permutations $\tau_{\rX}$ and $\tau_{\rY}$ such that $(\xv_{1:\krX }, \yv_{1:\krY })= (\xvt_{\tau_{\rX}(1)}, \allowbreak\dots, \allowbreak\xvt_{\tau_{\rX}(\krX )},  \yvt_{\tau_{\rY}(1)},   \allowbreak\dots,  \allowbreak \yvt_{\tau_{\rY}(\krY )})$, i.e., $(\xv_{1:\krX }, \yv_{1:\krY })\in \psi_{\tau_{\rX}, \tau_{\rY}}(A)$.

It remains to show $\phi_{\krX , \krY }^{-1}(\phi_{\krX , \krY }(A))
\supseteq \bigcup_{\tau_{\rX}, \tau_{\rY}} \psi_{\tau_{\rX}, \tau_{\rY}}(A)$ or, equivalently, $\phi_{\krX , \krY }^{-1}(\phi_{\krX , \krY }(A))
\supseteq   \psi_{\tau_{\rX}, \tau_{\rY}}(A)$ for all permutations $\tau_{\rX}$ and $\tau_{\rY}$.
To this end, let $(\xv_{1:\krX }, \yv_{1:\krY })\in \psi_{\tau_{\rX}, \tau_{\rY}}(A)$.
Thus, $(\xv_{1:\krX }, \yv_{1:\krY })= (\xvt_{\tau_{\rX}(1)},  \allowbreak  \dots, \allowbreak\xvt_{\tau_{\rX}(\krX )},  \yvt_{\tau_{\rY}(1)},  \allowbreak \dots, \allowbreak\yvt_{\tau_{\rY}(\krY )})$ for some $(\xvt_{1:\krX }, \yvt_{1:\krY })\in A$. 
In particular, this equality implies that $(\{\xvt_1, \dots, \xvt_{\krX }\}, \{\yvt_1, \dots, \yvt_{\krY }\})=(\{\xv_1, \allowbreak\dots, \allowbreak\xv_{\krX }\}, \{\yv_1, \allowbreak\dots, \allowbreak\yv_{\krY }\})$, or, equivalently, $\phi_{\krX , \krY }(\xvt_{1:\krX }, \yvt_{1:\krY })=\phi_{\krX , \krY }(\xv_{1:\krX }, \yv_{1:\krY })$.
The latter equality implies $(\xv_{1:\krX }, \yv_{1:\krY })\in \phi_{\krX , \krY }^{-1}(\phi_{\krX , \krY }(A))$.
\end{IEEEproof}

\section{Proof of Lemma~\ref{lem:rndrelations}} \label{app:rndrelations}
%

We first present a preliminary result.
\begin{lemma}
Let $(\rX, \rY)$ be a pair of PPs. 
For $\sA \in \mathfrak{S}\otimes  \mathfrak{S}$, we have
\ba
& P_{\rX}\times P_{\rY}(\sA)
\notag \\ 
& = p_{\card{\rX}}(0)p_{\card{\rY}}(0) \ind_{\sA}\big((\emptyset, \emptyset)\big)
 \notag \\* & \quad 
+ \sum_{\krX \in \N} p_{\card{\rX}}(\krX )p_{\card{\rY}}(0) P_{\rX}^{(\krX )}\big(\phi_{\krX , 0}^{-1}(\sA)\big) 
\notag \\* & \quad 
+ \sum_{\krY \in \N} p_{\card{\rX}}(0)p_{\card{\rY}}(\krY ) P_{\rY}^{(\krY )}\big(\phi_{0,\krY }^{-1}(\sA)\big)  
\notag \\* & \quad 
+ \sum_{\krX  \in \N} \sum_{ \krY \in \N} p_{\card{\rX}}(\krX )p_{\card{\rY}}(\krY ) 
\big(P_{\rX}^{(\krX )}\times P_{\rY}^{(\krY )}\big)\big(\phi_{\krX ,\krY }^{-1}(\sA)\big) \,. \label{eq:pxpydecomp}
\ea
\end{lemma}
\begin{IEEEproof}
We first note that both sides of \eqref{eq:pxpydecomp} are finite measures on $\mathfrak{S}\otimes  \mathfrak{S}$.
Because finite measures can be uniquely extended to a product $\sigma$-algebra based on their values on rectangles, it suffices to consider sets $\sA= \sA_{\rX}\times \sA_{\rY}$ with $\sA_{\rX}, \sA_{\rY}\in \mathfrak{S}$.
For such $\sA$,
we have 
\ba
&  P_{\rX} \times  P_{\rY}(\sA_{\rX}\times \sA_{\rY}) \notag \\
& = P_{\rX}(\sA_{\rX}) P_{\rY}(\sA_{\rY}) \notag \\
& \stackrel{\hidewidth \eqref{eq:pxa} \hidewidth}= 
\bigg(p_{\card{\rX}}(0) \ind_{\sA_{\rX}}(\emptyset)
+ \sum_{\krX \in \N} p_{\card{\rX}}(\krX ) P_{\rX}^{(\krX )}\big(\phi_{\krX }^{-1}(\sA_{\rX})\big)\bigg)
\notag \\*
& \quad \times 
\bigg(p_{\card{\rY}}(0) \ind_{\sA_{\rY}}(\emptyset)
+ \sum_{\krY \in \N} p_{\card{\rY}}(\krY ) P_{\rY}^{(\krY )}\big(\phi_{\krY }^{-1}(\sA_{\rY})\big) \bigg)  \notag \\
& = 
 p_{\card{\rX}}(0) \ind_{\sA_{\rX}}(\emptyset) p_{\card{\rY}}(0) \ind_{\sA_{\rY}}(\emptyset) 
\notag \\*[1mm] & \quad 
+ p_{\card{\rY}}(0) \ind_{\sA_{\rY}}(\emptyset) \sum_{\krX \in \N} p_{\card{\rX}}(\krX ) P_{\rX}^{(\krX )}\big(\phi_{\krX }^{-1}(\sA_{\rX})\big)
\notag \\*
& \quad 
+ p_{\card{\rX}}(0) \ind_{\sA_{\rX}}(\emptyset) \sum_{\krY \in \N} p_{\card{\rY}}(\krY ) P_{\rY}^{(\krY )}\big(\phi_{\krY }^{-1}(\sA_{\rY})\big) 
\notag \\*
& \quad 
+ \sum_{\krX \in \N} \sum_{\krY \in \N} p_{\card{\rX}}(\krX ) p_{\card{\rY}}(\krY ) 
\notag \\*[-2mm]
&  \rule{20mm}{0mm}  \times 
P_{\rX}^{(\krX )}\big(\phi_{\krX }^{-1}(\sA_{\rX})\big)P_{\rY}^{(\krY )}\big(\phi_{\krY }^{-1}(\sA_{\rY})\big)  \notag \\
& \stackrel{\hidewidth  \eqref{eq:iotachar} \hidewidth}= \;p_{\card{\rX}}(0)p_{\card{\rY}}(0) \ind_{\sA_{\rX}\times \sA_{\rY}}\big((\emptyset, \emptyset)\big)
\notag \\*[1mm] & \quad 
 \;+ \sum_{\krX \in \N} p_{\card{\rX}}(\krX )p_{\card{\rY}}(0) P_{\rX}^{(\krX )}\big(\phi_{\krX , 0}^{-1}(\sA_{\rX}\times \sA_{\rY})\big) 
\notag \\* & \quad 
 \;+ \sum_{\krY \in \N} p_{\card{\rX}}(0)p_{\card{\rY}}(\krY ) P_{\rY}^{(\krY )}\big(\phi_{0,\krY }^{-1}(\sA_{\rX}\times \sA_{\rY})\big)  
\notag \\* & \quad 
 \;+ \sum_{\krX \in \N} \sum_{ \krY \in \N} p_{\card{\rX}}(\krX )p_{\card{\rY}}(\krY ) 
\notag \\*[-2mm]
&  \rule{20mm}{0mm} \times 
\big(P_{\rX}^{(\krX )}\times P_{\rY}^{(\krY )}\big)\big(\phi_{\krX ,\krY }^{-1}(\sA_{\rX}\times \sA_{\rY})\big)\notag \,.
\ea
This shows that \eqref{eq:pxpydecomp} holds for all rectangles and thus concludes the proof.
\end{IEEEproof}

\subsection{Equivalence of \ref{en:rdexists} and \ref{en:allrdexist}}

Next, we show that properties  \ref{en:rdexists} and \ref{en:allrdexist} in Lemma~\ref{lem:rndrelations} are equivalent.

\emph{\ref{en:rdexists} $\Rightarrow$ \ref{en:allrdexist}:}
We first assume that \ref{en:rdexists} holds, i.e.,  $P_{\rX,\rY}\ll P_{\rX}\times P_{\rY}$.
We want to show that this implies \ref{en:allrdexist}.
To this end, 
let $P_{\rX}^{(\krX )}\times P_{\rY}^{(\krY )}(A)=0$ for $\krX ,\krY  \in \N$ with $p_{\card{\rX},\card{\rY}}(\krX ,\krY )\neq 0$ and a Borel set $A \subseteq (\R^d)^{\krX +\krY }$.
Because $P_{\rX}^{(\krX )}$ and $P_{\rY}^{(\krY )}$ are symmetric measures, this implies (see~\eqref{eq:symmeasure}) $P_{\rX}^{(\krX )}\times P_{\rY}^{(\krY )}\big(\psi_{\tau_{\rX}, \tau_{\rY}}(A)\big)=0$ for all permutations $\tau_{\rX}, \tau_{\rY}$.
By~\eqref{eq:phiinvphieqpsi}, this implies 
\ba
P_{\rX}^{(\krX )}\times P_{\rY}^{(\krY )}\big(\phi_{\krX , \krY }^{-1}(\phi_{\krX , \krY }(A))\big)
& = \sum_{\tau_{\rX}, \tau_{\rY}}P_{\rX}^{(\krX )}\times P_{\rY}^{(\krY )}\big( \psi_{\tau_{\rX}, \tau_{\rY}}(A)\big) 
\notag \\ &
 = 0\,.\label{eq:phiphiinvzero1}
\ea
Because $\phi_{\krX' , \krY' }$ and $\phi_{\krX , \krY }$ have disjoint images for $(\krX' , \krY' )\neq (\krX , \krY )$, we obtain $\phi_{\krX' , \krY' }^{-1}(\phi_{\krX , \krY }(A))=\emptyset$, which 
implies
\begin{subequations}\label{eq:phiphiinvzero2}
\ba
P_{\rX}^{(\krX' )}\times P_{\rY}^{(\krY' )}\big(\phi_{\krX' , \krY' }^{-1}(\phi_{\krX , \krY }(A))\big)& =0 \quad \text{ if } \krX' , \krY' \in \N \text{ and } \notag  \\
& \hspace{11mm} (\krX' , \krY' )\neq (\krX , \krY ) \\
P_{\rX}^{(\krX' )} \big(\phi_{\krX' ,0}^{-1}(\phi_{\krX , \krY }(A))\big)& =0  \quad \text{ if } \krX'  \in \N \text{ and } \notag  \\
& \hspace{11mm} (\krX' , 0)\neq (\krX , \krY )\\
 P_{\rY}^{(\krY' )}\big(\phi_{0, \krY' }^{-1}(\phi_{\krX , \krY }(A))\big)& =0  \quad \text{ if }  \krY' \in \N \text{ and } \notag  \\
& \hspace{11mm} (0 , \krY' )\neq (\krX , \krY )\,.
\ea
Furthermore, $(\emptyset, \emptyset)\notin \phi_{\krX , \krY }(A)$ and 
thus
\be
\ind_{\phi_{\krX , \krY }(A)}\big((\emptyset, \emptyset)\big) = 0\,.
\ee
\end{subequations}
By~\eqref{eq:pxpydecomp} with $\sA=\phi_{\krX , \krY }(A)$, \eqref{eq:phiphiinvzero1} and \eqref{eq:phiphiinvzero2} imply
$P_{\rX}\times P_{\rY}(\phi_{\krX , \krY }(A))=0$.
Due to the assumed absolute continuity $P_{\rX,\rY}\ll P_{\rX}\times P_{\rY}$, this implies 
$P_{\rX,\rY}(\phi_{\krX , \krY }(A))=0$ and in turn, by~\eqref{eq:prodprob}  with $\sA=\phi_{\krX , \krY }(A)$, 
$P_{\rX, \rY}^{(\krX ,\krY )}\big(\phi_{\krX , \krY }^{-1}(\phi_{\krX , \krY }(A))\big)=0$ (recall that we assumed $p_{\card{\rX},\card{\rY}}(\krX ,\krY )\neq 0$).
Because $A \subseteq \phi_{\krX , \krY }^{-1}(\phi_{\krX , \krY }(A))$, we obtain $P_{\rX, \rY}^{(\krX ,\krY )}(A)=0$.
Thus, we showed that for $\krX ,\krY \in \N$ with $p_{\card{\rX},\card{\rY}}(\krX ,\krY )\neq 0$, $P_{\rX}^{(\krX )}\times P_{\rY}^{(\krY )}(A)=0$ implies $P_{\rX, \rY}^{(\krX ,\krY )}(A)=0$,
i.e., we have $P_{\rX,\rY}^{(\krX ,\krY )} \ll P_{\rX}^{(\krX )}\times P_{\rY}^{(\krY )}$.
If $\krX =0$ or $\krY =0$, the proof follows analogously.

\emph{\ref{en:allrdexist} $\Rightarrow$ \ref{en:rdexists}:}
For the converse direction, we assume that \ref{en:allrdexist} holds.
In order to show \ref{en:rdexists}, assume that $P_{\rX}\times P_{\rY}(\sA)=0$ for  $\sA \in \mathfrak{S}\otimes  \mathfrak{S}$.
By~\eqref{eq:pxpydecomp}, this implies 
\begin{subequations}
\label{eq:allkxkypzero}
\ba
p_{\card{\rX}}(0)p_{\card{\rY}}(0)\ind_{\sA}((\emptyset, \emptyset)) & =0   
\\
P_{\rX}^{(\krX )}\big(\phi_{\krX , 0}^{-1}(\sA)\big) & =0 
 \quad  \text{ if } \krX \in \N \text{ and } \notag \\*[-1mm]
& \hspace{13mm}  p_{\card{\rX}}(\krX )p_{\card{\rY}}(0)\neq 0   
\\
P_{\rY}^{(\krY )}\big(\phi_{0,\krY }^{-1}(\sA)\big) & =0 
 \quad \text{ if } \krY \in \N \text{ and } \notag \\*[-1mm]
& \hspace{13mm}p_{\card{\rX}}(0)p_{\card{\rY}}(\krY )\neq 0  
\\
P_{\rX}^{(\krX )}\times P_{\rY}^{(\krY )}\big(\phi_{\krX ,\krY }^{-1}(\sA)\big)& = 0 
 \quad \text{ if } \krX ,\krY \in \N \text{ and } \notag \\*[-1mm]
& \hspace{13mm}p_{\card{\rX}}(\krX )p_{\card{\rY}}(\krY ) \neq 0   
\,.
\ea
\end{subequations}
By~\eqref{eq:margcarddist}, we have that for any $(\krX ,\krY )\in \N_0^2$, both $p_{\card{\rX}}(\krX )$ and $p_{\card{\rY}}(\krY )$ are nonzero  if $p_{\card{\rX},\card{\rY}}(\krX ,\krY )\neq 0$.
Thus, the conditions 
in \eqref{eq:allkxkypzero} are implied by corresponding conditions on $p_{\card{\rX},\card{\rY}}(\krX ,\krY )$, and we obtain
\ba
p_{\card{\rX},\card{\rY}}(0,0) \ind_{\sA}((\emptyset, \emptyset)) & =0 
\notag 
\\
P_{\rX}^{(\krX )}\big(\phi_{\krX , 0}^{-1}(\sA)\big) & =0 \quad \text{ if } \krX \in \N \text{ and } 
p_{\card{\rX},\card{\rY}}(\krX ,0)\neq 0   
\notag \\
P_{\rY}^{(\krY )}\big(\phi_{0,\krY }^{-1}(\sA)\big) & =0  \quad \text{ if } \krY \in \N \text{ and } p_{\card{\rX},\card{\rY}}(0,\krY )\neq 0   
\notag 
\\
P_{\rX}^{(\krX )}\times P_{\rY}^{(\krY )}\big(\phi_{\krX ,\krY }^{-1}(\sA)\big)& = 0 
\quad \text{ if } \krX ,\krY \in \N \text{ and } 
\notag \\*[-1mm]
& \hspace{13mm}
p_{\card{\rX},\card{\rY}}(\krX ,\krY )\neq 0\,.  \notag 
\ea
By the absolute continuity assumptions in \ref{en:allrdexist}, these equations imply $P_{\rX, \rY}^{(\krX ,\krY )}\big(\phi_{\krX ,\krY }^{-1}(\sA)\big)=0$ for any $(\krX ,\krY )\in \N_0^2\setminus \{(0,0)\}$ with $p_{\card{\rX},\card{\rY}}(\krX ,\krY )\neq 0$.
Thus, all summands on the right-hand side of \eqref{eq:prodprob} are zero, which implies $P_{\rX, \rY}(\sA)=0$.
Hence, we showed that  $P_{\rX}\times P_{\rY}(\sA)=0$ implies  $P_{\rX, \rY}(\sA)=0$, i.e., $P_{\rX, \rY} \ll P_{\rX}\times P_{\rY}$, which is \ref{en:rdexists}.

\subsection{Proof that $\frac{\mathrm{d}P_{\rX,\rY}}{\mathrm{d}(P_{\rX}\times P_{\rY})} = \theta_{\rX, \rY}$}

We have to show that for all $\sA\in \mathfrak{S}\otimes \mathfrak{S}$
\ba\label{eq:pxyeqinfdens}
P_{\rX,\rY}(\sA)
& = \int_{\sA}\theta_{\rX,\rY}(X,Y)\, \intd (P_{\rX}\times P_{\rY})(X,Y)\,.
\ea
Again, because finite measures can be uniquely extended to a product $\sigma$-algebra based on their values on rectangles, it suffices to consider sets $\sA= \sA_{\rX}\times \sA_{\rY}$ with $\sA_{\rX}, \sA_{\rY}\in \mathfrak{S}$.
With this choice, it follows from \eqref{eq:prodprob}  that the left-hand side in \eqref{eq:pxyeqinfdens} can be rewritten as 
\ba
& P_{\rX,\rY}(\sA_{\rX}\times \sA_{\rY}) \notag \\*
& =
p_{\card{\rX},\card{\rY}}(0,0) \ind_{\sA_{\rX}\times \sA_{\rY}}\big((\emptyset, \emptyset)\big) 
 \notag \\*
& \quad
+ \sum_{\substack{(\krX , \krY )\in \N_0^2 \\ 
(\krX , \krY )\neq (0,0)}} p_{\card{\rX},\card{\rY}}(\krX ,\krY ) P_{\rX,\rY}^{(\krX ,\krY )}\big(\phi_{\krX ,\krY }^{-1}(\sA_{\rX}\times \sA_{\rY})\big) 
\notag \\
& \stackrel{\hidewidth \eqref{eq:iotachar} \hidewidth}= \,p_{\card{\rX},\card{\rY}}(0,0) \ind_{\sA_{\rX}}(\emptyset) \ind_{\sA_{\rY}}(\emptyset)
 \notag \\*[1mm]
& \quad
+ \ind_{\sA_{\rY}}(\emptyset)
\sum_{\krX \in \N}  p_{\card{\rX},\card{\rY}}(\krX ,0) P_{\rX, \rY}^{(\krX ,0)}\big(\phi_{\krX }^{-1}(\sA_{\rX})\big) 
 \notag \\*
& \quad
+ \ind_{\sA_{\rX}}(\emptyset) 
\sum_{\krY \in \N} p_{\card{\rX},\card{\rY}}(0,\krY ) P_{\rX, \rY}^{(0,\krY )}\big(\phi_{\krY }^{-1}(\sA_{\rY})\big) 
\notag \\*
& \quad
+ \sum_{\krX \in \N} \sum_{\krY \in \N} p_{\card{\rX},\card{\rY}}(\krX ,\krY ) 
P_{\rX, \rY}^{(\krX ,\krY )}\big(\phi_{\krX }^{-1}(\sA_{\rX})\times \phi_{\krY }^{-1}(\sA_{\rY})\big) \,.\label{eq:rndlhs} 
\ea
The right-hand side in \eqref{eq:pxyeqinfdens} can be rewritten 
as 
\ba
 &  \int_{\sA_{\rX}\times \sA_{\rY}}\theta_{\rX,\rY}(X,Y)\,\intd (P_{\rX}\times P_{\rY})(X,Y) \notag \\ 
& =\int_{\sA_{\rY}} \int_{\sA_{\rX}}\theta_{\rX,\rY}(X,Y)\, \intd P_{\rX}(X) \,\intd P_{\rY}(Y)
\notag \\ 
& \stackrel{\hidewidth (a) \hidewidth}=
 \int_{\sA_{\rY}} \bigg( p_{\card{\rX}}(0) \ind_{\sA_{\rX}}(\emptyset)\theta_{\rX,\rY}(\emptyset,Y)  
+  \sum_{\krX \in \N}p_{\card{\rX}}(\krX ) 
\notag \\*
& \quad \times 
\int_{\phi_{\krX }^{-1}(\sA_{\rX})}
 \theta_{\rX,\rY}(\phi_{\krX }(\xv_{1:\krX }),Y)\,\intd P_{\rX}^{(\krX )}(\xv_{1:\krX }) \bigg)\,\intd P_{\rY}(Y)
\label{eq:rhsinthetaint}
\ea
where we used \eqref{eq:pxaintsa} with $\widetilde{g}(X)=\theta_{\rX,\rY}(X,Y)$ in $(a)$.
The integral over the first summand in \eqref{eq:rhsinthetaint} can be rewritten as
\ba
& \int_{\sA_{\rY}}   p_{\card{\rX}}(0) \ind_{\sA_{\rX}}(\emptyset)\theta_{\rX,\rY}(\emptyset,Y) \,\intd P_{\rY}(Y) 
\notag \\
& = p_{\card{\rX}}(0) \ind_{\sA_{\rX}}(\emptyset)\int_{\sA_{\rY}} \theta_{\rX,\rY}(\emptyset,Y)  \,\intd P_{\rY}(Y) \notag \\
& \stackrel{\hidewidth (a) \hidewidth}= p_{\card{\rX}}(0) \ind_{\sA_{\rX}}(\emptyset) \bigg(
p_{\card{\rY}}(0) \ind_{\sA_{\rY}}(\emptyset) \theta_{\rX,\rY}(\emptyset,\emptyset) 
\notag \\* & \quad 
+ 
\sum_{\krY \in \N}p_{\card{\rY}}(\krY ) 
\int_{\phi_{\krY }^{-1}(\sA_{\rY})}
\theta_{\rX,\rY}(\emptyset,\phi_{\krY }(\yv_{1:\krY })) \,\intd P_{\rY}^{(\krY )}(\yv_{1:\krY }) 
\bigg) \notag \\
& \stackrel{\hidewidth\eqref{eq:condixygen}\hidewidth} = p_{\card{\rX},\card{\rY}}(0,0)\ind_{\sA_{\rX}}(\emptyset) 
 \ind_{\sA_{\rY}}(\emptyset) 
\notag \\* & \quad 
+ 
 \ind_{\sA_{\rX}}(\emptyset)\sum_{\krY \in \N} 
\int_{\phi_{\krY }^{-1}(\sA_{\rY})} p_{\card{\rX},\card{\rY}}(0,\krY ) \,\intd P_{\rX,\rY}^{(0,\krY )}(\yv_{1:\krY })
\label{eq:intthetasum1}
\ea
where we used   \eqref{eq:pxaintsa} with $\widetilde{g}(Y)=\theta_{\rX,\rY}(\emptyset,Y)$  in $(a)$.
The integral over the remaining summands in \eqref{eq:rhsinthetaint} can be rewritten as 
\ba
& \int_{\sA_{\rY}}p_{\card{\rX}}(\krX )   
\int_{\phi_{\krX }^{-1}(\sA_{\rX})}
  \theta_{\rX,\rY}(\phi_{\krX }(\xv_{1:\krX }),Y)
\,\intd P_{\rX}^{(\krX )}(\xv_{1:\krX }) \,\intd P_{\rY}(Y) \notag \\
& =
\hspace{-1mm}\int_{\phi_{\krX }^{-1}(\sA_{\rX})}
\hspace{-1mm}p_{\card{\rX}}(\krX )  \hspace{-.5mm} \int_{\sA_{\rY}} \hspace{-1mm} \theta_{\rX,\rY}(\phi_{\krX }(\xv_{1:\krX }),Y) 
\,\intd P_{\rY}(Y) \,\intd P_{\rX}^{(\krX )}(\xv_{1:\krX }) 
\notag \\
& \stackrel{\hidewidth (a) \hidewidth}=
 \int_{\phi_{\krX }^{-1}(\sA_{\rX})}
p_{\card{\rX}}(\krX ) \bigg(
 p_{\card{\rY}}(0) \ind_{\sA_{\rY}}(\emptyset) 
\theta_{\rX,\rY}(\phi_{\krX }(\xv_{1:\krX }),\emptyset) 
\notag \\*
& \quad  
+ 
\sum_{\krY \in \N}p_{\card{\rY}}(\krY ) 
\int_{\phi_{\krY }^{-1}(\sA_{\rY})}
\theta_{\rX,\rY}(\phi_{\krX }(\xv_{1:\krX }),\phi_{\krY }(\yv_{1:\krY })) 
\notag \\*[-2mm]
&  \rule{40mm}{0mm} \times 
\,\intd P_{\rY}^{(\krY )}(\yv_{1:\krY }) 
\bigg)  \,\intd P_{\rX}^{(\krX )}(\xv_{1:\krX })
\notag \\
& \stackrel{\hidewidth (b) \hidewidth} =
\ind_{\sA_{\rY}}(\emptyset) 
\int_{\phi_{\krX }^{-1}(\sA_{\rX})} p_{\card{\rX},\card{\rY}}(\krX ,0)  \,\intd P_{\rX,\rY}^{(\krX ,0)}(\xv_{1:\krX })
\notag \\*
& \quad  
+ 
\sum_{\krY \in \N}
\int_{\phi_{\krX }^{-1}(\sA_{\rX}) \times \phi_{\krY }^{-1}(\sA_{\rY})}
p_{\card{\rX},\card{\rY}}(\krX ,\krY ) 
 \,\intd P_{\rX,\rY}^{(\krX ,\krY )}(\xv_{1:\krX },\yv_{1:\krY })
	\label{eq:intthetasum2}
\ea
\makebox[\linewidth][s]{where we used   \eqref{eq:pxaintsa} with $\widetilde{g}(Y)=\theta_{\rX,\rY}(\phi_{\krX }(\xv_{1:\krX }),Y)$ in $(a)$ and}
\newpage
\noindent plugged in \eqref{eq:condixygen} in $(b)$.
Inserting \eqref{eq:intthetasum1} and \eqref{eq:intthetasum2} into \eqref{eq:rhsinthetaint}, we obtain
\ba
& \int_{\sA_{\rX}\times \sA_{\rY}}\theta_{\rX,\rY}(X,Y)\,\intd (P_{\rX}\times P_{\rY})(X,Y)
\notag \\ 
& =
p_{\card{\rX},\card{\rY}}(0,0) \ind_{\sA_{\rX}}(\emptyset)\ind_{\sA_{\rY}}(\emptyset)  \notag \\* & \quad 
+ 
\ind_{\sA_{\rX}}(\emptyset)
\sum_{\krY \in \N}\int_{\phi_{\krY }^{-1}(\sA_{\rY})}
p_{\card{\rX},\card{\rY}}(0,\krY ) \,\intd P_{\rX,\rY}^{(0,\krY )}(\yv_{1:\krY }) 
\notag \\*
& \quad + 
\ind_{\sA_{\rY}}(\emptyset) \sum_{\krX \in \N}\int_{\phi_{\krX }^{-1}(\sA_{\rX})}
p_{\card{\rX},\card{\rY}}(\krX ,0)  \,\intd P_{\rX,\rY}^{(\krX ,0)}(\xv_{1:\krX })\notag \\*
& \quad + 
\sum_{\krX \in \N}\sum_{\krY \in \N}
\int_{\phi_{\krX }^{-1}(\sA_{\rX})\times \phi_{\krY }^{-1}(\sA_{\rY})}
p_{\card{\rX},\card{\rY}}(\krX ,\krY ) 
\notag \\*[-2mm] & \rule{45mm}{0mm} \times 
 \,\intd P_{\rX,\rY}^{(\krX ,\krY )}(\xv_{1:\krX },\yv_{1:\krY })
\,.
\notag
\ea
This is  seen to coincide with  \eqref{eq:rndlhs}, and thus the equality  \eqref{eq:pxyeqinfdens} holds.

\section{Proof of Theorem~\ref{th:mutinfgen}} \label{app:proofmutinfgen}
%
We first note that due to the equivalence of \ref{en:rdexists} and \ref{en:allrdexist} in Lemma~\ref{lem:rndrelations}, $P_{\rX,\rY}\ll P_{\rX}\times P_{\rY}$ if and only if $P_{\rX,\rY}^{(\krX ,0)} \ll P_{\rX}^{(\krX )}$, $P_{\rX,\rY}^{(0,\krY )}\ll P_{\rY}^{(\krY )}$,  and 
$P_{\rX,\rY}^{(\krX ,\krY )}\ll P_{\rX}^{(\krX )}\times P_{\rY}^{(\krY )}$ for all $(\krX , \krY )\in \N_0^2\setminus \{(0,0)\}$ with $p_{\card{\rX},\card{\rY}}(\krX ,\krY )\neq 0$.
Thus, if any of the aforementioned absolute continuities do not hold, \eqref{eq:gyptheorem} and \eqref{eq:kldiv} imply that both sides in \eqref{eq:sepmutinf} are infinite, which concludes the proof for this case.
Otherwise, \ref{en:rdexists} and \ref{en:allrdexist} in Lemma~\ref{lem:rndrelations} hold and we can express the mutual information and all relevant KLDs in \eqref{eq:sepmutinf} using Radon-Nikodym derivatives.
We recall from Lemma~\ref{lem:rndrelations} that in this case $\frac{\mathrm{d}P_{\rX,\rY}}{\mathrm{d}(P_{\rX}\times P_{\rY})}(X,Y)=\theta_{\rX,\rY}(X,Y)$, where $\theta_{\rX,\rY}$ satisfies \eqref{eq:condixygen}.
Thus, using \eqref{eq:gyptheorem}, we obtain 
$
I(\rX;\rY) = \int_{\sX^2}\log \theta_{\rX,\rY}(X,Y) \, \mathrm{d}P_{\rX,\rY}(X,Y)
$.
Using  \eqref{eq:prodprobint} with $g(X,Y)=\log \theta_{\rX,\rY}(X,Y)$, we obtain further
\ba
& I(\rX;\rY) 
\notag \\
& = p_{\card{\rX},\card{\rY}}(0,0) \log \theta_{\rX,\rY}(\emptyset, \emptyset) 
+
\hspace{-1mm}
\sum_{\substack{(\krX , \krY )\in \N_0^2 \\ 
(\krX , \krY )\neq (0,0)}} \hspace{-1mm} p_{\card{\rX},\card{\rY}}(\krX ,\krY ) 
\notag \\[-1mm]
& \quad \times
\int_{(\R^{d})^{\krX +\krY }}
\log \theta_{\rX,\rY}\big(\phi_{\krX , \krY }(\xv_{1:\krX }, \yv_{1:\krY })\big) 
\,\intd P_{\rX, \rY}^{(\krX , \krY )}(\xv_{1:\krX }, \yv_{1:\krY })\,. \notag
\ea
Inserting for $\theta_{\rX,\rY}$ the expressions \eqref{eq:condixygen} 
yields
\ba
& I(\rX;\rY) \notag \\[-1mm]
&  = 
p_{\card{\rX},\card{\rY}}(0,0) \log 
\bigg(\frac{p_{\card{\rX},\card{\rY}}(0,0)}{p_{\card{\rX}}(0)p_{\card{\rY}}(0)}  \bigg) 
+ 
\sum_{\krX \in \N}p_{\card{\rX},\card{\rY}}(\krX ,0) 
\notag \\
& \quad  \times
\int_{(\R^d)^{\krX }}\hspace{-1mm}\log \bigg(\frac{p_{\card{\rX},\card{\rY}}(\krX ,0)}{p_{\card{\rX}}(\krX )p_{\card{\rY}}(0)}
\frac{\intd P_{\rX,\rY}^{(\krX ,0)}}{\intd P_{\rX}^{(\krX )}}(\xv_{1:\krX })\bigg) 
\,\intd  P_{\rX, \rY}^{(\krX ,0)}(\xv_{1:\krX }) \notag \\
& \quad
+ \sum_{\krY \in \N}p_{\card{\rX},\card{\rY}}(0,\krY ) 
\notag \\*
& \quad \times
\int_{(\R^d)^{\krY }}\log \bigg(\frac{p_{\card{\rX},\card{\rY}}(0,\krY )}{p_{\card{\rX}}(0)p_{\card{\rY}}(\krY )}
\frac{\intd P_{\rX,\rY}^{(0,\krY )}}{\intd P_{\rY}^{(\krY )}}(\yv_{1:\krY })\bigg) 
\,\intd P_{\rX, \rY}^{(0,\krY )}(\yv_{1:\krY }) 
\notag \\ & \quad
+  \sum_{\krX \in \N} \sum_{\krY \in \N}p_{\card{\rX},\card{\rY}}(\krX ,\krY ) 
\int_{(\R^d)^{\krX +\krY }}\log \bigg(\frac{p_{\card{\rX},\card{\rY}}(\krX ,\krY )}{p_{\card{\rX}}(\krX )p_{\card{\rY}}(\krY )}
\notag \\* & \quad \times
\frac{\intd P_{\rX,\rY}^{(\krX ,\krY )}}{\intd \big(P_{\rX}^{(\krX )}\times P_{\rY}^{(\krY )}\big)}(\xv_{1:\krX }, \yv_{1:\krY })\bigg)
 \,\intd P_{\rX, \rY}^{(\krX ,\krY )}(\xv_{1:\krX }, \yv_{1:\krY })
\notag \\
& = 
\sum_{\krX \in \N_0} \sum_{\krY \in \N_0}
p_{\card{\rX},\card{\rY}}(\krX ,\krY ) \log 
\bigg( \frac{p_{\card{\rX},\card{\rY}}(\krX ,\krY )}{p_{\card{\rX}}(\krX )p_{\card{\rY}}(\krY )} \bigg) \notag \\*[1mm]
& \quad + 
\sum_{\krX \in \N}p_{\card{\rX},\card{\rY}}(\krX ,0) 
\notag \\*[-3mm] & \rule{20mm}{0mm} \times
\int_{(\R^d)^{\krX }}\log \bigg(
\frac{\intd P_{\rX,\rY}^{(\krX ,0)}}{\intd P_{\rX}^{(\krX )}}(\xv_{1:\krX })\bigg) 
\,\intd  P_{\rX, \rY}^{(\krX ,0)}(\xv_{1:\krX }) \notag \\
& \quad
+ \sum_{\krY \in \N}p_{\card{\rX},\card{\rY}}(0,\krY ) 
\notag \\[-3mm] & \rule{20mm}{0mm} \times
\int_{(\R^d)^{\krY }}\log \bigg(
\frac{\intd P_{\rX,\rY}^{(0,\krY )}}{\intd P_{\rY}^{(\krY )}}(\yv_{1:\krY })\bigg) 
\,\intd P_{\rX, \rY}^{(0,\krY )}(\yv_{1:\krY }) 
\notag \\
& \quad
+ \sum_{\krX \in \N} \sum_{\krY \in \N}p_{\card{\rX},\card{\rY}}(\krX ,\krY ) 
\notag \\ & \quad \times
\int_{(\R^d)^{\krX +\krY }}  \log \bigg(\frac{\intd P_{\rX,\rY}^{(\krX ,\krY )}}{\intd \big(P_{\rX}^{(\krX )} \times  P_{\rY}^{(\krY )}\big)}(\xv_{1:\krX }, \yv_{1:\krY })\bigg) 
\notag \\[-1mm] & \rule{50mm}{0mm} \times
\,\intd P_{\rX, \rY}^{(\krX ,\krY )}(\xv_{1:\krX }, \yv_{1:\krY })\,.
\notag 
\ea
The result \eqref{eq:sepmutinf} now follows by recognizing that  
\be
\sum_{\krX \in \N_0} \sum_{\krY \in \N_0}
p_{\card{\rX},\card{\rY}}(\krX ,\krY ) \log 
\bigg( \frac{p_{\card{\rX},\card{\rY}}(\krX ,\krY )}{p_{\card{\rX}}(\krX )p_{\card{\rY}}(\krY )} \bigg)  = I(\card{\rX}; \card{\rY}) \notag 
\ee
and by using \eqref{eq:kldiv} in the remaining terms.

\section{Proof of Lemma~\ref{lem:eqospabcap}} \label{app:eqospabcap}
%
We have to show that $X^*_j$ as defined in \eqref{eq:ospacenterpp}, \eqref{eq:ospacenterasap} satisfies $\sum_{X\in\sA_j}\dist_2(X,X^*_j)\leq \sum_{X\in\sA_j}\dist_2(X,\widetilde{X})$ for all $\widetilde{X}\in \sX_k$.
To this end, we first construct an upper bound on $\sum_{X\in\sA_j}\dist_2(X,X^*_j)$ based on \eqref{eq:ospacenterasap}.
According to \eqref{eq:ospacenterasap},  the collection of permutations $\{\tau^*_{X}\}^{}_{X\in \sA_j}$ satisfies
\ba
& \sum_{i=1}^k\sum_{X\in\sA_j}\sum_{X'\in\sA_j}\frac{1}{2\card{\sA_j}}\bignorm{\xv^{(X)}_{\tau^*_{X}(i)}-\xv^{(X')}_{\tau^*_{X'}(i)}}^2 
\notag \\
& =\min_{\{\tau_{X}\}^{}_{X\in \sA_j}} \sum_{i=1}^k\sum_{X\in\sA_j}\sum_{X'\in\sA_j}\frac{1}{2\card{\sA_j}}\bignorm{\xv^{(X)}_{\tau_{X}(i)}-\xv^{(X')}_{\tau_{X'}(i)}}^2 \,.\label{eq:taustarprop}
\ea
Setting 
\be\label{eq:arithmean}
\bar{\xv}_i \triangleq \frac{1}{\card{\sA_j}}\sum_{X\in \sA_j}\xv^{(X)}_{\tau_{X}(i)} \qquad \text{ for } i\in \{1, \dots, k\}
\ee
 we can rewrite the two inner sums on the right-hand side of \eqref{eq:taustarprop} as
\ba
 &\frac{1}{2\card{\sA_j}}\sum_{X\in\sA_j} \sum_{X'\in\sA_j}
\bignorm{\xv^{(X)}_{\tau_{X}(i)}- \xv^{(X')}_{\tau_{X'}(i)}}^2 
\notag \\
& =
\frac{1}{2\card{\sA_j}}\sum_{X\in\sA_j} \sum_{X'\in\sA_j}
\bignorm{\big(\xv^{(X)}_{\tau_{X}(i)}-\bar{\xv}_i\big) + \big(\bar{\xv}_i- \xv^{(X')}_{\tau_{X'}(i)}\big)}^2 
\notag \\
& = 
\frac{1}{2\card{\sA_j}}\sum_{X'\in\sA_j}\sum_{X\in\sA_j}\bignorm{\xv^{(X)}_{\tau_{X}(i)}-\bar{\xv}_i}^2 
\notag \\*[0mm] & \quad 
+ 
 \frac{1}{\card{\sA_j}}\sum_{X\in\sA_j} \sum_{X'\in\sA_j} \big(\xv^{(X)}_{\tau_{X}(i)}-\bar{\xv}_i\big)^{\trans} \big(\bar{\xv}_i-\xv^{(X')}_{\tau_{X'}(i)}\big) 
\notag \\*
& \quad
 + \frac{1}{2\card{\sA_j}}\sum_{X\in\sA_j} \sum_{X'\in\sA_j}\bignorm{\bar{\xv}_i- \xv^{(X')}_{\tau_{X'}(i)}}^2
\notag \\
& = 
\frac{1}{2}\sum_{X\in\sA_j}\bignorm{\xv^{(X)}_{\tau_{X}(i)}-\bar{\xv}_i}^2 
\notag \\*[0mm] & \quad 
+ \bigg(\hspace{-.5mm}\underbrace{\frac{1}{\card{\sA_j}}\sum_{X\in\sA_j} \hspace{-1mm}\xv^{(X)}_{\tau_{X}(i)}-\bar{\xv}_i}_{=\0v}\bigg)^{\trans} \bigg(\sum_{X'\in\sA_j} \hspace{-1mm}\big(\bar{\xv}_i-\xv^{(X')}_{\tau_{X'}(i)}\big)\bigg) \notag \\*[0mm]
& \quad + \frac{1}{2}\sum_{X'\in\sA_j}\bignorm{\bar{\xv}_i- \xv^{(X')}_{\tau_{X'}(i)}}^2
\notag \\[1mm]
& = 
\sum_{X\in\sA_j}\bignorm{\xv^{(X)}_{\tau_{X}(i)} -\bar{\xv}_i}^2 
\,. \label{eq:sampvarrhs}
\ea
Similarly, using $\xv^*_i=\frac{1}{\card{\sA_j}}\sum_{X\in \sA_j}\xv^{(X)}_{\tau^*_{X}(i)}$ (see~\eqref{eq:ospacenterpp}), we can rewrite the two inner sums on the left-hand side of~\eqref{eq:taustarprop} as
\be
\frac{1}{2\card{\sA_j}}\sum_{X\in\sA_j} \sum_{X'\in\sA_j}
\hspace{-1mm}\bignorm{\xv^{(X)}_{\tau^*_{X}(i)}- \xv^{(X')}_{\tau^*_{X'}(i)}}^2 
 = \hspace{-1mm}
\sum_{X\in\sA_j} \hspace{-1mm}\bignorm{\xv^{(X)}_{\tau^*_{X}(i)} -\xv^*_i}^2 
\,.\label{eq:sampvarlhs}
\ee
Inserting \eqref{eq:sampvarrhs} and \eqref{eq:sampvarlhs} into \eqref{eq:taustarprop}  yields
\be\label{eq:ospaasapstar}
\sum_{i=1}^k\sum_{X\in\sA_j}\hspace{-1mm}\bignorm{\xv^{(X)}_{\tau^*_{X}(i)} -\xv^*_i}^2
=\hspace{-1mm}
\min_{\{\tau_{X}\}^{}_{X\in \sA_j}} \sum_{i=1}^k\sum_{X\in\sA_j}\hspace{-1mm}\bignorm{\xv^{(X)}_{\tau_{X}(i)} -\bar{\xv}_i}^2 \,.
\ee
Let us recall \eqref{eq:ospafc} in our setting, i.e., 
\be \notag 
\dist_2(X,X^*_j)= \min_{\tau_{X}}\sum_{i=1}^k\bignorm{\xv^{(X)}_{\tau_{X}(i)}-\xv^*_i}^2
\ee
and thus 
\ba
	\sum_{X\in\sA_j}\dist_2(X,X^*_j)
	& = \sum_{X\in\sA_j}\min_{\tau_{X}}\sum_{i=1}^k\bignorm{\xv^{(X)}_{\tau_{X}(i)}-\xv^*_i}^2 
	\notag \\
	& \leq \sum_{X\in\sA_j} \sum_{i=1}^k\bignorm{\xv^{(X)}_{\tau^*_{X}(i)}-\xv^*_i}^2
	\notag \\
	& \stackrel{\hidewidth \eqref{eq:ospaasapstar} \hidewidth}= 
	\min_{\{\tau_{X}\}^{}_{X\in \sA_j}}\sum_{X\in\sA_j} \sum_{i=1}^k\bignorm{\xv^{(X)}_{\tau_{X}(i)} -\bar{\xv}_i}^2 \,. \label{eq:distxxstarbound}
\ea
We next  want to relate the upper bound \eqref{eq:distxxstarbound} on $\sum_{X\in\sA_j}\dist_2(X,X^*_j)$ to $\sum_{X\in\sA_j}\dist_2(X,\widetilde{X})$ for an arbitrary $\widetilde{ X}=\{\widetilde{\xv}_1, \dots, \widetilde{\xv}_k\}\in \sX_k$.
For any permutations $\{\tau_{X}\}_{X\in \sA_j}$ and vector $\widetilde{\xv}_{1:k}\in (\R^d)^k$, the sum 
$
\sum_{X\in\sA_j}\sum_{i=1}^k\bignorm{\xv^{(X)}_{\tau_{X}(i)}-\widetilde{\xv}_i}^2
$
is a sum of squared-error distortions of $kd$-dimensional vectors.
The minimum of that sum with respect to  $\widetilde{\xv}_{1:k}$ is easily seen to be achieved by the arithmetic mean of $\big\{\big(\xv^{(X)}_{\tau_{X}(1)}, \dots,  \xv^{(X)}_{\tau_{X}(k)}\big)\big\}_{X\in \sA_j}$, i.e., by $\bar{\xv}_{1:k}$ (see~\eqref{eq:arithmean}).
Thus, we obtain  for any $\widetilde{\xv}_{1:k}\in (\R^d)^k$
\be\label{eq:sumospafcexchanged3}
	 \sum_{X\in\sA_j}\sum_{i=1}^k\bignorm{\xv^{(X)}_{\tau_{X}(i)} -\bar{\xv}_i}^2
	\leq \sum_{X\in\sA_j}\sum_{i=1}^k\bignorm{\xv^{(X)}_{\tau_{X}(i)}-\widetilde{\xv}_i}^2\,.
\ee
Using \eqref{eq:sumospafcexchanged3} in  \eqref{eq:distxxstarbound}, we have for any $\widetilde{\xv}_{1:k}\in (\R^d)^k$
\be\label{eq:sumospafcexchanged2}
	\sum_{X\in\sA_j}\dist_2(X,X^*_j)
	\leq \min_{\{\tau_{X}\}_{X\in \sA_j}} \sum_{X\in\sA_j}\sum_{i=1}^k\bignorm{\xv^{(X)}_{\tau_{X}(i)}-\widetilde{\xv}_i}^2\,.
\ee
Because each summand $\sum_{i=1}^k\bignorm{\xv^{(X)}_{\tau_{X}(i)}-\widetilde{\xv}_i}^2$ on the right-hand side of \eqref{eq:sumospafcexchanged2} depends only on one permutation $\tau_{X}$, we can exchange the outer sum and the minimization and obtain
\be\label{eq:sumospafcexchanged1}
	\sum_{X\in\sA_j}\dist_2(X,X^*_j)
	\leq \sum_{X\in\sA_j} \min_{\tau_{X}}\sum_{i=1}^k\bignorm{\xv^{(X)}_{\tau_{X}(i)}-\widetilde{\xv}_i}^2\,.
\ee
According to \eqref{eq:ospafc}, the right-hand side of \eqref{eq:sumospafcexchanged1} is equal to $\sum_{X\in\sA_j}\dist_2(X,\widetilde{X})$ for $\widetilde{X}=\{\widetilde{\xv}_1, \dots, \widetilde{\xv}_k\}$.
Thus, we have
$
\sum_{X\in\sA_j}\dist_2(X,X^*_j)
	\leq \sum_{X\in\sA_j}\dist_2(X,\widetilde{X})
	$
for any set $\widetilde{X}\in \sX_k$.
This proves that 
$X^*_j=\argmin_{\widetilde{X}\in \sX_k} \sum_{X\in\sA_j}\dist_2(X,\widetilde{X})$.

\section{Bound on $H\big( \rt^{(k)}_{\ryv} \bcondi \phi_k(\ryv^{(k)}), \rxv^{(k)}\big)$ in \eqref{eq:boundhtcondphiyx} } \label{app:boundent}
%
We recall from Section~\ref{sec:gaussdist} that $\rxv^{(k)}=\ryv^{(k)}+ \rwv^{(k)}$, where $\rwv^{(k)}$ has i.i.d.\ zero-mean Gaussian entries with variance $\sigma^2<1$, $\ryv^{(k)}$ has i.i.d.\ zero-mean Gaussian entries with variance $1-\sigma^2$, and $\ryv^{(k)}$ and $\rwv^{(k)}$ are independent.
We now have
\ba
 & H\big( \rt^{(k)}_{\ryv}\bcondi \phi_k(\ryv^{(k)}), \rxv^{(k)}\big) \notag \\
& =\E_{\phi_k(\ryv^{(k)}), \rxv^{(k)}}
\bigg[
 {-} \sum_{\tau} p_{\rt^{(k)}_{\ryv}|\phi_k(\ryv^{(k)}), \rxv^{(k)}}\big(\tau \bcondi \phi_k(\ryv^{(k)}), \rxv^{(k)}\big) 
\notag \\* & \quad \times  
\log p_{\rt^{(k)}_{\ryv}|\phi_k(\ryv^{(k)}), \rxv^{(k)}}\big(\tau \bcondi \phi_k(\ryv^{(k)}), \rxv^{(k)}\big)
\bigg] \notag \\
& =\int_{(\R^{d})^k} f_{\ryv^{(k)}}(\yv_{1:k}) \int_{(\R^{d})^k} f_{\rxv^{(k)}|\phi_k(\ryv^{(k)})}( \xv_{1:k} \condi \phi_k(\yv_{1:k})) 
\notag \\* & \quad \times  
\bigg({-}\sum_{\tau}  p_{\rt^{(k)}_{\ryv}|\phi_k(\ryv^{(k)}), \rxv^{(k)}}\big( \tau \bcondi \phi_k(\yv_{1:k}), \xv_{1:k}\big) 
\notag \\* & \quad \times  
\log p_{\rt^{(k)}_{\ryv}|\phi_k(\ryv^{(k)}), \rxv^{(k)}}\big( \tau \bcondi \phi_k(\yv_{1:k}), \xv_{1:k}\big) \bigg) \, \intd \xv_{1:k}  \, \intd \yv_{1:k} \,.  \label{eq:mycent1}
\ea
Using  Bayes' rule and  the law of total probability, we 
obtain
\ba
& p_{\rt^{(k)}_{\ryv}|\phi_k(\ryv^{(k)}), \rxv^{(k)}}(\tau \condi Y, \xv_{1:k})
\notag \\
& \quad  = \frac%
{p_{\rt^{(k)}_{\ryv}| \phi_k(\ryv^{(k)})}(\tau\condi Y) f_{\rxv^{(k)}| \rt^{(k)}_{\ryv}, \phi_k(\ryv^{(k)})}( \xv_{1:k} \condi \tau, Y)}%
{ f_{\rxv^{(k)}| \phi_k(\ryv^{(k)})}( \xv_{1:k} \condi   Y)}  \notag \\
& \quad    \stackrel{\hidewidth(a)\hidewidth}= \frac%
{p_{\rt^{(k)}_{\ryv}}(\tau) f_{\rxv^{(k)}| \rt^{(k)}_{\ryv}, \phi_k(\ryv^{(k)})}( \xv_{1:k} \condi \tau, Y)}%
{ \sum_{\tau'} p_{\rt^{(k)}_{\ryv}}(\tau') f_{\rxv^{(k)}| \rt^{(k)}_{\ryv}, \phi_k(\ryv^{(k)})}( \xv_{1:k} \condi \tau', Y)}   \notag \\
& \quad  \stackrel{\hidewidth(b)\hidewidth}= \frac%
{  f_{\rxv^{(k)}| \rt^{(k)}_{\ryv}, \phi_k(\ryv^{(k)})}( \xv_{1:k} \condi \tau, Y)}%
{ \sum_{\tau'}  f_{\rxv^{(k)}| \rt^{(k)}_{\ryv}, \phi_k(\ryv^{(k)})}( \xv_{1:k} \condi \tau', Y)}
\label{eq:proptcond}
\ea
where 
$(a)$ holds because, as discussed in Section~\ref{sec:gaussdist}, $\rt^{(k)}_{\ryv}$ is independent of $\phi_k(\ryv^{(k)})$ and 
$(b)$ holds because $p_{\rt^{(k)}_{\ryv}}(\tau)=1/k!$ for all $\tau$.
Recalling that $\ryv^{(k)}$ can be equivalently represented by $\rt^{(k)}_{\ryv}$ and $\phi_k(\ryv^{(k)})$, we have
\ba 
f_{\rxv^{(k)}| \rt^{(k)}_{\ryv}, \phi_k(\ryv^{(k)})}( \xv_{1:k} \condi \tau, Y)
& = f_{\rxv^{(k)}| \ryv^{(k)}}( \xv_{1:k} \condi \tau(Y)) \notag \\
& = f_{\rwv^{(k)}}( \xv_{1:k} - \tau(Y)) \,. \label{eq:fxcondty}
\ea
Inserting \eqref{eq:fxcondty} 
into \eqref{eq:proptcond}, we obtain 
\be \label{eq:ptcondyx}
p_{\rt^{(k)}_{\ryv}|\phi_k(\ryv^{(k)}), \rxv^{(k)}}(\tau \condi Y, \xv_{1:k})
= \frac%
{  f_{\rwv^{(k)}}( \xv_{1:k} - \tau(Y))}%
{ \sum_{\tau'}  f_{\rwv^{(k)}}( \xv_{1:k} - \tau'(Y))}\,.
\ee
Furthermore, we have
\ba
f_{\rxv^{(k)}|\phi_k(\ryv^{(k)})}( \xv_{1:k} \condi Y)
& = \frac{1}{k!} \sum_{\tilde{\tau}} f_{\rxv^{(k)}| \rt^{(k)}_{\ryv}, \phi_k(\ryv^{(k)})}( \xv_{1:k} \condi \tilde{\tau}, Y) 
\notag \\
& = \frac{1}{k!} \sum_{\tilde{\tau}} f_{\rwv^{(k)}}( \xv_{1:k} - \tilde{\tau}(Y))\,.\label{eq:fxcondy}
\ea

Inserting \eqref{eq:ptcondyx} and \eqref{eq:fxcondy} into \eqref{eq:mycent1}, we obtain
\ba
 &H\big( \rt^{(k)}_{\ryv} \bcondi \phi_k(\ryv^{(k)}), \rxv^{(k)}\big)  \notag \\
& = \int_{(\R^{d})^k} f_{\ryv^{(k)}}(\yv_{1:k}) 
\notag \\* & \quad \times  
\int_{(\R^{d})^k} \frac{1}{k!} \bigg(\sum_{\tilde{\tau}} f_{\rwv^{(k)}}( \xv_{1:k} - \tilde{\tau}(\phi_k(\yv_{1:k}))) \bigg) 
\notag \\* & \quad \times  
\bigg( {-}\sum_{\tau}  \frac%
{  f_{\rwv^{(k)}}( \xv_{1:k} - \tau(\phi_k(\yv_{1:k})))}%
{ \sum_{\tau'}  f_{\rwv^{(k)}}( \xv_{1:k} - \tau'(\phi_k(\yv_{1:k})))} 
\notag \\* & \quad \times  
\log \bigg( \frac%
{  f_{\rwv^{(k)}}( \xv_{1:k} - \tau(\phi_k(\yv_{1:k})))}%
{ \sum_{\tau'}  f_{\rwv^{(k)}}( \xv_{1:k} - \tau'(\phi_k(\yv_{1:k})))} \bigg) \bigg) \, \intd \xv_{1:k} \, \intd \yv_{1:k}  \notag \\[1mm]
& \stackrel{\hidewidth(a)\hidewidth}= \sum_{\tilde{\tau}} \frac{1}{k!} \int_{(\R^{d})^k} f_{\ryv^{(k)}}(\yv_{1:k}) \int_{(\R^{d})^k} f_{\rwv^{(k)}}( \xv_{1:k} - \tilde{\tau}(\yv_{1:k})) 
\notag \\* 
& \quad \times  \bigg({-} \sum_{\tau}    \frac%
{  f_{\rwv^{(k)}}( \xv_{1:k} - \tau(\yv_{1:k}))}%
{ \sum_{\tau'}  f_{\rwv^{(k)}}( \xv_{1:k} - \tau'(\yv_{1:k}))} 
\notag \\* & \quad \times  
\log \bigg( \frac%
{  f_{\rwv^{(k)}}( \xv_{1:k} - \tau(\yv_{1:k}))}%
{ \sum_{\tau'}  f_{\rwv^{(k)}}( \xv_{1:k} - \tau'(\yv_{1:k}))} \bigg) \bigg) \, \intd \xv_{1:k} \, \intd \yv_{1:k}  \notag \\[1mm]
& \stackrel{\hidewidth(b)\hidewidth}= \sum_{\tilde{\tau}} \frac{1}{k!} \int_{(\R^{d})^k} f_{\ryv^{(k)}}(\yv_{1:k}) \int_{(\R^{d})^k} f_{\rwv^{(k)}}( \wv_{1:k})  
\notag \\* & \quad \times  
 \bigg( {-}\sum_{\tau}   
 \frac%
{  f_{\rwv^{(k)}}( \wv_{1:k} + \tilde{\tau}(\yv_{1:k}) - \tau(\yv_{1:k}))}%
{ \sum_{\tau'}  f_{\rwv^{(k)}}( \wv_{1:k} + \tilde{\tau}(\yv_{1:k})  - \tau'(\yv_{1:k}))} 
 \notag \\* & \quad \times 
\log \bigg( \frac%
{  f_{\rwv^{(k)}}( \wv_{1:k} + \tilde{\tau}(\yv_{1:k})  - \tau(\yv_{1:k}))}%
{ \sum_{\tau'}  f_{\rwv^{(k)}}( \wv_{1:k} + \tilde{\tau}(\yv_{1:k})  - \tau'(\yv_{1:k}))} \bigg) \bigg) 
\notag \\* & \hspace{60mm} \times  
\, \intd \wv_{1:k} \, \intd \yv_{1:k}  \notag \\[-1mm]
& \stackrel{\hidewidth(c)\hidewidth}= \int_{(\R^{d})^k} f_{\ryv^{(k)}}(\yvt_{1:k}) \int_{(\R^{d})^k} f_{\rwv^{(k)}}( \wv_{1:k})  
\notag \\* & \quad \times  
 \bigg( {-}\sum_{\tau}    \frac%
{  f_{\rwv^{(k)}}( \wv_{1:k} +  \yvt_{1:k}  - \tau(\yvt_{1:k}))}%
{ \sum_{\tau'}  f_{\rwv^{(k)}}( \wv_{1:k} +  \yvt_{1:k} - \tau'(\yvt_{1:k}))} 
 \notag \\* & \quad \times 
\log \hspace{-.3mm}\bigg( \frac%
{  f_{\rwv^{(k)}}( \wv_{1:k} + \yvt_{1:k}  - \tau(\yvt_{1:k}))}%
{ \sum_{\tau'}  f_{\rwv^{(k)}}( \wv_{1:k} +  \yvt_{1:k}  - \tau'(\yvt_{1:k}))} \bigg) \hspace{-.3mm}\bigg)\hspace{-.3mm} \, \intd \wv_{1:k} \, \intd \yvt_{1:k} \label{eq:ourcondent}
\ea
where in $(a)$ 
we used that summation over all orderings of the elements of the set $\phi_k(\yv_{1:k})$ is the same as summation over all permutations of the subvectors of the  vector $\yv_{1:k}$, 
in $(b)$ we used the substitution $\wv_{1:k}=\xv_{1:k} - \tilde{\tau}(\yv_{1:k})$, 
and $(c)$ holds by substituting $\yvt_{1:k}=\tilde{\tau}(\yv_{1:k})$ and noting that $f_{\ryv^{(k)}}(\tilde{\tau}^{-1}(\yvt_{1:k}))=f_{\ryv^{(k)}}(\yvt_{1:k})$ and   that due to the summation over all permutations $\tau$ we can omit the additional permutation  $\tilde{\tau}^{-1}$.
The right-hand side in \eqref{eq:ourcondent} is a Gaussian expectation over the entropy of a discrete random variable $\rt$---depending on $\wv_{1:k}$ and $\yvt_{1:k}$---with $k!$ possible realizations and probability mass function 
\be \label{eq:ptdef}
p_{\rt}(\tau; \wv_{1:k}, \yvt_{1:k}) = \frac%
{  f_{\rwv^{(k)}}( \wv_{1:k} + \yvt_{1:k}  - \tau(\yvt_{1:k}))}%
{ \sum_{\tau'}  f_{\rwv^{(k)}}( \wv_{1:k} +  \yvt_{1:k}  - \tau'(\yvt_{1:k}))}\,.
\ee
Thus, \eqref{eq:ourcondent} can be rewritten as
\ba
&\! H\big( \rt^{(k)}_{\ryv} \bcondi \phi_k(\ryv^{(k)}), \rxv^{(k)}\big)
\notag \\ 
&\! =  \int_{(\R^{d})^k} f_{\ryv^{(k)}}(\yvt_{1:k}) \int_{(\R^{d})^k} f_{\rwv^{(k)}}( \wv_{1:k})  \,
 H(p_{\rt}(\,\cdot\,; \wv_{1:k}, \yvt_{1:k})) 
\notag \\* & \rule{53mm}{0mm} \times  
\, \intd \wv_{1:k} \, \intd \yvt_{1:k}\,. \label{eq:ourcondentred}
\ea
We next split the domains of integration in \eqref{eq:ourcondentred}: for $\wv_{1:k}$ into 
$\sW_{\varepsilon}\triangleq \{\wv_{1:k}\in (\R^{d})^k: \norm{\wv_{1:k}}<\varepsilon\}$ and $\sW_{\varepsilon}^c$ with $\varepsilon>0$, 
and for $\yvt_{1:k}$ into  $\sY_{\delta}\triangleq \{\yvt_{1:k}\in (\R^{d})^k: \norm{\yvt_i-\yvt_j}>\delta \text{ for all } i\neq j\}$ and $\sY_{\delta}^c$ with $\delta>0$.
Using $H(p_{\rt}(\,\cdot\,; \wv_{1:k}, \yvt_{1:k}))
\leq \log k!$ and $(\R^d)^k \times(\R^d)^k= \big(\sY_{\delta}^c\times (\R^d)^k\big) \cup \big((\R^d)^k \times \sW_{\varepsilon}^c \big) \cup \big(\sY_{\delta} \times \sW_{\varepsilon}\big)$, this leads to the following bound:
\ba
 & H\big(\rt^{(k)}_{\ryv} \bcondi \phi_k(\ryv^{(k)}), \rxv^{(k)}\big) \notag \\
& \leq \int_{\sY_{\delta}^c} f_{\ryv^{(k)}}(\yvt_{1:k}) \int_{(\R^d)^k} f_{\rwv^{(k)}}( \wv_{1:k})  
 \log k! \, \intd \wv_{1:k} \, \intd \yvt_{1:k} 
\notag \\* & \quad +
\int_{(\R^d)^k} f_{\ryv^{(k)}}(\yvt_{1:k}) \int_{\sW_{\varepsilon}^c} f_{\rwv^{(k)}}( \wv_{1:k})  
 \log k! \, \intd \wv_{1:k} \, \intd \yvt_{1:k} 
\notag \\* & \quad +
\int_{\sY_{\delta}} f_{\ryv^{(k)}}(\yvt_{1:k}) \int_{\sW_{\varepsilon}} f_{\rwv^{(k)}}( \wv_{1:k})\,  
H(p_{\rt}(\,\cdot\,; \wv_{1:k}, \yvt_{1:k}))
\notag \\*[-2mm] & \hspace{63mm} \times  
\, \intd \wv_{1:k} \, \intd \yvt_{1:k} 
\notag \\
& \leq \bigg(\int_{\sY_{\delta}^c} f_{\ryv^{(k)}}(\yvt_{1:k}) \, \intd\yvt_{1:k} 
+ \int_{\sW_{\varepsilon}^c} f_{\rwv^{(k)}}( \wv_{1:k}) \, \intd \wv_{1:k} \bigg)
\log k!  
\notag \\* & \quad 
+ \sup_{\wv_{1:k}\in \sW_{\varepsilon},\, \yvt_{1:k}\in \sY_{\delta}}
H(p_{\rt}(\,\cdot\,; \wv_{1:k}, \yvt_{1:k}))\,.
\label{eq:ourcondentbound}
\ea

Now we bound successively the terms on the right-hand side of \eqref{eq:ourcondentbound}.
For the first term, we have
\ba
\int_{\sY_{\delta}^c} f_{\ryv^{(k)}}& (\yvt_{1:k}) \, \intd\yvt_{1:k}
\notag \\
& = \Pr[\ryv^{(k)}\in \sY_{\delta}^c] \notag \\
& = \Pr\Big[\min_{i\neq j} \bignorm{\ryv^{(k)}_i-\ryv^{(k)}_j}\leq \delta\Big] \notag \\
& \stackrel{\hidewidth(a)\hidewidth}\leq \frac{k(k-1)}{2} \Pr\big[\bignorm{\ryv^{(k)}_1-\ryv^{(k)}_2}\leq \delta\big] \notag \\
& =  \frac{k(k-1)}{2} \Pr\bigg[\frac{\norm{\ryv^{(k)}_1-\ryv^{(k)}_2}^2}{2(1-\sigma^2)} \leq \frac{\delta^2}{2(1-\sigma^2)}\bigg] \notag \\
& \stackrel{(b)}= \frac{k(k-1)}{2} F_{\chi^2}\bigg(\frac{\delta^2}{2(1-\sigma^2)}; d\bigg)
\label{eq:boundydeltac}
\ea
where $(a)$ holds by the union bound for the $\frac{k(k-1)}{2}$ events $\big\{\bignorm{\ryv^{(k)}_i-\ryv^{(k)}_j}\leq \delta\big\}$, $i< j$ and $(b)$ holds because $\ryv^{(k)}_i$ is Gaussian with zero mean and variance $1-\sigma^2$ and thus $\bignorm{\ryv^{(k)}_1-\ryv^{(k)}_2}^2/(2(1-\sigma^2))$ is  $\chi^2$ distributed with $d$ degrees of freedom.
Similarly, for the second term, we 
have
\ba
 \int_{\sW_{\varepsilon}^c} f_{\rwv^{(k)}}( \wv_{1:k}) \, \intd \wv_{1:k}
& 
= \Pr[\norm{\rwv^{(k)}}\geq \varepsilon] 
\notag \\ & 
= \Pr\bigg[\frac{\norm{\rwv^{(k)}}^2}{ \sigma^2 } \geq \frac{\varepsilon^2}{ \sigma^2 }\bigg]
 \notag \\ & 
= 1-F_{\chi^2}\bigg( \frac{\varepsilon^2}{ \sigma^2 }; kd\bigg)
\label{eq:boundwepsc}
\ea
where 
we used the fact that
$\norm{\rwv^{(k)}}^2 / \sigma^2$ is $\chi^2$ distributed with $kd$ degrees of freedom.

To bound the third term, i.e., $H(p_{\rt}(\,\cdot\,; \wv_{1:k}, \yvt_{1:k}))$ for $\wv_{1:k}\in \sW_{\varepsilon}$ and $\yvt_{1:k}\in \sY_{\delta}$, we first bound the probability 
$p_{\rt}(\tau; \wv_{1:k}, \yvt_{1:k})$ in \eqref{eq:ptdef} for $\tau$ equal to the identity permutation, denoted $\iota$, i.e., 
\ba
& p_{\rt}(\iota; \wv_{1:k}, \yvt_{1:k})
\notag \\
& = \frac{f_{\rwv^{(k)}}( \wv_{1:k})}{f_{\rwv^{(k)}}( \wv_{1:k}) + \sum_{\tau'\neq \iota}  f_{\rwv^{(k)}}( \wv_{1:k} +  \yvt_{1:k} - \tau'(\yvt_{1:k}))}\,. 
\label{eq:ptiota}
\\*[-8mm] \notag 
\ea
For $\wv_{1:k}\in \sW_{\varepsilon}$, we have
\ba
f_{\rwv^{(k)}}( \wv_{1:k})
& = \frac{1}{(2\pi\sigma^2)^{kd/2}}\exp\bigg({-}\frac{\norm{\wv_{1:k}}^2}{2\sigma^{2}}\bigg)
\notag \\
& > \frac{1}{(2\pi\sigma^2)^{kd/2}}\exp\bigg({-}\frac{\varepsilon^2}{2\sigma^{2}}\bigg)
\label{eq:fww}
\ea
and, if additionally $\yvt_{1:k}\in \sY_{\delta}$ with $\delta>\varepsilon$, we have for $\tau'\neq \iota$
\ba
& f_{\rwv^{(k)}}( \wv_{1:k} +  \yvt_{1:k}  - \tau'(\yvt_{1:k}))
\notag \\
& = \frac{1}{(2\pi\sigma^2)^{kd/2}}\exp\bigg({-}\frac{\norm{ \wv_{1:k} +  \yvt_{1:k}  - \tau'(\yvt_{1:k})}^2}{2\sigma^{2}}\bigg) \notag \\
& \stackrel{\hidewidth(a)\hidewidth}\leq \frac{1}{(2\pi\sigma^2)^{kd/2}}\exp\bigg({-}\frac{(\norm{ \wv_{1:k}} -  \norm{\yvt_{1:k}  - \tau'(\yvt_{1:k})})^2}{2\sigma^{2}}\bigg) \notag \\
&  \stackrel{\hidewidth(b)\hidewidth}\leq \frac{1}{(2\pi\sigma^2)^{kd/2}}\exp\bigg({-}\frac{(\delta-\varepsilon)^2}{2\sigma^{2}}\bigg)\label{eq:fwwyty}
\ea
where $(a)$ holds by the reverse triangle inequality and $(b)$ holds because the difference between $\norm{\yvt_{1:k}  - \tau'(\yvt_{1:k})}$ and $\norm{ \wv_{1:k}}$ is larger than $\delta-\varepsilon$ due to $\norm{ \wv_{1:k}}<\varepsilon$ and $\norm{\yvt_{1:k}  - \tau'(\yvt_{1:k})}\geq \norm{\yvt_i  - \yvt_j}>\delta>\varepsilon$ for some $i,j$ with $i\neq j$.
We specifically choose $\delta=3\varepsilon$, for which \eqref{eq:fwwyty} yields 
$
f_{\rwv^{(k)}}( \wv_{1:k} +  \yvt_{1:k}  - \tau'(\yvt_{1:k}))
\leq \frac{1}{(2\pi\sigma^2)^{kd/2}}\exp\big({-}\frac{(2\varepsilon)^2}{2\sigma^{2}}\big)
$.
Inserting this bound and the bound \eqref{eq:fww} into  \eqref{eq:ptiota}, we obtain
\ba
p_{\rt}(\iota; \wv_{1:k}, \yvt_{1:k})
& \stackrel{\hidewidth(a)\hidewidth}\geq 
\frac%
{  \exp\big({-}\frac{\varepsilon^2}{2\sigma^{2}}\big)}%
{  \exp\big({-}\frac{\varepsilon^2}{2\sigma^{2}}\big) + (k!-1)  \exp\big({-}\frac{(2\varepsilon)^2}{2\sigma^{2}}\big)} 
\notag \\ & 
= \frac%
{ 1}%
{ 1 + (k!-1)  \exp\big({-}\frac{3\varepsilon^2}{2\sigma^{2}}\big)}
 \notag \\ & 
=: p_0(\varepsilon) \notag
\ea
where in $(a)$ we used that $\overline{a}/(\overline{a}+\underline{b})\geq a/(a+b)$ for $\overline{a}\geq a$ and $\underline{b} \leq b$.
Thus, we bounded the probability that $\rt=\iota$ (namely, $p_{\rt}(\iota; \wv_{1:k}, \yvt_{1:k})$) from below.
By the variation of Fano's inequality presented in  \cite[eq.~(2.143)]{Cover91}, this implies the following bound on the entropy:
\ba
H(p_{\rt}(\,\cdot\,; \wv_{1:k}, \yvt_{1:k}))
& \leq H_2(p_0(\varepsilon)) + (1-p_0(\varepsilon))\log (k!-1)\,.
\label{eq:boundenttau}
\ea
Finally, 
inserting \eqref{eq:boundydeltac}, \eqref{eq:boundwepsc}, and \eqref{eq:boundenttau} into \eqref{eq:ourcondentbound}, we obtain~\eqref{eq:boundhtcondphiyx}.

\section{Proof of Lemma~\ref{lem:usospabounds}} \label{app:proofusospabounds}
%
\textit{Case  $\kX  \geq \kY $:} 
According to \eqref{eq:ospagen}, we have 
\be\label{eq:concdist2}
\dist_2^{  (c) }(X,Y)= (\kX -\kY )\,c^2 + \sum_{i=1}^{\kY }\min\big\{\bignorm{\xv_{\tau_{X,Y}(i)}-\yv_i}^2, c^2\big\}
\ee
for some  permutation $\tau_{X,Y}$.
Representing the product $(\kX -\kY )\,c^2$ as a sum, we can rewrite \eqref{eq:concdist2} 
as
\be\label{eq:concdistexpandsum}
\dist_2^{  (c) }(X,Y)= \sum_{i=1}^{\kY }\min\big\{\bignorm{\xv_{\tau_{X,Y}(i)}-\yv_i}^2, c^2\big\} +
\sum_{i=\kY +1}^{\kX }c^2\,.
\ee
We proceed by bounding each summand in \eqref{eq:concdistexpandsum} for $i\in \{1, \dots, \kX \}$.
For $i\in \{1, \dots, \kY \}$, we have 
\ba
& \min\big\{\bignorm{\xv_{\tau_{X,Y}(i)} -\yv_i}^2, c^2\big\}
\notag \\
& \qquad \geq \min_{j=1}^{\kY }\min\big\{\bignorm{\xv_{\tau_{X,Y}(i)}-\yv_j}^2, c^2\big\}\,.
\label{eq:boundmimintau}
\ea
For the remaining $i\in \{\kY +1, \dots, \kX \}$, we have trivially
\be\label{eq:boundc2}
c^2
\geq \min_{j=1}^{\kY }\min\big\{\bignorm{\xv_{\tau_{X,Y}(i)}-\yv_j}^2, c^2\big\}\,.
\ee
Inserting \eqref{eq:boundmimintau} and \eqref{eq:boundc2} into \eqref{eq:concdistexpandsum}, we obtain
\ba
\dist_2^{  (c) }(X,Y) 
& \geq \sum_{i=1}^{\kY }\min_{j=1}^{\kY }\min\big\{\bignorm{\xv_{\tau_{X,Y}(i)}-\yv_j}^2, c^2\big\} 
\notag \\* & \quad
+
\sum_{i=\kY +1}^{\kX }\min_{j=1}^{\kY }\min\big\{\bignorm{\xv_{\tau_{X,Y}(i)}-\yv_j}^2, c^2\big\} \notag \\
& = \sum_{i=1}^{\kX }\min_{j=1}^{\kY }\min\big\{\bignorm{\xv_{\tau_{X,Y}(i)}-\yv_j}^2, c^2\big\} \notag  
\ea
which, due to the bijectivity of $\tau_{X,Y}$,  is equivalent to \eqref{eq:bounddist2a}.

\textit{Case  $\kX  \leq \kY $:} 
Inserting $\norm{\xv_i-\yv_{\tau(i)}}^2\geq \min_{j=1}^{\kY }\norm{\xv_i-\yv_j}^2$ into 
\eqref{eq:ospagen} yields
\be
\dist_2^{  (c) }(X,Y)  
\geq (\kY -\kX )\,c^2  + \sum_{i=1}^{\kX }\min\Big\{\min_{j=1}^{\kY }\,\norm{\xv_i-\yv_j}^2, c^2\Big\} \notag 
\ee
which is equivalent to \eqref{eq:bounddist2b}.

\section{Proof of Theorem~\ref{th:ppplowerbound}} \label{app:proofppplowerbound}
%
%
According to  \eqref{eq:denspppk} and the discussion preceding it,  the probability measures $P_{\rX}^{(k)}$ are  absolutely continuous with respect to $(\Leb^d)^k$ with probability density function $f^{(k)}_{\rX}(\xv_{1:k})=\prod_{i=1}^k g_{\rX}(\xv_i)$.
Furthermore, by the assumption $g_{\rX}(\xv)=0$ for $\Leb^{d}$-almost all $\xv\in A^c$, we obtain 
$f^{(k)}_{\rX}(\xv_{1:k})=0$  for $(\Leb^{d})^k$-almost all $\xv_{1:k}\in (A^k)^c$.
Hence, the conditions in Theorem~\ref{th:shlbpp} are satisfied,
and we can rewrite the bound \eqref{eq:rdlbac} as
\ba
 R(D)
& \stackrel{\hidewidth (a) \hidewidth}\geq
  \sum_{k\in \N} \frac{e^{-\expcard}\expcard^k}{k!} h\bigg(\prod_{i=1}^k g_{\rX}(\xv_i)\bigg) 
	\notag \\* & \quad 
	+ \max_{s\geq 0} \bigg({-}\sum_{k\in \N_0} \frac{e^{-\expcard}\expcard^k}{k!}\log \gamma_k(s) -sD\bigg) \notag \\
& =  \sum_{k\in \N}\frac{e^{-\expcard}\expcard^k}{k!} k\, h(g_{\rX}) 
\notag \\* & \quad 
+ \max_{s\geq 0} \bigg({-}\sum_{k\in \N_0} \frac{e^{-\expcard}\expcard^k}{k!}\log \gamma_k(s)-sD \bigg) 
\notag \\ 
& =\expcard \,h(g_{\rX}) + \max_{s\geq 0} \bigg({-}\sum_{k\in \N_0} \frac{e^{-\expcard}\expcard^k}{k!}\log \gamma_k(s)-sD \bigg) \label{eq:rdlbppppr}
\ea
where  $(a)$ holds due to \eqref{eq:pppcard}.
The functions $\gamma_k(s)$ in \eqref{eq:rdlbppppr} have to satisfy%
\footnote{For $k=0$, \eqref{eq:choosegamma} gives $\gamma_0(s)\geq e^{-s \dist_2^{  (c) }(\emptyset,Y)}$.
If  $Y=\emptyset$, this simplifies to  $\gamma_0(s)\geq 1$
because $\dist_2^{  (c) }(\emptyset, \emptyset)=0$. For all other $Y\in \sX$, we trivially have $e^{-s \dist_2^{  (c) }(\emptyset,Y)}\leq 1$. 
Hence, $\gamma_0(s)\geq 1$ is equivalent to $\gamma_0(s)\geq e^{-s \dist_2^{  (c) }(\emptyset,Y)}$ for all $Y \in \sX$.}
 (see~\eqref{eq:choosegamma})
\be\label{eq:choosegammappp}
\gamma_k(s)\geq 
\begin{cases}
1 & \text{ if } k=0 \\
 \int_{A^k} e^{-s \dist_2^{  (c) }(\phi_k(\xv_{1:k}),Y)}\, \mathrm{d}
\xv_{1:k}
& \text{ if } k\in \N 
\end{cases}
\ee
for all $Y\in \sX$.
The constant functions $\gamma_k(s)= (\Leb^d(A))^k$  satisfy \eqref{eq:choosegammappp} because
\ba
(\Leb^d(A))^k
& = (\Leb^{d})^k (A^k)
= \int_{A^k} 1 \, \mathrm{d}
\xv_{1:k}
\notag \\ & 
\geq \int_{A^k} e^{-s \dist_2^{  (c) }(\phi_k(\xv_{1:k}),Y)}\, \mathrm{d}
\xv_{1:k}\label{eq:trivgammahold}
\ea
for $k\in \N$ and $\gamma_0(s)=1$ for $k=0$.
The following lemma, proved further below, states that also the functions $\gammat_k$ defined in \eqref{eq:mygammappp} satisfy \eqref{eq:choosegammappp}.
\begin{lemma}\label{lem:gammatsat}
Let $A\subseteq \R^d$ be  a Borel set and $s\geq 1/c^2$. 
Then \eqref{eq:choosegammappp} holds for 
\ba
& \gamma_k(s) 
= \gammat_k(s) 
\notag \\
& =\hspace{-.5mm} \bigg(e^{-s c^2} \Leb^{d}(A)
+k\bigg(\hspace{-.5mm}{-}e^{-s c^2}\Leb^d(U_c) + \hspace{-.5mm}\int_{U_c}\hspace{-.5mm} e^{-s  \norm{\xv}^2}  \mathrm{d}
\xv\hspace{-.5mm}
\bigg) \hspace{-.5mm}\bigg)^k \hspace{-.5mm}. \label{eq:mygammappp2}
\ea
\end{lemma}
 By Lemma~\ref{lem:gammatsat} and \eqref{eq:trivgammahold}, we have that \eqref{eq:rdlbppppr} holds with $\gamma_k(s)=\min\{(\Leb^d(A))^k, \gammat_k(s)\}$ if we ad\-di\-tion\-al\-ly restrict the maximization to $s\geq 1/c^2$.
With these modifications, \eqref{eq:rdlbppppr} is equal to \eqref{eq:rdlbppp} up to the summand for $k=0$, namely $e^{-\expcard}\log (\min\{(\Leb^d(A))^0, \gammat_0(s)\})$, which is zero because $\min\{(\Leb^d(A))^0, \gammat_0(s)\}=1$.
Thus, \eqref{eq:rdlbppp} has been proved.

It remains to prove Lemma~\ref{lem:gammatsat}.
To this end, we will need the following technical result.
\begin{lemma}
Let $A\subseteq \R^d$ be  a Borel set, $s> 0$,  and $c>0$. 
Then
\ba
& \int_{A} \exp\Big({-}s\min_{j=1}^{\kY }\min\{\norm{\xv-\yv_j}^2, c^2\}\Big)\, \mathrm{d}
\xv
\notag \\ & \rule{5mm}{0mm} 
\leq   e^{-s c^2}   \Leb^{d}(A)+ \kY  \bigg({-}e^{-s c^2} \Leb^d(U_c) 
+  \int_{U_c} e^{-s  \norm{\xv}^2} \mathrm{d}
\xv
\bigg) 
\label{eq:boundonefac}
\\*[-8mm] \notag 
\ea
for any point pattern  $Y=\{\yv_1, \dots, \yv_{\kY }\}$.
\end{lemma}
\begin{IEEEproof}
For 
$\xv\in A$, 
\ba\label{eq:splitmin}
\min_{j=1}^{\kY }\min\{\norm{\xv-\yv_j}^2, c^2\}
\!=\!
\begin{cases}
\norm{\xv-\yv_i}^2 & \text{if } \xv \in B_i \\
c^2 & \text{if } \xv \in A\setminus \bigcup_{i=1}^{\kY }B_i
\end{cases}
\ea 
where $B_i\subseteq A$, $i\in \{1, \dots,  \kY \}$ is given by
$
B_i\triangleq \big\{\xv\in A: \big(\norm{\xv-\yv_i} \leq c\big) \wedge \big(\norm{\xv-\yv_i}\leq \norm{\xv-\yv_j}\; \forall j\in \{1, \dots, \kY \}\setminus \{i\}\big)\big\}
$.
The sets $B_i$, $i\in \{1, \dots, \kY \}$ and $A\setminus \bigcup_{i=1}^{\kY }B_i$ are (up to intersections of measure zero) a partition of the set $A$.
Thus, we can rewrite the integral on the left-hand side of \eqref{eq:boundonefac} as the following sum of integrals:
\ba
&  \int_{A} \exp\Big({-}s\min_{j=1}^{\kY }\min\{\norm{\xv-\yv_j}^2, c^2\}\Big)\, \mathrm{d} 
\xv
\notag \\
& 
= \int_{A\setminus \bigcup_{i=1}^{\kY }B_i}\exp\Big({-}s \min_{j=1}^{\kY }\min\{\norm{\xv-\yv_j}^2, c^2\}\Big)\, \mathrm{d}
\xv
   \notag \\*
& 
\quad
+ \sum_{i=1}^{\kY }\int_{B_i} \exp\Big({-}s \min_{j=1}^{\kY }\min\{\norm{\xv-\yv_j}^2, c^2\} \Big)\, \mathrm{d}
\xv
\notag \\
& 
\stackrel{\hidewidth \eqref{eq:splitmin} \hidewidth }= \int_{A\setminus \bigcup_{i=1}^{\kY }B_i}e^{-s c^2} \mathrm{d}
\xv   
+ \sum_{i=1}^{\kY }\int_{B_i} e^{-s \norm{\xv-\yv_i}^2 } \mathrm{d}
\xv
\notag \\
& 
=  e^{-s c^2}  \Leb^{d}\bigg(A \setminus \bigcup_{i=1}^{\kY }B_i\bigg)
+ \sum_{i=1}^{\kY }\int_{B_i-\yv_i} e^{-s  \norm{\xv}^2} \mathrm{d}
\xv
\,. \label{eq:sepminintintosum}
\ea
We have $B_i\subseteq \big\{\xv\in A: \norm{\xv-\yv_i} \leq c\big\}$ and hence $B_i-\yv_i\subseteq \big\{\xv\in (A-\yv_i):  \norm{\xv} \leq c \} \subseteq \{\xv\in \R^d:\norm{\xv}\leq  c\}=U_c$;
furthermore, $e^{-s  \norm{\xv}^2}\geq e^{-s c^2}$ for all $\xv\in U_c$.
Thus,
\ba
& 
\int_{B_i-\yv_i} e^{-s  \norm{\xv}^2} \mathrm{d}
\xv 
\notag \\
& 
\stackrel{\hidewidth (a) \hidewidth}= \int_{B_i-\yv_i} e^{-s  \norm{\xv}^2}  \mathrm{d}
\xv
+ \int_{U_c\setminus (B_i-\yv_i)}  e^{-s c^2} \mathrm{d}
\xv 
\notag \\* & \quad 
 -   e^{-s c^2} \Leb^{d}\big(U_c\setminus (B_i-\yv_i)\big) \notag \\
& 
\leq \int_{B_i-\yv_i} e^{-s  \norm{\xv}^2}  \mathrm{d}
\xv
+ \int_{U_c\setminus (B_i-\yv_i)}  e^{-s  \norm{\xv}^2}  \mathrm{d}
\xv
\notag \\* & \quad 
  -   e^{-s c^2} \Leb^{d}\big(U_c\setminus (B_i-\yv_i)\big)  \notag \\
& 
= \int_{U_c} e^{-s  \norm{\xv}^2}  \mathrm{d}
\xv
 -  e^{-s c^2} \Leb^{d}\big(U_c\setminus (B_i-\yv_i)\big)  \label{eq:boundexparoundzero}
\ea
where in $(a)$ we added and subtracted $e^{-s c^2} \Leb^{d}\big(U_c\setminus (B_i-\yv_i)\big)$.
Inserting  \eqref{eq:boundexparoundzero} into  \eqref{eq:sepminintintosum}, we obtain
\ba
& \int_{A} \exp\Big(-s\min_{j=1}^{\kY }\min\{\norm{\xv-\yv_j}^2, c^2\}\Big)\, \mathrm{d}
\xv
 \notag \\
&  \leq  e^{-s c^2} \Bigg( \Leb^{d}\bigg(A\setminus \bigcup_{i=1}^{\kY }B_i\bigg) - \sum_{i=1}^{\kY }  \Leb^d\big(U_c\setminus(B_i-\yv_i)\big)
\Bigg)
\notag \\* & \quad 
+ \kY \int_{U_c}e^{-s  \norm{\xv}^2} \mathrm{d}
\xv
\notag \\
&  \stackrel{\hidewidth (a) \hidewidth}= e^{-s c^2} \Bigg( \Leb^{d}(A)- \sum_{i=1}^{\kY }\Leb^{d}(B_i) 
\notag \\* & \quad 
- \sum_{i=1}^{\kY } \big(\Leb^d(U_c)-\Leb^d(B_i)\big)
\Bigg)+ \kY \int_{U_c}e^{-s  \norm{\xv}^2} \mathrm{d}
\xv 
\notag \\
&  = e^{-s c^2}   \Leb^{d}(A)+ \kY  \bigg({-}e^{-s c^2} \Leb^d(U_c) 
+  \int_{U_c}e^{-s  \norm{\xv}^2} \mathrm{d}
\xv\bigg) \notag
\ea
where in $(a)$ we used $\Leb^{d}\big(A\setminus \bigcup_{i=1}^{\kY }B_i\big)=
\Leb^{d}(A)- \sum_{i=1}^{\kY }\Leb^{d}(B_i)$ and $\Leb^d(U_c\setminus(B_i-\yv_i))=\Leb^d(U_c)-\Leb^d(B_i-\yv_i)=\Leb^d(U_c)-\Leb^d(B_i)$.
\end{IEEEproof}

\begin{IEEEproof}[Proof of Lemma~\ref{lem:gammatsat}]
We first note that $\gammat_0(s)=1$ and thus in the case $k=0$, \eqref{eq:choosegammappp} is trivially satisfied.
It remains to show that for $k\in \N$ and for all $Y\in \sX$, 
\be\label{eq:gammattoshow}
\gammat_k(s)\geq 
 \int_{A^k} e^{ -s \dist_2^{  (c) }(\phi_k(\xv_{1:k}),Y)}  \mathrm{d}
\xv_{1:k}
\,.
\ee
To this end, we set  $Y=\{\yv_1, \dots, \yv_{\kY }\}$ and
consider the cases  $\kY \leq k$ and $\kY  >k$ separately.

\emph{Case $\kY \leq k$:}
Because $\phi_k(\xv_{1:k})=\{\xv_1, \dots, \xv_k\}$, we have
\ba
e^{-s \dist_2^{  (c) }(\phi_k(\xv_{1:k}),Y)}
& \stackrel{\hidewidth \eqref{eq:bounddist2a} \hidewidth }\leq \exp\bigg(\hspace{-0.3mm}{-}s\sum_{i=1}^{k}\min_{j=1}^{\kY }\min\{\norm{\xv_{i}-\yv_j}^2, c^2\}\hspace{-0.3mm}\bigg) 
\notag \\
& =
 \prod_{i=1}^{k}\exp\Big(\hspace{-0.3mm}{-}s\min_{j=1}^{\kY }\min\{\norm{\xv_{i}-\yv_j}^2, c^2\}\hspace{-0.3mm}\Big). \label{eq:boundexpsdist2}
\ea
Using \eqref{eq:boundexpsdist2}, we can bound the integral on the right-hand side of \eqref{eq:gammattoshow} as follows:
\ba
& \int_{A^k} e^{-s \dist_2^{  (c) }(\phi_k(\xv_{1:k}),Y)} \,\mathrm{d}
\xv_{1:k}
\notag \\ 
& \leq \int_{A^k} \prod_{i=1}^{k}\exp\Big({-}s\min_{j=1}^{\kY }\min\{\norm{\xv_{i}-\yv_j}^2, c^2\}\Big)\, \mathrm{d}
\xv_{1:k}
\notag \\
&= \bigg(\int_{A} \exp\Big({-}s\min_{j=1}^{\kY }\min\{\norm{\xv-\yv_j}^2, c^2\}\Big)\, \mathrm{d}
\xv
 \bigg)^k \notag \\
& \stackrel{\hidewidth \eqref{eq:boundonefac} \hidewidth}\leq  
\bigg(\hspace{-.5mm} e^{-s c^2}   \Leb^{d}(A)+ \kY  \bigg(\hspace{-.5mm}{-} e^{-s c^2} \Leb^d(U_c) 
+  \int_{U_c}\hspace{-.7mm}e^{-s  \norm{\xv}^2} \mathrm{d}
\xv
\bigg) \hspace{-.5mm} \bigg)^k
\hspace{-.5mm}.\label{eq:boundindpower}
\ea
Because $\norm{\xv}\leq c$ for $\xv \in U_c$, we have $ e^{-s c^2}\Leb^d(U_c) \leq \int_{U_c}e^{-s  \norm{\xv}^2} \mathrm{d}
\xv
$ and hence the  right-hand side in \eqref{eq:boundindpower} is monotonically increasing in $\kY $.
Thus, due to $\kY \leq k$, we can further upper-bound \eqref{eq:boundindpower} by
\ba
& \int_{A^k} e^{-s \dist_2^{  (c) }(\phi_k(\xv_{1:k}),Y)}\, \mathrm{d}
\xv_{1:k} 
\notag \\
&  \leq 
\hspace{-.5mm} \bigg(e^{-s c^2} \Leb^{d}(A)
+k\bigg(\hspace{-.5mm}{-}e^{-s c^2}\Leb^d(U_c) + \hspace{-.5mm}\int_{U_c}\hspace{-.6mm} e^{-s  \norm{\xv}^2}  \mathrm{d}
\xv\hspace{-.5mm}
\bigg) \hspace{-.5mm}\bigg)^k 
\notag \\ & \,
\stackrel{\hidewidth \eqref{eq:mygammappp2} \hidewidth}= \;\gammat_k(s) \notag
\ea
which is \eqref{eq:gammattoshow}.

\emph{Case $\kY > k$:}
We have
\ba
& e^{-s \dist_2^{  (c) }(\phi_k(\xv_{1:k}),Y)} \,
\notag \\
& \quad \stackrel{\hidewidth \eqref{eq:bounddist2b} \hidewidth}\leq  \,
e^{-s c^2 (\kY -k)} \, \exp\bigg({-}s\sum_{i=1}^{k}\min_{j=1}^{\kY }\min\{\norm{\xv_{i}-\yv_j}^2, c^2\}\bigg) \notag \\[-1mm]
&  \quad =  \, e^{-s c^2 (\kY -k)} \, \prod_{i=1}^{k}\exp\Big({-}s\min_{j=1}^{\kY }\min\{\norm{\xv_{i}-\yv_j}^2, c^2\}\Big)\,. \notag
\ea
Thus, we obtain
\ba
& \int_{A^k} e^{-s \dist_2^{  (c) }(\phi_k(\xv_{1:k}),Y)}  \mathrm{d}
\xv_{1:k}
 \notag \\
&  \leq e^{-s c^2 (\kY -k)} 
\notag \\* & \quad \times 
 \int_{A^k} \prod_{i=1}^{k}\exp\Big({-}s\min_{j=1}^{\kY }\min\{\norm{\xv_{i}-\yv_j}^2, c^2\}\Big)\, \mathrm{d}
\xv_{1:k}
\notag \\
& = e^{-s c^2 (\kY -k)} \bigg(\hspace{-0.5mm}\int_{A} \hspace{-0.5mm} \exp\Big({-}s\min_{j=1}^{\kY }\min\{\norm{\xv-\yv_j}^2, c^2\}\hspace{-0.5mm}\Big)  \mathrm{d}
\xv \hspace{-0.5mm}
 \bigg)^k\notag \\
&  \stackrel{\hidewidth \eqref{eq:boundonefac} \hidewidth}\leq e^{-s c^2 (\kY -k)}  \bigg(e^{-s c^2}   \Leb^{d}(A)
\notag \\* & \hspace{20mm}
+ \kY  \bigg({-}e^{-s c^2} \Leb^d(U_c) 
+  \int_{U_c}e^{-s  \norm{\xv}^2} \mathrm{d}
\xv
\bigg) \bigg)^k 
\notag \\
&  = \bigg( e^{-s c^2 \kY /k}  \Leb^{d}(A) 
+ e^{-s c^2 (\kY -k)/k}\,\kY 
\notag \\* & \hspace{11mm} \times
\bigg({-}e^{-s c^2} \Leb^d(U_c) + \int_{U_c} e^{-s  \norm{\xv}^2} \mathrm{d}
\xv 
\bigg) \bigg)^k \,.\label{eq:boundexpyconstky}
\ea
For $s c^2\geq 1$, the functions $e^{-s c^2 \kY /k}$ and $e^{-s c^2 (\kY -k)/k}\kY $ are monotonically decreasing in $\kY $.
Thus, recalling that $ e^{-s c^2}\Leb^d(U_c) \leq \int_{U_c}e^{-s  \norm{\xv}^2} \mathrm{d}
\xv
$,  the entire right-hand side in \eqref{eq:boundexpyconstky} is monotonically decreasing in $\kY $.
Hence, because $\kY > k$, we can further upper-bound \eqref{eq:boundexpyconstky} as
\ba 
& \int_{A^k} e^{-s \dist_2^{  (c) }(\phi_k(\xv_{1:k}),Y)}  \mathrm{d}
\xv_{1:k}
\notag \\
& \leq 
\hspace{-.5mm} \bigg(e^{-s c^2} \Leb^{d}(A)
+k\bigg(\hspace{-.5mm}{-}e^{-s c^2}\Leb^d(U_c) + \hspace{-.5mm}\int_{U_c}\hspace{-.6mm} e^{-s  \norm{\xv}^2} \mathrm{d}
\xv\hspace{-.5mm}
\bigg) \hspace{-.5mm}\bigg)^k \notag 
\ea
which is \eqref{eq:gammattoshow}.
\end{IEEEproof}

\section{Proof of Theorem~\ref{th:ubppp}} \label{app:proofubppp}
%
We will use   Corollary~\ref{cor:upperbounddpi}  with random vectors $(\rxv^{(k)}, \ryv^{(k)})$ that are given by the probability measure
\ba
& P_{\rxv^{(k)}, \ryv^{(k)}}(A_1 \times A_2) 
\notag \\
& \quad \triangleq 
\int_{A_1 } \int_{A_2}  g_{\rxv^{(k)}\mid\ryv^{(k)}}(\xv_{1:k}\condi \yv_{1:k}) \,\intd \lambda_{\rY}^k(\yv_{1:k})\, \intd \xv_{1:k} \notag
\ea
for $A_1, A_2 \subseteq (\R^d)^k$.
The marginal $\rxv^{(k)}$  is distributed according to the probability measure
\ba
P_{\rxv^{(k)}}(A_1)
& =
P_{\rxv^{(k)}, \ryv^{(k)}}(A_1 \times (\R^d)^k) \notag \\[1mm]
& =
\int_{A_1 } \int_{(\R^d)^k}  g_{\rxv^{(k)}\mid\ryv^{(k)}}(\xv_{1:k}\condi \yv_{1:k}) \,\intd \lambda_{\rY}^k(\yv_{1:k})\, \intd \xv_{1:k} \notag \\
& 
\stackrel{\hidewidth \eqref{eq:asscorrmarg} \hidewidth} =
\int_{A_1 } \Bigg( \prod_{i=1}^k g_{\rX}(\xv_i)\Bigg) \, \intd \xv_{1:k}\notag \\
& 
\stackrel{\hidewidth \eqref{eq:denspppk} \hidewidth}= 
P_{\rX}^{(k)}(A_1)\,. \notag
\ea
Thus, $\rxv^{(k)}$ has the same distribution as $\rxv_{\rX}^{(k)}$ and, hence, $\phi_k(\rxv^{(k)})$ has the same distribution as $\phi_k(\rxv_{\rX}^{(k)})$.
The assumption \eqref{eq:explconddistbound} is a specialization of \eqref{eq:assexbounded} for the case of a Poisson PP, USOSPA distortion,  and the vectors $(\rxv^{(k)}, \ryv^{(k)})$ constructed above.
Therefore, all assumptions in   Corollary~\ref{cor:upperbounddpi}  are satisfied and we obtain (see~\eqref{eq:rdboundvectors})
\ba 
R(D)
& \leq H(\card{\rX})
+ \sum_{k\in \N} p_{\card{\rX}}(k) \,
\Big( 
I\big(\rxv^{(k)}; \ryv^{(k)}\big)
\notag \\
& \quad 
- I\big(\rt^{(k)}_{\rxv}; \rY^{(k)} \bcondi \rX^{(k)}\big)
- I\big(\rxv^{(k)}; \rt^{(k)}_{\ryv} \bcondi \rY^{(k)}\big)
\Big) \notag \\
& \leq H(\card{\rX})
+ \sum_{k\in \N} p_{\card{\rX}}(k) \, 
I\big(\rxv^{(k)}; \ryv^{(k)}\big)\,. \label{eq:imuboundppp}
\ea
Here, omitting the conditional mutual informations does not loosen the bound if the function  $g_{\rxv^{(k)}\mid\ryv^{(k)}}$ is symmetric with respect to permutations, 
i.e., $g_{\rxv^{(k)}\mid\ryv^{(k)}}(\xv_{1:k}\condi \yv_{1:k})=g_{\rxv^{(k)}\mid\ryv^{(k)}}(\tau_{\xv}(\xv_{1:k})\condi \tau_{\yv}(\yv_{1:k}))$ for any permutations $\tau_{\xv}$ and $\tau_{\yv}$.
(Indeed,  it is easy to see that in this case the ordering of either $\rxv^{(k)}$ or $\ryv^{(k)}$ does not provide any information about the ordering of the respective other random variable.)
Because $\card{\rX}$ is given by a Poisson distribution, we have $p_{\card{\rX}}(k)= e^{-\expcard} \expcard^k / k!$ and 
\ba\label{eq:entboundxyn}
H(& \card{\rX}) 
\notag \\[-2mm]
& = - \sum_{k\in \N_0} \frac{e^{-\expcard} \expcard^k}{k!} \log \bigg( \frac{e^{-\expcard} \expcard^k}{k!} \bigg) \notag \\
& = \sum_{k\in \N_0} \frac{e^{-\expcard} \expcard^k}{k!} \big({-}\log e^{-\expcard} - \log \expcard^k + \log k! \big) \notag \\
& = \expcard \sum_{k\in \N_0} \frac{e^{-\expcard} \expcard^k}{k!} - \sum_{k\in \N} \frac{e^{-\expcard} \expcard^k k}{k!} \log \expcard + \sum_{k\in \N} \frac{e^{-\expcard} \expcard^k}{k!} \log k! \notag \\
& = \expcard - \expcard \log \expcard + \sum_{k\in \N} \frac{e^{-\expcard} \expcard^k}{k!} \log k! \,.
\ea
To derive the mutual informations $I\big(\rxv^{(k)}; \ryv^{(k)}\big)$, we first note that $\rxv^{(k)}$ is a continuous random vector and the same holds for $\rxv^{(k)}$ conditioned on $\ryv^{(k)}=\yv_{1:k}$. 
Thus, according to \cite[eq.~(8.48)]{Cover91}, we can calculate the mutual information as 
\ba
& I\big(\rxv^{(k)}; \ryv^{(k)}\big) 
\notag \\
& \quad  = h(\rxv^{(k)}) - \int_{(\R^d)^k}  h\big(\rxv^{(k)} \condi \ryv^{(k)}=\yv_{1:k}\big) \,\intd \lambda_{\rY}^k(\yv_{1:k}) \notag \\
& \quad  = k \,h(g_{\rX}) + \int_{(\R^d)^k} \int_{(\R^d)^k}  g_{\rxv^{(k)}\mid\ryv^{(k)}}(\xv_{1:k}\condi \yv_{1:k}) 
\notag \\* & \quad  \quad \times 
\log g_{\rxv^{(k)}\mid\ryv^{(k)}}(\xv_{1:k}\condi \yv_{1:k}) \, \intd \xv_{1:k}\,\intd \lambda_{\rY}^k(\yv_{1:k})\,. \label{eq:splitmutinfxyn}
\ea
Here, we used $h(\rxv^{(k)})=h(\rxv_{\rX}^{(k)})= k\, h(g_{\rX})$, which holds because of  \eqref{eq:denspppk} and because $\rxv^{(k)}$ has the same distribution as  $\rxv_{\rX}^{(k)}$. 
Inserting \eqref{eq:entboundxyn}, $p_{\card{\rX}}(k)= e^{-\expcard} \expcard^k / k!$, and \eqref{eq:splitmutinfxyn} into \eqref{eq:imuboundppp}
gives 
\ba
& R(D) 
\notag \\
& \leq
\expcard - \expcard \log \expcard + \sum_{k\in \N} \frac{e^{-\expcard} \expcard^k}{k!} \log k!
\notag \\ & \quad 
+ \sum_{k\in \N} \frac{e^{-\expcard} \expcard^k}{k!}\bigg(
k \,h(g_{\rX}) 
+ \hspace{-.5mm}\int_{(\R^d)^k} \hspace{-.5mm}\int_{(\R^d)^k} \hspace{-.7mm} g_{\rxv^{(k)}\mid\ryv^{(k)}}(\xv_{1:k}\condi \yv_{1:k}) 
\notag \\*[-2mm] & \hspace{23mm} \times 
\log g_{\rxv^{(k)}\mid\ryv^{(k)}}(\xv_{1:k}\condi \yv_{1:k}) \, \intd \xv_{1:k}\,\intd \lambda_{\rY}^k(\yv_{1:k})
\bigg)
\notag \\
& = 
\expcard - \expcard \log \expcard + \expcard \,h(g_{\rX})    
\notag \\ & \quad 
+ \sum_{k\in \N} \frac{e^{-\expcard} \expcard^k}{k!}\bigg(
 \log k! 
+ \int_{(\R^d)^k} \int_{(\R^d)^k}  g_{\rxv^{(k)}\mid\ryv^{(k)}}(\xv_{1:k}\condi \yv_{1:k}) 
\notag \\*[-2mm] & \hspace{23mm} \times 
\log g_{\rxv^{(k)}\mid\ryv^{(k)}}(\xv_{1:k}\condi \yv_{1:k}) \, \intd \xv_{1:k}\,\intd \lambda_{\rY}^k(\yv_{1:k})
\bigg) \notag
\ea
which is 
\eqref{eq:rdubppp}--\eqref{eq:condentgiven}.

\section{Lemmata for Example~\ref{ex:unifppp}} \label{app:proofexunifppp}
%
We consider the setting of Example~\ref{ex:unifppp}, i.e., $\rX$ is a Poisson PP  on $\R^2$ with intensity measure $\lambda=\expcard\Leb^2|_{[0,1)^2}$.
Furthermore,  
$N\geq 1/(\sqrt{2}c)$,
$\lambda_{\rY}$ is  defined by \eqref{eq:defyn},
and $g_{\rxv^{(k)}\mid\ryv^{(k)}}(\xv_{1:k}\condi \yv_{1:k})$ 
 by \eqref{eq:deftransprop}.

\begin{lemma}\label{lem:glemcor}
Equation \eqref{eq:concmargdist} is satisfied, i.e., 
\be  
\int_{(\R^2)^k}g_{\rxv^{(k)}\mid\ryv^{(k)}}(\xv_{1:k}\condi \yv_{1:k})\,\intd \lambda_{\rY}^k(\yv_{1:k})
= 
\prod_{i=1}^k \ind_{[0,1)^2}(\xv_i)\,.
\label{eq:exlemres}
\ee
\end{lemma}
\begin{IEEEproof}
Inserting \eqref{eq:defyn} and \eqref{eq:deftransprop} into the left-hand side of 
\eqref{eq:exlemres}, we obtain
\ba
& \int_{(\R^2)^k}g_{\rxv^{(k)}\mid\ryv^{(k)}}(\xv_{1:k}\condi \yv_{1:k})\,\intd \lambda_{\rY}^k(\yv_{1:k}) \notag \\
& =  \int_{\R^2} 
\dots
\int_{\R^2}  
g_{\rxv^{(k)}\mid\ryv^{(k)}}(\xv_{1:k}\condi \yv_{1:k}) \,\intd \lambda_{\rY}(\yv_{k}) \cdots
\intd \lambda_{\rY}(\yv_{1})     \notag \\
& = \bigg(\frac{1}{ N^{2}}\bigg)^k\sum_{j_1^{(1)}=1}^N \sum_{j_2^{(1)}=1}^N   
\dots
\sum_{j_1^{(k)}=1}^N \sum_{j_2^{(k)}=1}^N  
\frac{1}{k!}
\notag \\* & \hspace{35mm} \times
\sum_{\tau} \prod_{i=1}^k  N^2   \ind_{\squr_{j_1^{(\tau(i))}, j_2^{(\tau(i))}}} (\xv_i)   \notag \\
& =  \frac{1}{k!}\sum_{\tau} \prod_{i=1}^k \Bigg( \sum_{j_1=1}^N \sum_{j_2=1}^N
\ind_{\squr_{j_1 , j_2 }} (\xv_i)  \Bigg) \notag \\
& =  \prod_{i=1}^k \Bigg( \sum_{j_1=1}^N \sum_{j_2=1}^N
\ind_{\squr_{j_1 , j_2 }} (\xv_i) \Bigg) \notag \\
& \stackrel{\hidewidth (a) \hidewidth}= \prod_{i=1}^k  
\ind_{[0,1)^2} (\xv_i) \notag
\ea
where we used in $(a)$ that $\{\squr_{j_1 , j_2}\}_{j_1,j_2=1, \dots, N}$ is a partition of $[0,1)^2$.
\end{IEEEproof}

\begin{lemma}\label{lem:expecdistunifppp}
Equation \eqref{eq:expecdistunifppp} holds, i.e.,
\be \notag 
\sum_{k\in \N} \frac{e^{-\expcard}\expcard^k}{k!} \E\big[\dist_2^{  (c) }\big(\phi_k(\rxv^{(k)}), \phi_k(\ryv^{(k)})\big)\big]
 = \frac{\expcard}{6 N^{2}}\,.
\ee
\end{lemma}
\begin{IEEEproof}
%
Using \eqref{eq:defyn} and \eqref{eq:deftranspropalt}, we obtain
\ba
&\E\big[\dist_2^{  (c) }\big(\phi_k(\rxv^{(k)}), \phi_k(\ryv^{(k)})\big)\big] 
\notag \\
& = \int_{(\R^2)^k} \bigg(\frac{1}{ N^{2}}\bigg)^k\sum_{j_1^{(1)}=1}^N \sum_{j_2^{(1)}=1}^N   
\dots
\sum_{j_1^{(k)}=1}^N \sum_{j_2^{(k)}=1}^N 
\frac{N^{2k}}{k!} 
\notag \\* & \quad \times
\bigg(\sum_{\tau}   \ind_{\prod_{i=1}^k \squr_{j_1^{(\tau(i))}, j_2^{(\tau(i))}}} (\xv_{1:k})  \bigg)
\notag \\* & \quad \times
 \dist_2^{  (c) }\big(\phi_k(\xv_{1:k}),\phi_k\big(\qv_{j_1^{(1)}, j_2^{(1)}}, \dots, \qv_{j_1^{(k)}, j_2^{(k)}}\big)\big)  \, \intd 
\xv_{1:k}
 \notag \\
& = \sum_{j_1^{(1)}=1}^N \sum_{j_2^{(1)}=1}^N   
\dots
\sum_{j_1^{(k)}=1}^N \sum_{j_2^{(k)}=1}^N \frac{1}{k!}
\notag \\* & \quad \times
\sum_{\tau}
\int_{(\R^2)^k} 
  \ind_{\prod_{i=1}^k \squr_{j_1^{(\tau(i))}, j_2^{(\tau(i))}}} (\xv_{1:k})
\notag \\* 
& \quad \times
 \dist_2^{  (c) }\big(\phi_k(\xv_{1:k}),\phi_k\big(\qv_{j_1^{(1)}, j_2^{(1)}}, \dots, \qv_{j_1^{(k)}, j_2^{(k)}}\big)\big)    \, \intd 
\xv_{1:k}
\,.
\label{eq:explconddistboundspec}
\ea
Using the short-hand notation $\squr_{j_1,j_2}^{(\tau)}\triangleq\prod_{i=1}^k   \squr_{j_1^{(\tau(i))}, j_2^{(\tau(i))}}$, the integrals in this expression can be rewritten as
\ba
 & \int_{\squr_{j_1,j_2}^{(\tau)}}
 \dist_2^{  (c) }\big(\phi_k(\xv_{1:k}),\phi_k\big(\qv_{j_1^{(1)}, j_2^{(1)}}, \dots, \qv_{j_1^{(k)}, j_2^{(k)}}\big)\big)   \, \intd \xv_{1:k}
\notag \\
&  \stackrel{\hidewidth (a) \hidewidth }=  
\int_{\squr_{j_1,j_2}^{(\tau)}} 
\min_{\tau'} \sum_{i'=1}^k \min\big\{\bignorm{\xv_{i'}-\qv_{j_1^{(\tau'(i'))}, j_2^{(\tau'(i'))}}}^2, c^2\big\}   \, \intd \xv_{1:k}
\notag \\
&  \stackrel{\hidewidth (b) \hidewidth }=  
\int_{\squr_{j_1,j_2}^{(\tau)}} 
 \sum_{i'=1}^k \min\big\{\bignorm{\xv_{i'}-\qv_{j_1^{(\tau(i'))}, j_2^{(\tau(i'))}}}^2, c^2\big\}   \, \intd \xv_{1:k}
\notag \\
&  =  
\sum_{i'=1}^k \int_{\squr_{j_1,j_2}^{(\tau)}} 
 \min\big\{\bignorm{\xv_{i'}-\qv_{j_1^{(\tau(i'))}, j_2^{(\tau(i'))}}}^2, c^2\big\}   \, \intd \xv_{1:k}
\notag \\
&  =  
\sum_{i'=1}^k  
\int_{\prod\limits_{\substack{i=1\\ i\neq i'}}^k   \squr_{j_1^{(\tau(i))}, j_2^{(\tau(i))}}}  
\notag \\*
&   \quad \times 
\int_{\squr_{j_1^{(\tau(i'))}, j_2^{(\tau(i'))}}}
 \min\big\{\bignorm{\xv_{i'}-\qv_{j_1^{(\tau(i'))}, j_2^{(\tau(i'))}}}^2, c^2\big\}  \, \intd \xv_{i'}
\notag \\*
&   \quad \times 
  \, \intd 
	(\xv_1, \dots, \xv_{i'-1}, \xv_{i'+1}, \dots, \xv_k) 
\notag \\
&  =  
\sum_{i'=1}^k  \int_{\prod\limits_{\substack{i=1\\ i\neq i'}}^k   \squr_{j_1^{(\tau(i))}, j_2^{(\tau(i))}}} 
  1  \, \intd 
	(\xv_1, \dots, \xv_{i'-1}, \xv_{i'+1}, \dots, \xv_k)  
\notag \\
&   \quad \times \int_{\squr_{j_1^{(\tau(i'))}, j_2^{(\tau(i'))}}} 
 \min\big\{\bignorm{\xv_{i'}-\qv_{j_1^{(\tau(i'))}, j_2^{(\tau(i'))}}}^2, c^2\big\}   \, \intd \xv_{i'}
\notag \\
&   =  
\sum_{i'=1}^k \Bigg( \prod_{\substack{i=1\\ i\neq i'}}^k  \underbrace{ \Leb^2\big(\squr_{j_1^{(\tau(i))}, j_2^{(\tau(i))}}\big)}_{=1/N^2} \Bigg)
\notag \\* & \quad \times 
\int_{   \squr_{j_1^{(\tau(i'))}, j_2^{(\tau(i'))}}} 
 \min\big\{\bignorm{\xv_{i'}-\qv_{j_1^{(\tau(i'))}, j_2^{(\tau(i'))}}}^2, c^2\big\}   \, \intd  \xv_{i'}
\notag \\
&  =  
\bigg(\frac{ 1 }{ N^2 }\bigg)^{k-1}
\sum_{i'=1}^k  \int_{  [-\frac{1}{2N},\frac{1}{2N} )^2} 
 \min\big\{\norm{\xv_{i'} }^2, c^2\big\}   \, \intd  \xv_{i'}
\notag \\
&  \stackrel{\hidewidth (c) \hidewidth }=  
\frac{1}{N^{2(k-1)}}
\sum_{i'=1}^k  \int_{  [-\frac{1}{2N},\frac{1}{2N} )^2} 
  \norm{\xv_{i'} }^2   \, \intd  \xv_{i'}
\notag \\
&  =  
\frac{1}{N^{2(k-1)}}
  \frac{k }{6 N^4}
\notag \\
&  =  
\frac{k }{6 N^{2(k+1)}}\label{eq:oneint}
\ea
where $(a)$ is due to \eqref{eq:ospagen} (with $\krX =\krY$); 
$(b)$ holds because for $\xv_{i'}\in  \squr_{j_1^{(\tau(i'))}, j_2^{(\tau(i'))}}$, the closest $\qv_{j_1, j_2}$ for $j_1, j_2\in \{1, \dots, k\}$ is $\qv_{j_1^{(\tau(i'))}, j_2^{(\tau(i'))}}$, and thus  $\tau'=\tau$ is the minimizing permutation;
and $(c)$ holds because $\norm{\xv_{i'} }^2\leq  1/(2N^2)\leq c^2$ for $\xv_{i'}\in \big[-\frac{1}{2N},\frac{1}{2N} \big)^2$ (recall our assumption $N\geq 1/(\sqrt{2}c)$).
Inserting \eqref{eq:oneint} into \eqref{eq:explconddistboundspec}, we obtain
\ba
& \E\big[\dist_2^{  (c) }\big(\phi_k(\rxv^{(k)}), \phi_k(\ryv^{(k)})\big)\big] 
 \notag \\ & \qquad  
= \sum_{j_1^{(1)}=1}^N \sum_{j_2^{(1)}=1}^N   
\dots
\sum_{j_1^{(k)}=1}^N \sum_{j_2^{(k)}=1}^N \frac{1}{k!}\sum_{\tau}
\frac{k }{6 N^{2(k+1)}}
\notag \\ & \qquad
= \frac{k }{6 N^{2}}\label{eq:expdistfixk}
\ea
and inserting \eqref{eq:expdistfixk} into the left-hand side of \eqref{eq:expecdistunifppp} yields
\ba
\sum_{k\in \N} \frac{e^{-\expcard}\expcard^k}{k!} \E\big[\dist_2^{  (c) }\big(\phi_k(\rxv^{(k)}), \phi_k(\ryv^{(k)})\big)\big]
 & =  \frac{1}{6 N^{2}}\sum_{k\in \N} \frac{e^{-\expcard}\expcard^k}{k!} k
\notag \\ & 
 = \frac{\expcard}{6 N^{2}}\,.\hspace{16mm}\IEEEQEDhere
\notag
\ea
\end{IEEEproof}

\begin{lemma}\label{lem:condentbound}
Equation \eqref{eq:mutinfboundxyn} holds, i.e.,
$
h\big(\rxv^{(k)}\bcondi\ryv^{(k)}\big) 
 \geq  \binom{N^2}{k}\frac{k!}{N^{2k}} \log k!
- k \log N^{2} 
$.
\end{lemma}
\begin{IEEEproof}
We obtain from \eqref{eq:condent} and \eqref{eq:defyn}
\ba 
 h\big(\rxv^{(k)}\bcondi\ryv^{(k)}\big) 
& 
= 
\bigg(\frac{1}{ N^{2}}\bigg)^k
\sum_{j_1^{(1)}=1}^N \sum_{j_2^{(1)}=1}^N   
\dots
\sum_{j_1^{(k)}=1}^N \sum_{j_2^{(k)}=1}^N 
\notag \\* & \quad \times
h\big(\rxv^{(k)}\bcondi\ryv^{(k)}=\big(\qv_{j_1^{(1)}, j_2^{(1)}}, \dots, \qv_{j_1^{(k)}, j_2^{(k)}}\big)\big) \label{eq:negcondent}
\\*[-8mm] \notag 
\ea
where, by \eqref{eq:condentgiven} and \eqref{eq:deftranspropalt},
\ba
h\big(\rxv^{(k)}& \bcondi\ryv^{(k)}=\big(\qv_{j_1^{(1)}, j_2^{(1)}}, \dots, \qv_{j_1^{(k)}, j_2^{(k)}}\big)\big) 
\notag \\
& =
-\int_{(\R^2)^k} \bigg( \frac{ N^{2k}}{k!}\sum_{\tau} \ind_{ \prod_{i=1}^k \squr_{j_1^{(\tau(i))}, j_2^{(\tau(i))}}} (\xv_{1:k}) \bigg) \notag \\*
& \quad \times
\log \bigg(\frac{ N^{2k}}{k!}\sum_{\tau} \ind_{ \prod_{i=1}^k \squr_{j_1^{(\tau(i))}, j_2^{(\tau(i))}}} (\xv_{1:k})\bigg) \, \intd \xv_{1:k}\,.
\label{eq:condentspec}
\ea
We distinguish two cases:
If 
$\big(j_1^{(i)}, j_2^{(i)}\big)\neq\big(j_1^{(i')}, j_2^{(i')}\big)$ for all $i\neq i'$, then the sets $\squr_{j_1^{(\tau(i))}, j_2^{(\tau(i))}}$ are all disjoint.
This implies that the Cartesian products $\prod_{i=1}^k \squr_{j_1^{(\tau(i))}, j_2^{(\tau(i))}}$ are disjoint for different permutations $\tau$ and hence the probability density function in the differential entropy in \eqref{eq:condentspec} simplifies to 
\ba
&\frac{ N^{2k}}{k!}\sum_{\tau} \ind_{ \prod_{i=1}^k \squr_{j_1^{(\tau(i))}, j_2^{(\tau(i))}}} (\xv_{1:k})
\notag \\
& \quad
= 
\frac{ N^{2k}}{k!}   \ind_{\bigcup_{\tau}\prod_{i=1}^k \squr_{j_1^{(\tau(i))}, j_2^{(\tau(i))}}} (\xv_{1:k})  \notag
\ea
i.e., the probability density function of a uniform distribution on $\bigcup_{\tau}\prod_{i=1}^k \squr_{j_1^{(\tau(i))}, j_2^{(\tau(i))}}$.
Because a uniform distribution on a Borel set $A\subseteq \R^d$ has differential entropy $\log(\Leb^d(A))$ (see \cite[eq.~(8.2)]{Cover91}), the  differential entropy in \eqref{eq:condentspec} is given 
by
\ba
h\big(\rxv^{(k)}\bcondi\ryv^{(k)}=\big(& \qv_{j_1^{(1)}, j_2^{(1)}}, \dots, \qv_{j_1^{(k)}, j_2^{(k)}}\big)\big)
\notag \\
& =
\log \Bigg( (\Leb^2)^k\bigg( \bigcup_{\tau}    \prod_{i=1}^k \squr_{j_1^{(\tau(i))}, j_2^{(\tau(i))}} \bigg)\Bigg) \notag \\
& =
\log \Bigg( \sum_{\tau }(\Leb^2)^k\bigg( \prod_{i=1}^k \squr_{j_1^{(\tau(i))}, j_2^{(\tau(i))}} \bigg)\Bigg) \notag \\
& = \log \bigg(\sum_{\tau } \frac{1}{N^{2k}}\bigg) \notag \\
& = \log \bigg( \frac{k!}{N^{2k}}\bigg)\,. \label{eq:condentdistji}
\ea
On the other hand, if there exist indices $\big(j_1^{(i)}, j_2^{(i)}\big)=\big(j_1^{(i')}, j_2^{(i')}\big)$ for some $i\neq i'$, we can still bound the differential entropy in \eqref{eq:condentspec}. 
In fact, we can trivially upper-bound the corresponding probability density function in \eqref{eq:deftranspropalt} by $N^{2k}$
because $\sum_{\tau}   \ind_{\prod_{i=1}^k \squr_{j_1^{(\tau(i))}, j_2^{(\tau(i))}}} (\xv_{1:k})\leq k!$.
Using this bound in the argument of the logarithm in  \eqref{eq:condentspec}, we obtain
\ba
&  h\big(\rxv^{(k)}\bcondi\ryv^{(k)}=\big(\qv_{j_1^{(1)}, j_2^{(1)}}, \dots, \qv_{j_1^{(k)}, j_2^{(k)}}\big)\big) 
\notag \\
& \quad 
\geq 
-\int_{(\R^2)^k} \bigg( \frac{ N^{2k}}{k!}\sum_{\tau} \ind_{ \prod_{i=1}^k \squr_{j_1^{(\tau(i))}, j_2^{(\tau(i))}}} (\xv_{1:k}) \bigg) 
\notag \\*
& \hspace{20mm} \times
\log N^{2k} \, \intd \xv_{1:k}
\notag \\
& \quad =
- \log N^{2k}\,. \label{eq:condenteqji}
\ea

Inserting for all $\big(j_1^{(i)}, j_2^{(i)}\big)$, $i\in \{1, \dots, k\}$ either \eqref{eq:condentdistji} or \eqref{eq:condenteqji} into \eqref{eq:negcondent} 
(depending on whether the $\big(j_1^{(i)}, j_2^{(i)}\big)$ are pairwise distinct or not), we obtain
\ba
h\big(\rxv^{(k)}\bcondi\ryv^{(k)}\big) 
& 
\stackrel{\hidewidth (a) \hidewidth }\geq  
\frac{1}{N^{2k}}\Bigg( \binom{N^2}{k}k! \log \bigg( \frac{k!}{N^{2k}}\bigg)
\notag \\* & \quad 
-  \bigg(N^{2k} - \binom{N^2}{k}k!\bigg)  \log  N^{2k}  \Bigg) \notag \\
& = \frac{1}{N^{2k}}\Bigg( \binom{N^2}{k}k! \log k!- \binom{N^2}{k}k! \log N^{2k} 
\notag \\* & \quad 
- N^{2k} \log  N^{2k}  + \binom{N^2}{k}k!  \log  N^{2k}  \Bigg) \notag \\
& =  \frac{\binom{N^2}{k}k!}{N^{2k}} \log k!
- k \log  N^{2} 
\notag 
\ea
where we used in $(a)$ that there are $\binom{N^2}{k}k!$  choices of pairwise distinct $\big(j_1^{(i)}, j_2^{(i)}\big)$ and, hence, $N^{2k} - \binom{N^2}{k}k!$  choices of  $\big\{\big(j_1^{(i)}, j_2^{(i)}\big)\big\}_{i=1, \dots, k}$ where there exist indices $\big(j_1^{(i)}, j_2^{(i)}\big)=\big(j_1^{(i')}, j_2^{(i')}\big)$ for some $i\neq i'$.
\end{IEEEproof}

\begin{lemma}\label{lem:boundrdnusixn}
For $\widetilde{N}\in \N$ with $\widetilde{N}\leq N$, we have
\ba 
\sum_{k\in \N} \frac{e^{-\expcard} \expcard^k}{k!} & \big(
\log k!
+ k \log N^{2} 
\big)  -  \sum_{k=1}^{\widetilde{N}^2}  \frac{e^{-\expcard} \expcard^k \binom{N^2}{k} }{N^{2k}} \log k! \notag \\[-1mm] 
& \leq
  \expcard \log N^2  + \sum_{k=1}^{\widetilde{N}^2} e^{-\expcard} \expcard^k \log k! \bigg(\frac{1}{k!}- \frac{\binom{N^2}{k}}{N^{2k}}\bigg)
\notag \\* & \quad 
 + \bigg(1-\sum_{k=0}^{\widetilde{N}^2-2} \frac{e^{-\expcard} \expcard^k}{k!} \bigg) \expcard^2\,. \label{eq:boundrdnusixn}
\ea
\end{lemma}
\begin{IEEEproof}
We have
\ba
 \sum_{k\in \N}  &\frac{e^{-\expcard} \expcard^k}{k!} \big(
\log k!
+ k \log N^{2} 
\big)  -  \sum_{k=1}^{\widetilde{N}^2}  \frac{e^{-\expcard} \expcard^k \binom{N^2}{k} }{N^{2k}} \log k!  \notag \\
&   =  \, \sum_{k\in \N} \frac{e^{-\expcard} \expcard^k}{k!} k \log N^{2}  
 + \sum_{k=1}^{\widetilde{N}^2} \frac{e^{-\expcard} \expcard^k}{k!}  \log k!  
\notag \\* & \quad 
 + \sum_{k=\widetilde{N}^2+1}^{\infty} \frac{e^{-\expcard} \expcard^k}{k!} \log k!  -  \sum_{k=1}^{\widetilde{N}^2}  \frac{e^{-\expcard} \expcard^k \binom{N^2}{k} }{N^{2k}} \log k!  \notag \\
&  \stackrel{\hidewidth (a) \hidewidth }=   \expcard \log  N^2  + \sum_{k=1}^{\widetilde{N}^2} e^{-\expcard} \expcard^k \log k! \bigg(\frac{1}{k!}- \frac{\binom{N^2}{k}}{N^{2k}}\bigg)
\notag \\* & \quad 
 + \sum_{k=\widetilde{N}^2+1}^{\infty} \frac{e^{-\expcard} \expcard^k}{k!} \log k!  \label{eq:rateubunifprep}
\ea  
where $(a)$ holds because $\sum_{k\in \N} \frac{e^{-\expcard} \expcard^k}{k!} k=  \expcard$.
The infinite sum on the right-hand side of \eqref{eq:rateubunifprep} can be bounded by
\ba
 \sum_{k=\widetilde{N}^2+1}^{\infty} \frac{e^{-\expcard} \expcard^k}{k!} \log k!
& \stackrel{\hidewidth (a) \hidewidth }\leq  \sum_{k=\widetilde{N}^2+1}^{\infty} \frac{e^{-\expcard} \expcard^k}{(k-1)!} \log k \notag \\
& \stackrel{\hidewidth (b) \hidewidth }\leq   \sum_{k=\widetilde{N}^2+1}^{\infty} \frac{e^{-\expcard} \expcard^k}{(k-2)!} \notag \\
&  =   \sum_{k=0}^{\infty} \frac{e^{-\expcard} \expcard^{k+2}}{k!}- \sum_{k=0}^{\widetilde{N}^2-2} \frac{e^{-\expcard} \expcard^{k+2}}{k!} \notag \\
& =  \bigg(1-\sum_{k=0}^{\widetilde{N}^2-2} \frac{e^{-\expcard} \expcard^k}{k!} \bigg) \expcard^2 \label{eq:boundinfsum}
\ea
where $(a)$ holds because $\log k!\leq k \log k$ and $(b)$ holds because $\log k \leq k-1$.
Inserting \eqref{eq:boundinfsum} into \eqref{eq:rateubunifprep}, we obtain
\eqref{eq:boundrdnusixn}.
\end{IEEEproof}

%

\section*{Acknowledgment} 

We would like to thank the Associate Editor, Prof.\ David L.\ Neuhoff,  for suggesting the lower bound in Theorem~\ref{th:lbspecific}.
We would also like to thank the anonymous reviewers,
whose comments helped us improve  our results and their presentation.





\vspace{-15mm}
\begin{IEEEbiographynophoto}{G\"unther Koliander}
 received the Master degree in Technical Mathematics
(with distinction) in 2011 and the PhD degree in Electrical Engineering (with
distinction) in 2015 from TU Wien, Vienna, Austria. 
From 2015 to 2017, he was a postdoctoral researcher with the Institute of Telecommunications, TU Wien, Vienna, Austria. 
He twice held visiting researcher positions
at Chalmers University of Technology, Gothenburg, Sweden. 
Since 2017, he has been a postdoctoral researcher with the Acoustics Research Institute, Austrian Academy of Sciences, Vienna, Austria.
His research
interests are in the areas of  information theory,
geometric measure theory, and point processes.
\end{IEEEbiographynophoto}
\vspace{-15mm}
\begin{IEEEbiographynophoto}{Dominic Schuhmacher}
received the Diploma and Ph.D. degrees in mathematics from the University of Z\"urich, Switzerland, in 2000 and 2005, respectively, 
and the Habilitation degree in stochastics from the University of Bern, Switzerland, in 2013. 
He is currently a full professor of stochastics at the University of G\"ottingen, Germany.
His research interests include point process theory, spatial statistics, Stein's method for distributional approximation, and computational methods for spatial problems, including optimal transport.
\end{IEEEbiographynophoto}

\begin{IEEEbiographynophoto}{Franz~Hlawatsch}(S'85--M'88--SM'00--F'12) received the Diplom-Ingenieur, Dr. techn., and Univ.-Dozent (habilitation) degrees in electrical engineering/signal processing from TU Wien, Vienna, Austria in 1983, 1988, and 1996, respectively. 
Since 1983, he has been with the Institute of Telecommunications, TU Wien, where he is currently an Associate Professor. 
During 1991--1992, as a recipient of an Erwin Schr\"odinger Fellowship, he spent a sabbatical year with the Department of Electrical Engineering, University of Rhode Island, Kingston, RI, USA. 
In 1999, 2000, and 2001, he held one-month Visiting Professor positions with INP/ENSEEIHT, Toulouse, France and IRCCyN, Nantes, France. 
He (co)authored a book, three review papers that appeared in the {\sc IEEE Signal Processing Magazine}, about 200 refereed scientific papers and book chapters, and three patents. 
He coedited three books. 

Dr. Hlawatsch was a member of the IEEE SPCOM Technical Committee from 2004 to 2009. 
He was a Technical Program Co-Chair of EUSIPCO 2004 and served on the technical committees of numerous IEEE conferences. 
He was an Associate Editor for the {\sc IEEE Transactions on Signal Processing} from 2003 to 2007, for the {\sc IEEE Transactions on Information Theory} from 2008 to 2011, and for the {\sc IEEE Transactions on Signal and Information Processing over Networks} from 2014 to 2017. 
He coauthored papers that won an IEEE Signal Processing Society Young Author Best Paper Award and a Best Student Paper Award at IEEE ICASSP 2011. 
His research interests include statistical and compressive signal processing methods and their application to sensor networks and wireless communications.
\end{IEEEbiographynophoto}

\end{document}